\documentclass[11pt]{article}
\usepackage[utf8]{inputenc}
\usepackage[T1]{fontenc}
\pdfoutput = 1
\usepackage{dsdshorthand}
\usepackage{myoptions}
\usepackage{upgreek}

\usepackage{amsmath}
\usepackage{amssymb}
\usepackage{bbm}
\usepackage{bbold}
\usepackage{esint}
\usepackage{braket}
\allowdisplaybreaks
\usepackage{physics}
\usepackage{caption}
\usepackage{cite}
\usepackage{color}
    \definecolor{darkgreen}{rgb}{0,0.5,0}
    \definecolor{darkred}{rgb}{0.5,0,0}
    \definecolor{darkblue}{rgb}{0,0,0.6}
    \definecolor{purple}{rgb}{0.4,.2,0.7}
\usepackage[mathcal]{eucal}
\usepackage{float}
\usepackage{graphicx}

\usepackage{tikz}
\usetikzlibrary{arrows,positioning} 
\tikzset{
    >=stealth',
    punkt/.style={
           rectangle,
           rounded corners,
           draw=black, very thick,
           text width=15em,
           minimum height=2em,
           text centered},
    pil/.style={
           ->,
           thick,
           shorten <=2pt,
           shorten >=2pt,}
}

\usepackage[margin = 2.2cm]{geometry}
\renewcommand{\d}{\mathrm{d}}
\renewcommand{\i}{\mathrm{i}}
\newcommand{\Ss}{\textsf{S}_0}
\newcommand{\Hh}{\textsf{H}_0}

\newcommand{\average}[1]{\left\langle #1 \right\rangle}

\def\nn{\nonumber}

\def\ba{\begin{array}}
\def\ea{\end{array}}

\def\td{\tilde}

\def\dalemb#1#2{{\vbox{\hrule height .#2pt
      \hbox{\vrule width.#2pt height#1pt \kern#1pt
              \vrule width.#2pt}
      \hrule height.#2pt}}}

\def\cont{{{\mathcal{C}}}}
\def\mo{{{\mathcal{O}}}}

\def\rconn{\color{red}{\text{conn.}}}

\def\bn{\boldsymbol{n}}

\let\tilde=\widetilde

\def\brane{\text{brane}}

\def\S{\textsf{S}_0}
\def\rL{{\color{red}\langle}}
\def\rR{{\color{red}\rangle}}
\renewcommand{\d}{\mathrm{d}}
\renewcommand{\i}{\mathrm{i}}
\renewcommand{\tilde}{\widetilde}

\numberwithin{equation}{section}

\begin{document}

\thispagestyle{empty}
\begin{center}
    ~\vspace{5mm}
    
    {\LARGE \bf Gravity factorized  
    }\\

    \vspace{0.4in}
    
    {\bf Andreas Blommaert$^1$, Luca V. Iliesiu$^2$ and Jorrit Kruthoff$\,^2$}

    \vspace{0.4in}
    {$^1$SISSA, Via Bonomea 265, 34127 Trieste, Italy\\$^2$Department of Physics, Stanford University, Stanford, CA 94305, USA}
    \vspace{0.1in}
    
    {\tt ablommae@sissa.it, liliesiu@stanford.edu, kruthoff@stanford.edu}
\end{center}

\vspace{0.4in}

\begin{abstract}
\noindent 
We find models of two dimensional gravity that resolve the factorization puzzle and have a discrete spectrum, whilst retaining a semiclassical description. A novelty of these models is that they contain non-trivially correlated spacetime branes or, equivalently, nonlocal interactions in their action. Such nonlocal correlations are motivated in the low-energy gravity theory by integrating out UV degrees of freedom. Demanding factorization fixes almost all brane correlators, and the exact geometric expansion of the partition function collapses to only two terms: the black hole saddle and a subleading ``half-wormhole'' geometry, whose sum yields the desired discrete spectrum. By mapping the insertion of correlated branes to a certain double-trace deformation in the dual matrix integral, we show that factorization and discreteness also persist non-perturbatively.  
While in our model all wormholes completely cancel, they are still computationally relevant: self-averaging quantities, like the Page curve, computed in the original theory with wormholes, accurately approximate observables in our theory, which accounts for UV corrections. Our models emphasize the importance of correlations between different disconnected components of spacetime, providing a possible resolution to the factorization puzzle in any number of dimensions.
\end{abstract}

\pagebreak
\setcounter{page}{1}
\tableofcontents

\newpage

\section{Introduction}
\label{sec:intro}
Understanding the duality between black holes and conventional quantum mechanical systems remains an important problem in quantum gravity. The idea behind this holographic program \cite{tHooft:1993dmi, Susskind:1994vu, Maldacena:1997re} is to consider known features of quantum mechanical systems, and to determine whether black holes follow similar rules. Much progress has been made in this direction in recent years by studying non-perturbative wormhole contributions to the Euclidean gravitational path integral. For example, trying to replicate the Page curve \cite{Page:1993df} has thought us the importance of considering spacetime wormhole contributions in the gravitational path integral in order to find a unitary process of black hole evaporation \cite{Almheiri:2019qdq,Penington:2019kki, Almheiri:2020cfm}.

Whilst the existence of spacetime wormholes has shed light on some important problems in quantum gravity, it has also introduced puzzles. The goal of this paper is to address two such puzzles: 
\begin{enumerate}
    \item \textbf{The  factorization puzzle.} The existence of wormholes in quantum gravity immediately raises a tension with the dual quantum theory \cite{Maldacena:2004rf}. In the dual theory, when we want to calculate moments of the partition function $Z(\b)$, we trivially find the product of partitions functions, thus obtaining a factorizing answer. On the bulk side, there is an asymptotic boundary for each $Z(\b)$ and we are instructed to integrate over all spacetimes consistent with these boundary conditions. This includes spacetimes with wormholes that connect different asymptotic boundaries, which results in a non-factorizing answer $Z(\beta_1,\beta_2)\neq Z(\beta_1)Z(\beta_2)$.\footnote{In supergravity there are exceptions where the wormhole contribution to various supersymmetry-protected observables happens to vanish \cite{Iliesiu:2021are}. } Thus the bulk and the boundary appear to be in disagreement.
    
    \item \textbf{The discreteness puzzle.} In known models in which the sum over bulk geometries can be computed exactly, the gravitational partition function does not produce a discrete energy spectrum.\footnote{Here, one should distinguish two toy models in which the geometric expansion is (at least somewhat) under control. The first is JT gravity where one can include the contribution of all wormholes (and geometries) to still find a continuous density of states for the single-boundary partition function. The second is pure 3d gravity, where all black hole geometries can be included, but some more complicated topologies have yet to be accounted for \cite{Maloney:2007ud,Cotler:2020ugk,Maxfield:2020ale}. Both models appear to have continuous spectra.}  A consequence of this lack of discreteness can be observed for  boundary correlation functions. Correlation functions in the dual quantum mechanics with a discrete spectrum oscillate heavily around a non-zero averaged value \cite{Maldacena:2001kr, Cotler:2016fpe}. Connected geometries and in particular spacetime wormholes explain this non-zero averaged value \cite{Saad:2018bqo,Saad:2019pqd,Blommaert:2019hjr,Blommaert:2020seb} but do not explain the erratic oscillations in the dual quantum mechanics. So again, the bulk and the boundary appear to disagree.
\end{enumerate}

One way around both puzzles is to interpret gravity as dual to an ensemble of quantum mechanical systems, instead of a single boundary theory \cite{Saad:2019lba,Saad:2019pqd,Almheiri:2019qdq,Penington:2019kki,Marolf:2020xie,Stanford:2020wkf,Blommaert:2019wfy,Blommaert:2020seb,Pollack:2020gfa,Afkhami-Jeddi:2020ezh,Maloney:2020nni,Belin:2020hea,Cotler:2020ugk,Anous:2020lka,Chen:2020tes,Liu:2020jsv,Marolf:2021kjc,Meruliya:2021utr,Giddings:2020yes,Stanford:2019vob,Okuyama:2019xbv,Belin:2020jxr,Verlinde:2021jwu,Collier:2021rsn,Betzios:2021fnm,Belin:2021ryy,Saad:2021uzi}. Simple models of gravity, like JT gravity and generalizations thereof, and pure three-dimensional gravity or even supergravity, are all seemingly dual to ensembles.

However these simple models are not realistic theories of the universe.
 There is a widely held believe that all UV complete theories of quantum gravity in AdS do factorize, and that they are each dual to one discrete quantum mechanical systems. For example the Strominger-Vafa microstate counting \cite{Strominger:1996sh} is evidence of the discreteness of the spectrum, and recent work of Eberhardt \cite{Eberhardt:2018ouy, Eberhardt:2020bgq, Eberhardt:2021jvj} addresses the factorization puzzle in a particular tensionless AdS$_3$ string theory. Furthermore, in AdS/CFT, starting from the boundary perspective, in conventional CFT examples whose bulk duals are better understood (such as $\mathcal N=4$ super Yang-Mills) it is unclear what couplings one could even average over in order to get answers consistent with multi-boundary wormholes.  

We would like to understand how factorization and discreteness arise in the UV, by modifying the tractable toy models discussed above. For this, we consider JT gravity enriched with additional effective UV ingredients. 

Suppose we start with some UV complete theory of quantum gravity that has a low energy JT sector. For instance, we can consider one realization of the SYK model \cite{Saad:2021rcu}  which, in its full glory, has string corrections in the bulk \cite{Maldacena:2016hyu}. Alternatively, we can consider the AdS$_2$ near-horizon region of a near-extremal black hole in a higher dimensional UV complete theory, for which JT gravity is a good effective theory \cite{Almheiri:2016fws,Nayak:2018qej,Castro:2018ffi,Moitra:2019bub,Sachdev:2019bjn,Yang:2018gdb,Iliesiu:2020qvm, Heydeman:2020hhw} but string corrections need to be considered in order to determine the full spectrum of black hole microstates. Now, instead of making any approximation that results in JT, suppose one can integrate out exactly all the degrees of freedom, except for the metric and dilaton. For low enough temperatures one should find JT gravity but with complicated deformations in the action. Those deformations would include deformations in the dilaton potential like those considered in \cite{Mertens:2019tcm,Maxfield:2020ale,Witten:2020wvy}, but, in general, one also expects to get nonlocal deformations, such as nonlocal dilaton potential terms. We can then apply a form of open-closed string duality (but applied to spacetimes) where we interpret the deformed dilaton gravity as JT gravity with spacetime branes \cite{Marolf:2020xie} inserted.\footnote{For example, JT gravity with FZZT branes (open string picture) is equivalent to JT gravity with a deformed potential (closed string picture) \cite{Blommaert:2021gha}. } Similarly, nonlocal deformations turn out to correspond to inserting correlated (or coupled) spacetime branes in JT theory.\footnote{As compared to \cite{JafferisEOW}, who also introduced branes with the motivation to  address the factorization puzzle, the critical new ingredient that we introduced, and which actually result in a factorizing theory of gravity, is precisely  this nontrivial coupling between the spacetime branes.}

The key point is that in the above thought experiment we have made no approximation whatsoever. Therefore, if it is true that UV theories do factorize and are discrete, then there must be versions of JT gravity with correlated spacetime branes or nonlocal corrections in the action which also factorize, and have a discrete energy spectrum. Because UV complete theories are quite scarce and special, we only expect this to be true for very specific values for the correlation between spacetime branes and equivalently, for the nonlocal deformations of the action.

\subsection*{Summary and structure}
We thus set out to find spacetime brane correlators which result in a factorizing and discrete boundary dual. Our main results are the following:

\begin{enumerate}
\item In \textbf{section \ref{sect:branes}}, we introduce JT gravity with correlated spacetime branes and explain why these correlated branes are equivalent to nonlocal deformations in the action which, in turn, can arise when integrating-out UV degrees of freedom. We spell out the precise rules for including the contribution of (correlated) branes to the gravitational partition function. The point is there are additional boundaries for spacetimes to end on, and there can be correlation between those additional boundaries. For example, the following surfaces are a representative sample for the contribution to the partition function
\be 
Z(\beta) = \;\; \begin{tikzpicture}[baseline={([yshift=5.6ex]current bounding box.center)}, scale=0.7]
 \pgftext{\includegraphics[scale=0.55]{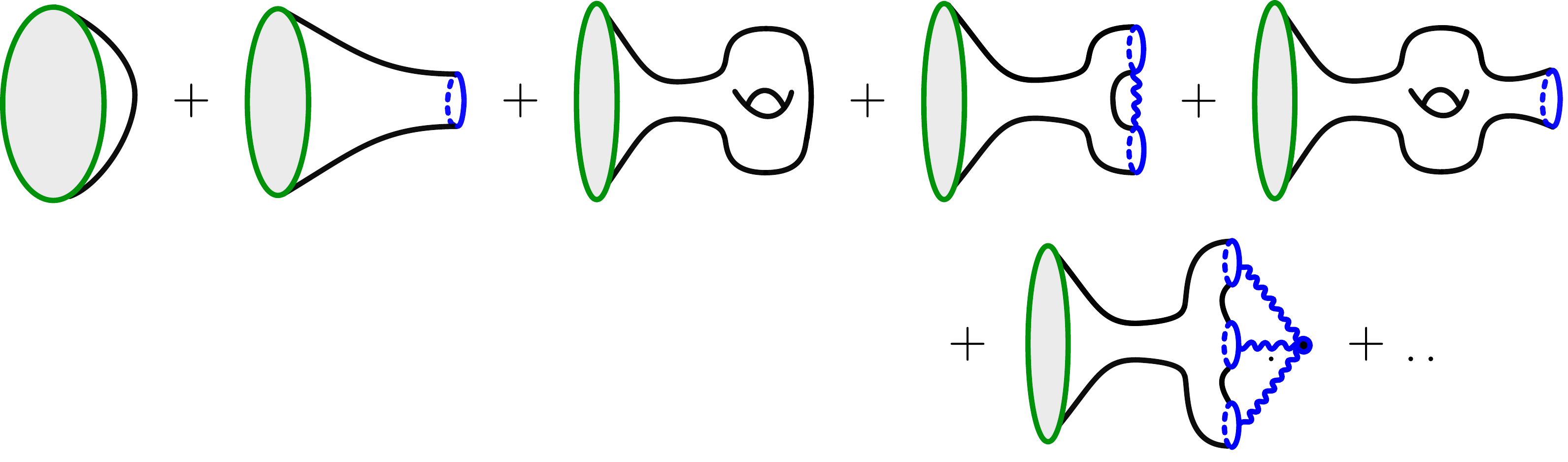}} at (0,0);
  \end{tikzpicture} 
  \label{eq:brane-example-intro}
\ee
where the blue wiggly lines represent the correlations (or couplings) between the additional brane boundaries.

    \item Factorization happens because the brane interactions can cancel the contribution of wormholes. For instance, in \textbf{section \ref{sect:allorders}} we find that the brane two-point interaction must be tuned such that
    \begin{equation}
    \label{eq:factorization-leading-order}
         \begin{tikzpicture}[baseline={([yshift=-.5ex]current bounding box.center)}, scale=0.7]
 \pgftext{\includegraphics[scale=0.55]{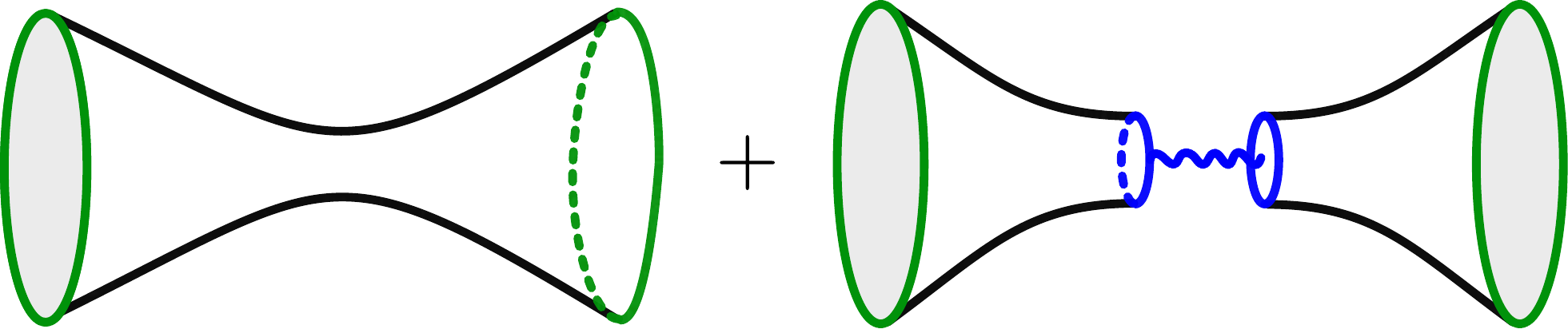}} at (0,0);
  \end{tikzpicture} \,\,=\,\, 0\,.
    \end{equation}
    This phenomenon was also observed by Saad, Shenker, Stanford and Yao \cite{Saad:2021rcu,Saad:2021uzi} who investigated factorization to leading order in the genus expansion.\footnote{Leading order factorization was also discussed in \cite{Mukhametzhanov:2021hdi}.} In our setup, this brane two-point function implies factorization to all orders in the genus expansion and for an arbitrary number of boundaries. Additionally, we show that three or higher-point interactions between branes need to vanish. 
 In the UV setting, these brane interactions or more generally, the interactions between disconnected components of the spacetime, can be thought of as highly stringy, non-geometric connected configurations, non-geometric wormholes. These cancel the geometric wormholes so that the full answer factorizes. 
    \item The brane two-point interaction \eqref{eq:factorization-leading-order} results in massive cancellations in the $e^{\Ss}$ expansion. The only contributions to the one-boundary gravitational partition function which do not cancel, are the disk, and the geometry with precisely one brane boundary inserted
    \begin{equation}
        Z(\beta) = \Tr e^{-\beta \Hh}=
  \begin{tikzpicture}[baseline={([yshift=1.7ex,xshift=0ex]current bounding box.center)}, scale=0.7 ]
 \pgftext{\includegraphics[scale=0.45]{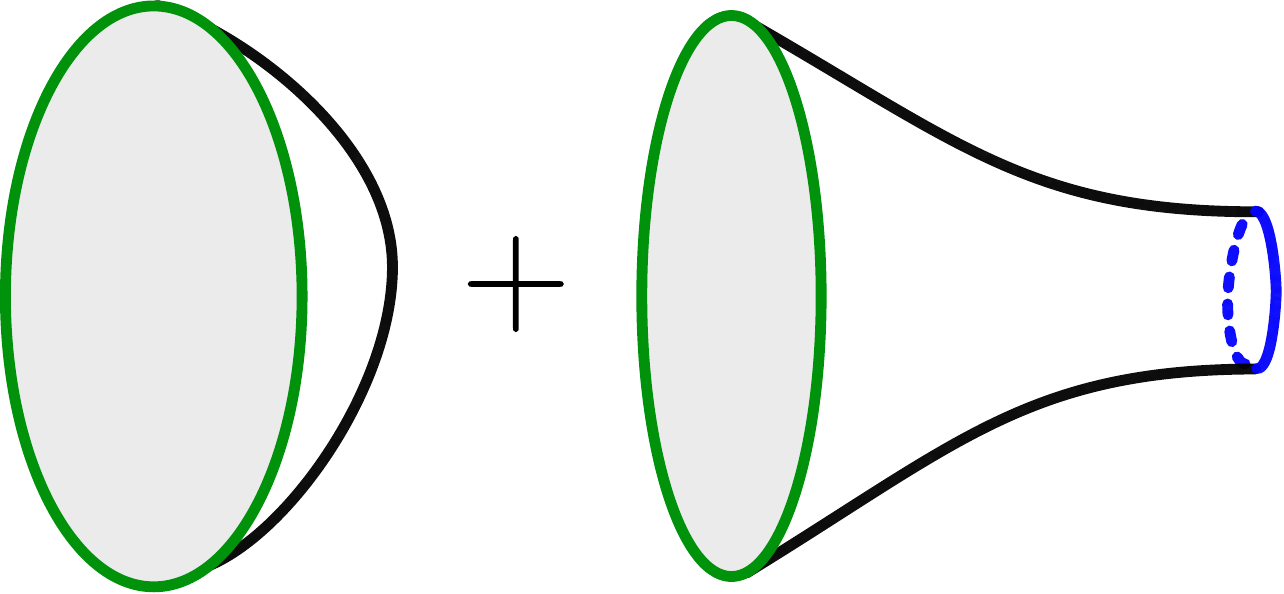}} at (0,0);
     \draw (-1.9, -2) node {black hole};
         \draw (2.3, -2) node {one-point function};
  \end{tikzpicture}\,.\label{simple}
    \end{equation}
The one-point function of this brane encodes the microstructure of the dual quantum mechanics: each Hamiltonian $\Hh$  determines  the one-point function such that the above equation is exactly satisfied.
As shown in \textbf{section \ref{sect:onepoint}}, this answer is very similar to the half-wormhole picture \cite{Saad:2021rcu}, with that crucial distinction that \eqref{simple} is exact, to all perturbative and non-perturbative orders in $e^{\S}$. For typical draws of $\Hh$, the contribution of the half-wormhole in \eqref{simple} is sub-leading compared to the black hole. This is the reason why the classical black hole is oftentimes a good approximation. The partition function in \eqref{simple}  can be viewed as analogous to the whole expansion of strings on the disk background. This includes configurations that look like classical geometries (with wormholes) but also includes highly non-geometric configurations, some of which again cancel the geometric wormholes.

\item Wormholes re-emerge when we average over $\Hh$, since they are encoded in the statistical properties of the brane one-point function. Denoting the ensemble average over $\Hh$ by $\rL \dots \rR_{\rconn}$ we find that
\be 
\label{eq:red-average-intro}
\begin{tikzpicture}[baseline={([yshift=-.5ex]current bounding box.center)}, scale=0.6]
 \pgftext{\includegraphics[scale=0.7]{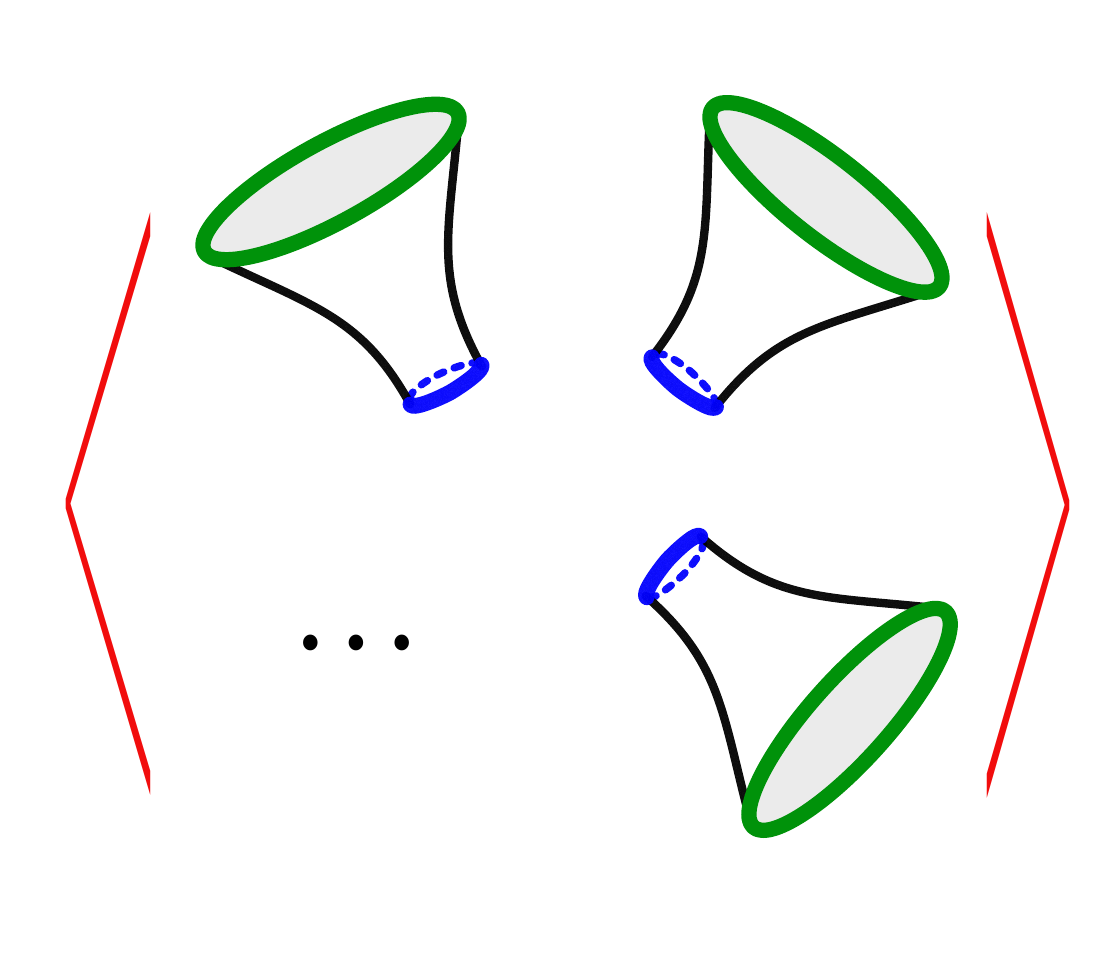}} at (0,0);
     \draw (3.8,-2.5) node  {$\color{red}{\text{conn}}$};
  \end{tikzpicture} \quad = \quad 
  \begin{tikzpicture}[baseline={([yshift=-.5ex]current bounding box.center)}, scale=0.6]
 \pgftext{\includegraphics[scale=0.7]{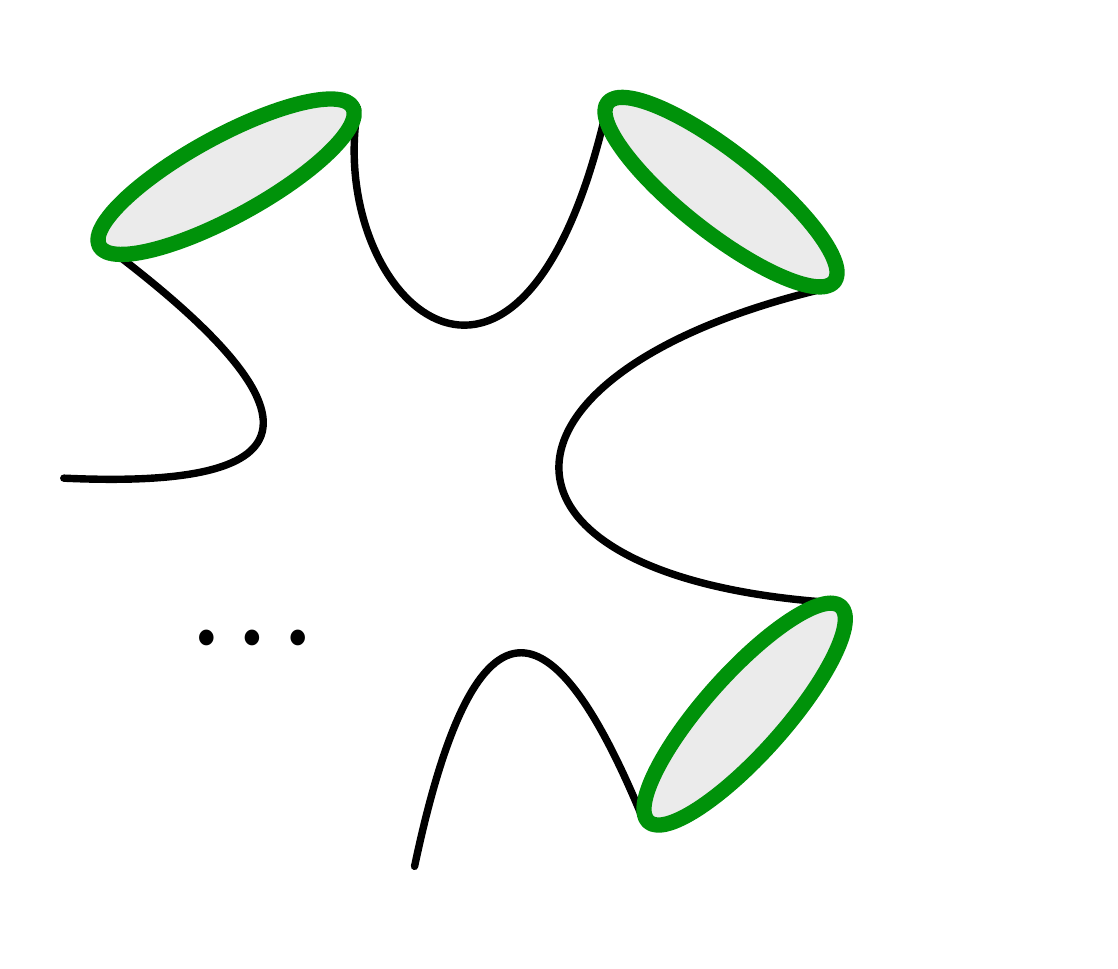}} at (0,0);
  \end{tikzpicture}  + \quad \text{higher genus}\,.
\ee 
This equality shows that the expectation value of self-averaging observables\footnote{Self-averaging quantities are defined by having variances that are much smaller than their expectation values.} in the theory \eqref{simple} (where wormholes cancel), are well approximated by computations in the original JT gravity theory, where branes are absent but wormholes are present. In \textbf{section \ref{sect:recovering} and \ref{sect:matter}}, we exemplify how this occurs for some known self-averaging quantities: the time-averaged spectral form factor and matter two-point function, the volume of the black hole interior, and the entanglement entropy of Hawking radiation (which follows the Page curve). 
    
\item In the dual matrix integral description of JT gravity \cite{Saad:2019lba}, we use the map between FZZT branes and determinants to investigate the effect of our correlated branes. We find that the branes deform the ensemble partition function to
    \begin{align}
        \mathcal{Z}(\Hh)&=\lim_{q\to\infty}\prod_{i=1}^L\int_\cont \d\l_i\,\exp\bigg(q\sum_{i,j=1}^L I(\l_i,E_j)\bigg)=\prod_{i=1}^L\int_\cont \d\l_i\,\bigg(\prod_{j=1}^L\delta(\l_j-E_j)+\text{permutations}\bigg)\,,\nn
    \end{align}
    where the parameter $q$ is found to be large from a solution to a simple Schwinger–Dyson equation. The large $q$ saddle-points of the emergent action that we find are indeed
    \begin{equation}
        \frac{\partial}{\partial \l_i}\sum_{i,j=1}^L I(\l_i ,E_j)= 0\quad\Leftrightarrow\quad \l_1\dots\lambda_L = \text{permutations of }E_1\dots E_L\,.
    \end{equation}
    This localizes the eigenvalues (up to permutations) to $\lambda_i=E_i$, the eigenvalues of the Hamiltonian $\Hh$ of the dual quantum system. Thus, JT gravity with correlated branes is discrete and factorizing, even non-perturbatively. We discuss this matrix integral perspective in \textbf{section \ref{sect:matrix-integral-localization}}.
    
    \item In \textbf{section \ref{sect:matter}} we include bulk matter and explain that factorization is again resolved by fixing the brane two-point function. UV divergences associated to wormholes with small necks are absent in our model. We then compute the probe matter correlators, accounting for the full backreaction on the metric and, for instance, find the expected two-point function for a theory with a discrete spectrum. This addresses the puzzle posed in \cite{Maldacena:2001kr} concerning the decay of the two-point function in the eternal black hole geometry.
    
    \item In  \textbf{section \ref{sect:discussion}}, we rewrite our model in several equivalent ways: as a theory with correlated branes, a theory with some nonlocal dilaton potential, or a theory in which the dilaton potential is picked from an ensemble average distribution. Through these re-writings we emphasize that the critical ingredient that is necessary to obtain a factorizing gravitational theory, is the correlations between different disconnected components of spacetime. This entices us to propose a similar mechanism for resolving the factorization puzzle in higher dimensional theories, and forces us to question what the origin of such correlations between disconnected components is in string theory.
\end{enumerate}

\section{JT gravity with correlated spacetime branes}\label{sect:branes}
In this section we introduce JT gravity with correlated spacetime branes, this sets the stage for section \ref{sect:factor} where we will fine-tune these brane correlations to ensure factorization and discreteness.

\subsection{JT gravity}

JT gravity is a two-dimensional gravity theory that involves a dilaton $\Phi$ and metric $g$, with action \cite{jackiw1985lower,teitelboim1983gravitation}
\begin{equation}
I[g,\Phi]=-\frac{1}{2} \int_{\Sigma} \d^2 x \sqrt{g}\, \Phi\,(R+2) - \int_{\partial \Sigma} \d u \sqrt{h}\, \Phi\,(K-1) -\S\chi(\Sigma)\,.
\end{equation}
With this boundary action one needs to impose Dirichlet boundary conditions on the metric $h_{u u}=1/\e^2$ and dilaton $\Phi=1/2\e$, which fixes the length of the Euclidean boundary circle to $\beta/\e$ \cite{Jensen:2016pah,Maldacena:2016upp,Engelsoy:2016xyb}. The extremal entropy $\S$ multiplies the Euler character $\chi(\Sigma)=2-2g-n$, and thus suppresses higher genus geometries. Path integrating out the bulk dilaton $\Phi$ (after a contour rotation $\Phi \to \i \Phi$) localizes the geometries to hyperbolic Riemann surfaces (with boundary) with constant curvature $R+2=0$. We must include all surfaces consistent with the boundary conditions \cite{Saad:2019lba}.

A convenient way to organize the calculations, say for the case with one asymptotic boundary, is by viewing the higher genus surface as consisting of two parts: the trumpet, which has one asymptotic fixed length boundary, and one geodesic boundary with fixed length $b$; and the remaining genus $g$ Riemann surface, which has one geodesic boundary of identical length $b$. We must then integrate over the moduli space of genus $g$ Riemann surfaces with geodesic boundary $b$, which produces the Weil-Petersson volume $V_{g,1}(b)$ \cite{Dijkgraaf:2018vnm,mirzakhani2007simple}, and glue the trumpet to the genus $g$ Riemann surfaces, by integrating over $b$. This gives the genus expansion of the JT gravity partition function \cite{Saad:2019lba}
\begin{align} 
Z(\b)= Z_\text{disk}(\b) + \sum_{g=1}^{\infty} e^{\S(1-2g)} \int_0^{\infty} \d b\, b \, V_{g,1}(b)\, Z_\text{trumpet}(\b,b),\label{Zbexpansion}
\end{align}
where the disk and trumpet partition function can be computed by integrating over boundary modes \cite{Stanford:2017thb,Jensen:2016pah,Maldacena:2016upp,Engelsoy:2016xyb}
\be\label{ExpansionZcorr}
Z_{\rm disk}(\b) = \frac{e^{\S}}{4 \pi^{1/2} \b^{3/2}} e^{\frac{\pi^2}{\b}},\quad Z_{\rm trumpet}(\b,b) = \frac{1}{2\pi^{1/2}\b^{1/2}} e^{-\frac{b^2}{4\b}}\,.
\end{equation}
This generalizes to the multi-boundary case (Weil-Petersson volumes count connected Riemann surfaces)
\begin{align}
Z(\b_1,\dots,\b_n)_\text{conn.} = \sum_{g=0}^{\infty} e^{\S(2-2g-n)} \int_0^{\infty} \d b_1 b_1\dots\int_0^\infty &d b_n b_n \, V_{g,n}(b_1\dots b_n)\nn\\ &\quad Z_{\rm trumpet}(\b_1,b_1)\cdots Z_{\rm trumpet}(\b_n,b_n)\,.
\end{align}

This decomposition also makes it clear that JT gravity has a dual description as a double scaled matrix integral \cite{Saad:2019lba}, because the $V_{g,n}(b_1\dots b_n)$ satisfy the topological recursion relations (or loop equations) for a double scaled Hermitian matrix integral with genus zero spectral density \cite{Eynard:2007fi,Dijkgraaf:2018vnm}
\begin{equation}\label{rhoLeading}
\rho_0(E) = \frac{e^{\Ss}}{4\pi^2} \sinh 2\pi \sqrt{E}\,,
\end{equation}
which crucially matches also the spectrum associated with the JT gravity disk partition function \eqref{ExpansionZcorr}. Multi-boundary partition functions are computed in the matrix integral as ensemble averages
\begin{equation}
    \average{Z(\b_1)\cdots Z(\b_n)}=\frac{1}{\mathcal{Z}}\int \d H\, e^{-L \Tr V(H)}\,\Tr e^{\b_1 H} \cdots \Tr e^{\b_n H}=Z(\b_1,\dots,\b_n)\,,
\end{equation}
where the last equality states the JT-matrix integral duality. The potential $V(H)$ will be specified when needed in section \ref{sect:matrix-integral-localization}.

As stressed in \cite{Saad:2019lba}, an important point is that we can really think of $H$ in the matrix integral as the Hamiltonian of the boundary quantum mechanics. This means that here we have a version of AdS/CFT that involves JT gravity theory in the bulk and an ensemble of boundary Hamiltonians on the boundary. Consequently, the JT gravity fixed thermal length gravitational partition function (with an arbitrary number of boundaries) are dual to ensemble averages of partition functions of quantum mechanical systems (with an arbitrary number of $\Tr e^{-\beta H}$ insertions within the ensemble).

As motivated in the introduction, we now introduce additional ingredients in gravity that eliminate this average, whilst retaining a geometric description. 

\subsection{Spacetime branes}
\label{sect:sdb}

Spacetime branes are natural candidates for these effective new UV ingredients. They introduce extra boundaries for spacetimes to end on, and (as mentioned in section \ref{sec:intro}) are known to have a dual description as deformations in the JT gravity action \cite{JafferisEOW,Blommaert:2021gha}. In this dual description, there are no extra boundaries, this is essentially open-closed universe duality (similar to open-closed string duality).

As in string theory, different boundary conditions define different spacetime branes (branes simply from hereon), see \cite{Goel:2020yxl} for a semiclassical classification. We will encounter two types in particular.

First we have FZZT branes, the most commonly studied brane in minimal string theory \cite{Maldacena:2004sn,Saad:2019lba,Fateev:2000ik,Ponsot:2001ng,Goel:2020yxl,Mertens:2020hbs,Mertens:2020pfe,Hosomichi:2008th,Kostov:2002uq,Okuyama:2021eju,Teschner:2000md}. Here we should view the worldsheets as spacetimes, so the FZZT branes are really spacetime branes. Classically in JT gravity they correspond to fixed energy boundaries (and fixed dilaton). In the dual matrix integral, one FZZT boundary corresponds to\footnote{The significant potential term is related with the small $\b$ divergence in the Laplace transform \cite{Saad:2019lba}.}
\be 
\mo_{\rm FZZT}(E)=-\int_{0}^{\infty} \frac{\d\b}{\b}\, e^{\b E}\, Z(\b)= \Tr \log(E - H)-\frac{L}{2}V(E)\, ,\label{fzzt}
\ee
and FZZT branes are exponentials of boundaries, and so correspond in the matrix ensemble to \cite{Maldacena:2004sn,Saad:2019lba,Blommaert:2019wfy,Okuyama:2021eju}
\be\label{brane}
\psi_{\rm FZZT} (E) = \exp\left(-\int \frac{\d\b}{\b} e^{\b E} Z(\b)\right) =\det(E-H)\,\exp\bigg(-\frac{L}{2}V(E)\bigg)\,.
\ee

Another semiclassical interpretation of FZZT boundaries, that we will use in this work, was worked out in \cite{Okuyama:2021eju}. The basic idea is to decompose the FZZT boundary in a complete set of geodesic boundaries. To do this, we need the expansion coefficients, which can be found by Laplace transforming the trumpet \cite{Okuyama:2019xbv}
\begin{align}
    -\int_{0}^{\infty} \frac{\d\b}{\b}\, e^{\b E}\, Z(\b,\b_1,\dots,\b_n)_\text{conn.}&=\sum_{g=1}^n e^{\Ss(1-2g-n)}\int_0^\infty \d b\, b\,M_E(b)\int_0^\infty \d b_1 b_1\dots \int_0^\infty \d b_n b_n\label{29}\\&\qquad\qquad\qquad V_{g,n+1}(b,b_1\dots b_n)\,Z_\text{trumpet}(b_1,\b_1)\dots Z_\text{trumpet}(b_n,\b_n)\,,\nn
\end{align}
where the expansion coefficient or wavefunction explicitly becomes
\begin{equation}
    M_E(b)=-\int_{0}^{\infty} \frac{\d\b}{\b}\, e^{\b E}\,Z_\text{trumpet}(b,\b)=-\frac{1}{b}e^{-b (-E)^{1/2}}\,.\label{210}
\end{equation}
This has a branchcut for positive real energies. The two values above and below the cut define the two different FZZT boundary conditions for positive energies, indeed this feature was already there in \eqref{fzzt}.

Formula \eqref{29} means that one can view the FZZT boundary as consisting of fixed geodesic length $b$ boundaries with a wavefunction $M_E(b)$. In semiclassical JT variables this wavefunction implements the Legendre transform from geodesic boundary conditions to fixed energy boundary conditions \cite{Goel:2020yxl}. The wavefunction can be thought of as coming from a cosmological constant on the geodesic boundary. Semiclassically, with this extra boundary term in the action, there is saddle for $b$ which depends on $E$.\footnote{The saddle-point can generically be at arbitrary complex $b$. This is an important observation that we discuss more in section \ref{sec:additional-comments}.}

The second type of branes in which we are interested are inserting geodesic boundaries. In the dual matrix integral, it was found in \cite{Goel:2020yxl} that inserting one geodesic boundary corresponds with the operator\footnote{Here we are using a hybrid notation, where the second term is a double scaled expression. However it is just a function of $b$ and can easily be written in a finite $L$ form just as the first term as we also do in section \ref{sect:matrix-integral-localization}.}
\be 
\mo_{\rm G}(b)=\frac{2}{b}\Tr \cos(b H^{1/2})-\int_0^\infty \d E\,\rho_0(E)\,\frac{2}{b}\cos( b E^{1/2})\,. \label{obdef}
\ee
The second term subtracts the naive disk contribution as there are indeed no disk shaped geometries with these geodesic boundary conditions \cite{Goel:2020yxl}, in particular
\begin{equation}
    \average{\mo_{\rm G}(b)}=\sum_{g=1}^\infty e^{\Ss(1-2g)}\,V_{g,1}(b)\,,\quad \average{\mo_{\rm G}(b_1)\dots\mo_{\rm G}(b_n)}_\text{conn}=\sum_{g=0}^\infty e^{\Ss(2-2g-n)}\,V_{g,n}(b_1\dots b_n)\,.\label{212}
\end{equation}
Formula \eqref{obdef} can be checked from the matrix side directly, by computing the inverse Laplace transform of the FZZT operator $\mo(E)$, see formula \eqref{fzzt}. Geodesic branes are exponentials of this operator
\be 
\psi_{\rm G}(b) = \exp\bigg(\frac{2}{b}\Tr \cos(b H^{1/2})-\int_0^\infty \d E\,\rho_0(E)\,\frac{2}{b}\cos( b E^{1/2}) \bigg)\,.
\ee
The fact that ensemble averages of \eqref{obdef} return connected geometries \eqref{212} will be important in section \ref{sect:recovering}.

So far we mostly focused on the definition of branes in the open-string picture and their natural appearance in the matrix integral. To get a better sense of these branes in the closed-string picture, we need to understand what their effect is on the dilaton potential. For the FZZT brane we already have formula \eqref{29}, which is a nice expansion in terms of additional geodesic boundaries with a wavefunction on them, but we can also recast it in terms of a change in the dilaton potential. The dilaton potential $U(\Phi)$ is defined in the usual way (ignoring the boundary and topological term),
\be
I[g,\Phi]= -\frac{1}{2}\int_\Sigma \d^2 x \sqrt{g} \left(\Phi(R + 2) + U(\Phi) \right)\,.
\ee
The effect of adding an FZZT spacetime brane on the dilaton potential was worked out in \cite{Blommaert:2021gha, Okuyama:2021eju}. They found
\be 
U_{\rm FZZT}(\Phi) = -e^{-\S}e^{-2\pi \Phi}\frac{2 z}{z^2 + \Phi^2}\, ,
\ee
with $z^2 = -E$ and where the non-zero shift in the threshold energy $E_0$ has been taken into account. For our purposes, we actually need to generalize the notion of an FZZT brane slightly to a smeared version (see \cite{Maxfield:2020ale, Witten:2020wvy} for the defect case), which in the matrix integral corresponds to inserting 
\be 
\psi_{\rm FZZT}(\l) = \exp \left( \int_\cont \d z\, \l(z)\, \mathcal{O}_\text{FZZT}(z) \right)\, ,
\ee
with $\l$ a function (or distribution) integrated along some contour $\mathcal{C}$ and again $z^2 = -E$. All we do here is smear the brane in target space, something we often do in higher-dimensional string theories as well. The deformation in the dilaton potential changes then to 
\be 
U_{\rm FZZT}(\Phi,\l(z)) = -e^{-\S} e^{-2\pi \Phi} \int_\cont \d z\, \l(z) \frac{2z}{z^2 + \Phi}\,,\label{217}
\ee

Throughout this work we will mostly focus on insertions of $\mathcal{O}_{\rm G}(b)$. Geometrically, these correspond to an expansion like \eqref{29}, but where the wavefunction $M_E(b)$ is replaced by a more general wavefunction that we denote by $Z_\text{brane}(b)$. This wavefunction is essentially the inverse Laplace transform of the FZZT smearing function $\l(z)$
\be 
Z_\text{brane}(b)= \int_\cont \d z\, M_z(b)\, \l(z) = -\frac{1}{b}\int_\cont \d z\, e^{-b z}\,\l(z)\,.\label{218}
\ee
The deformation in the dilaton potential corresponding to insertions of (smeared) $\mathcal{O}_{\rm G}(b)$ operators is then
\be 
U_{\rm G}(\Phi,Z_\text{brane}(b)) = e^{-\S}\int_0^{\infty} \d b\, b\, Z_\text{brane}(b)\, e^{-2\pi \Phi}\,2\cos(b\Phi)\,.
\ee
Indeed, inserting \eqref{218} and doing the integral over $b$ returns \eqref{217}.
In summary, inserting smeared geodesic branes in the gravitational path integral corresponds with the deformation
\begin{equation}
    \exp\bigg(\int_0^\infty \d b\, b\,Z_\text{brane}(b)\,\mo_\text{G}(b) \bigg) \Leftrightarrow \exp\bigg(e^{-\Ss}\int_0^\infty \d b\, b\,Z_\text{brane}(b)\int_\Sigma \d^2 x\sqrt{g(x)}\,e^{-2\pi \Phi(x)}\,\cos(b\Phi(x))\bigg)\,.\label{map}
\end{equation}
We expand on this relation in appendix \ref{apt:branes}.

Notice that this looks similar to inserting smeared defects \cite{Maxfield:2020ale, Witten:2020wvy}, but with the defect angle analytically continued, and including both analytic continuations to $\i b$ and to $-\i b$. This was to be expected, because defects and geodesic boundaries act as analytic continuations of one another, this can be appreciated from several angles \cite{Maxfield:2020ale,Witten:2020wvy,Mertens:2019tcm}. We elaborate on this further in Appendix \ref{apt:branes}. Notice also that we worked to order $e^{-\S}$, going to higher order would require working with the string equation machinery. Since this is beyond the scope of this paper, we will not pursue that here, but it would be interesting to consider smeared branes in that setting.\footnote{Furthermore, it would be worthwhile to understand these potentials in more detail from the geometric point of view. Their oscillatory behavior signals possible phase transitions \cite{Witten:2020ert}. }

\subsection{Why correlated spacetime branes?}\label{sect:why}

As explained in the introduction, we expect on general grounds that integrating out UV degrees of freedom introduces nonlocal deformations in our low-energy effective theory. In QFT, such nonlocalities can be typically neglected in the low-energy effective theory since their effect is suppressed by the energy scale of the original degrees of freedom; thus, at least in a perturbative expansion around each saddle in the low-energy effective QFT such nonlocal terms in the action play no role. In the low-energy effective theory of gravity however, such terms need a more careful treatment; as we will explain in the next section, the perturbative expansion of nonlocal terms in the black hole saddle can have a competing effect to the sum over topologies that occurs in the gravitational path integral. Thus, it will prove important to keep track  of such nonlocal deformations in the low-energy effective theory. Below we will exemplify how such deformations can be re-expressed as insertions of correlated or nonlocal spacetime branes. We now  give an intuitive explanation for why this is the case, and explain what we mean with correlated spacetime branes.

Consider introducing the simplest nonlocal interaction in the dilaton gravity path integral: a bilocal deformation of the dilaton potential. Making use of the mapping of $\mo_\text{G}(b)$ to metric variables one reads off from \eqref{map} that we have
\begin{align}
    &\exp\bigg(\frac{1}{2}\int_0^\infty \d b_1 b_1\,\mo_\text{G}(b_1)\int_0^\infty \d b_2 b_2\,\mo_\text{G}(b_2)\, Z_\text{brane}(b_1,b_2)\bigg)\nn\\&\Leftrightarrow \exp\bigg(-I_{\text{nonlocal}}[g,\Phi]\bigg) = \exp\bigg(\frac{1}{2}\,e^{-2\Ss}\int_0^\infty \d b_1 b_1\int_0^\infty \d b_2 b_2\,Z_\text{brane}(b_1,b_2) \nn\\ &\hspace{1cm}\int_\Sigma \d^2 x_1\sqrt{g_1(x_1)}\,e^{-2\pi \Phi(x_1)}\,\cos(b_1\Phi_1(x_1))\int_\Sigma \d^2 x_2\sqrt{g_2(x_2)}\,e^{-2\pi \Phi_2(x_2)}\,\cos(b_2\Phi(x_2)) \bigg)\,.\label{nonlocex}
\end{align}
for some arbitrary function $Z_\text{brane}(b_1,b_2)$ whose meaning we now discuss. 
When we expand out \eqref{nonlocex}, what happens is that we are inserting extra geodesic brane boundaries in the gravitational path integral, but now there is a two-brane wavefunction component, which correlates two different smeared spacetime branes through the function  $Z_\text{brane}(b_1,b_2)$. The function $Z_\text{brane}(b_1,b_2)$ can thus be viewed at the \textit{connected} component of the two-brane correlator, and the deformation to the JT action \eqref{nonlocex} is manifestly bilocal whenever this brane correlation is non-zero.\footnote{{The correlation we mention here is a correlation between boundaries or in the universe field theory language correlators of (geodesic) boundary creating operators, but we will often use the terminology \emph{brane} correlators, even though these are exponentials of boundaries and live in a more abstract target space.}}

The first effect of this deformation is that when computing the partition function we have geometries with two extra correlated geodesic boundaries
\begin{align}
    Z(\beta)\supset& \,\frac{1}{2}\int_0^\infty \d b_1 b_1\int_0^\infty \d b_2 b_2\,Z_\text{brane}(b_1,b_2)\,\,\, \begin{tikzpicture}[baseline={([yshift=-.5ex]current bounding box.center)}, scale=0.4]
 \pgftext{\includegraphics[scale=0.75]{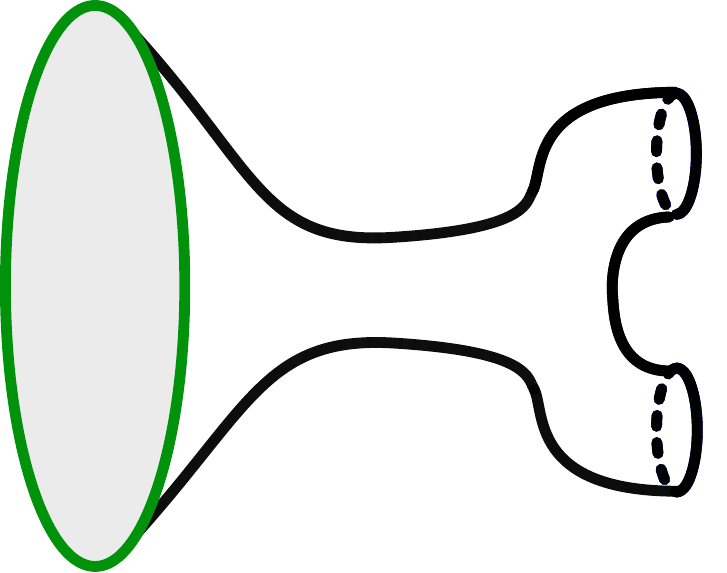}} at (0,0);
     \draw (3.5,1.1) node  {$b_1 $}; 
     \draw (3.5,-1.1) node  {$b_2 $}; 
  \end{tikzpicture}  \,= \,\,\frac{1}{2}\;\; \begin{tikzpicture}[baseline={([yshift=-.5ex]current bounding box.center)}, scale=0.4] 
  \pgftext{\includegraphics[scale=0.75]{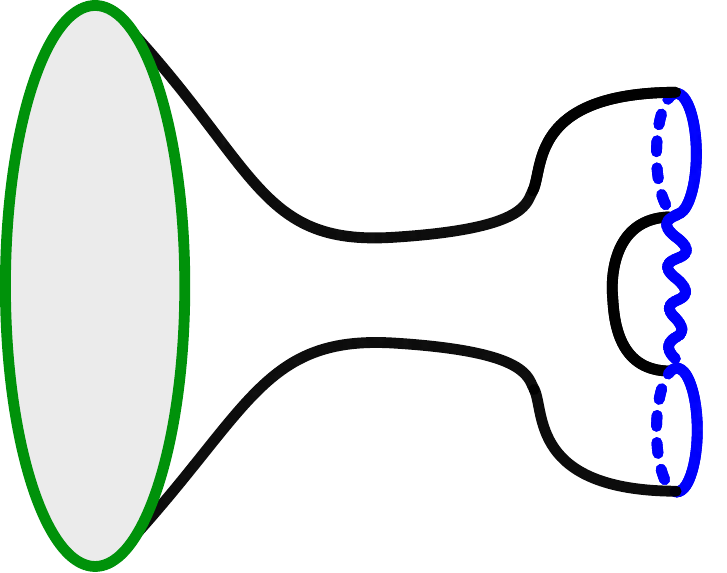}} at (0,0);
  \end{tikzpicture} 
    \nn\\&= \frac{1}{2}\int_0^\infty \d b_1 b_1\int_0^\infty \d b_2 b_2\,Z_\text{brane}(b_1,b_2) \int_0^\infty \d b\, b\,Z_\text{trumpet}(\beta,b)\,V_{0,3}(b_1,b_2,b)\,.
 \end{align}
In the second picture we introduced the notation that we will use for correlated branes henceforth. Because this bilocal operator is in the exponential in \eqref{nonlocex}, there will also be contributions where the coupled boundaries are inserted multiple times, for example
\begin{equation}
    Z(\beta)\supset \frac{1}{8}\;\; \begin{tikzpicture}[baseline={([yshift=-.5ex]current bounding box.center)}, scale=0.4]
 \pgftext{\includegraphics[scale=0.75]{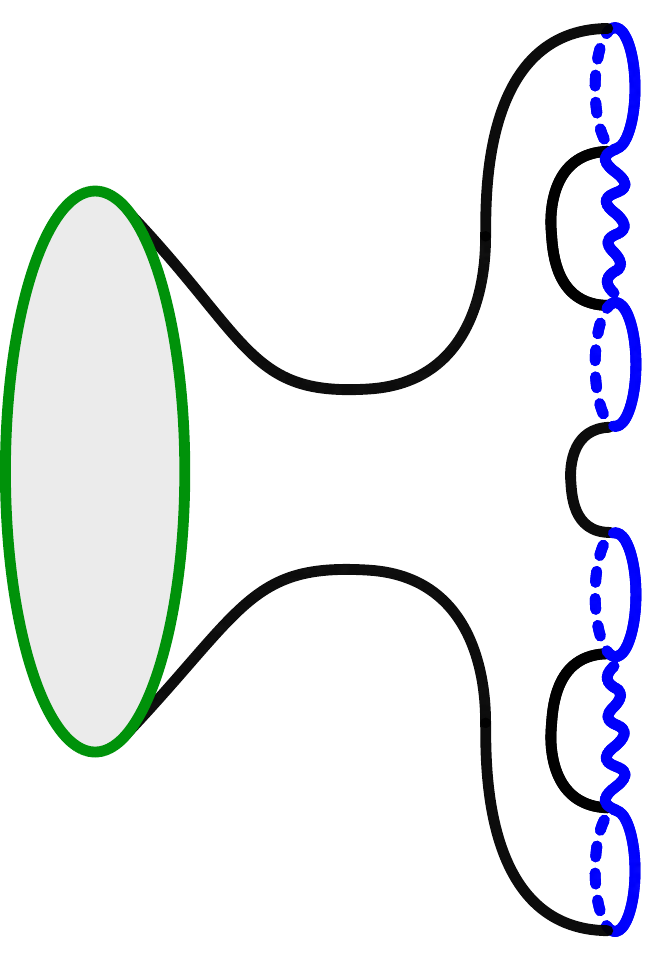}} at (0,0);
  \end{tikzpicture}  \quad\,.\label{twobilocal}
\end{equation}

One might expect that the UV theory generates all kinds of multi-local interactions, not just bilocal deformations \eqref{nonlocex} but also $n$-local deformations of the type
\be 
I_{\text{nonlocal}}[g,\Phi] \supset -\int_\Sigma \d^2x_1 \sqrt{g(x_1)} \dots \int_\Sigma \d^2 x_n \sqrt{g(x_n)}\, U_\text{nonlocal}(\Phi(x_1)\dots\Phi(x_n))\,.\label{multi}
\ee 
These correspond to higher-point brane interactions, which we should thus allow too. For example, the simplest effect of trilocal deformations are geometries with three extra correlated geodesic boundaries
\begin{align}
    Z(\beta)\supset& \,\frac{1}{6}\int_0^\infty d b_1 b_1\dots\int_0^\infty d b_3 b_3\,Z_\text{brane}(b_1,b_2,b_3)\,\, \begin{tikzpicture}[baseline={([yshift=-.5ex]current bounding box.center)}, scale=0.4]
 \pgftext{\includegraphics[scale=0.75]{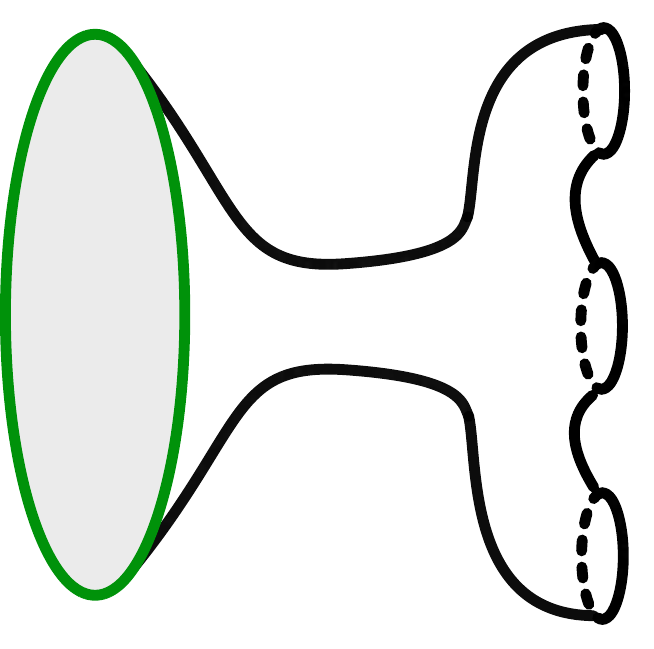}} at (0,0);
      \draw (3.1,1.95) node  {$b_1 \,$}; 
       \draw (3.1,0 ) node  {$b_2 \,$}; 
        \draw (3.1,-1.95) node  {$b_3 \,$}; 
  \end{tikzpicture}  =\,\,\frac{1}{6} \,\, \begin{tikzpicture}[baseline={([yshift=-.5ex]current bounding box.center)}, scale=0.4]
 \pgftext{\includegraphics[scale=0.75]{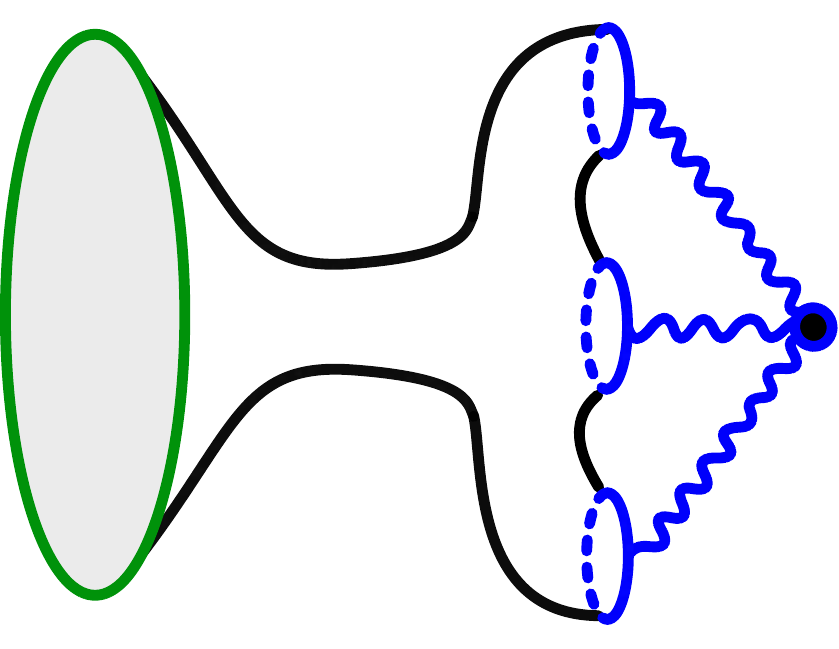}} at (0,0);
  \end{tikzpicture} \,\, \quad \nn\\&= \frac{1}{6}\int_0^\infty d b_1 b_1\dots\int_0^\infty d b_3 b_3\,Z_\text{brane}(b_1,b_2,b_3) \int_0^\infty \d b\, b\,Z_\text{trumpet}(\beta,b)\,V_{0,4}(b_1,b_2,b_3,b)\,.\label{3}
\end{align}
We will show in section \ref{sect:factor} that actually, factorization (to all orders in $e^{-\S}$) requires $Z_\text{brane}(b_1\dots b_n)=0$ whenever $n>2$, but the two-point function $Z_\text{brane}(b_1,b_2)$ crucially must be nonzero.

One important consequence of having nonlocal interactions in the gravitational path integral is that closed universes\footnote{{By closed universes we mean geometries without an asymptotic boundary. They could still have a geodesic boundary, like the torus with one geodesic boundary.}} no longer trivially cancel. For example, in \eqref{nonlocex} we could have $x_1$ being a point on the disk and $x_2$ being a point on a separate torus. Without this bilocal interaction the torus would be an irrelevant vacuum diagram, but here it can become part of the connected Feynman graph. The first correction of this type is
\begin{equation}
\label{eq:example-connection-to-closed-universe}
    Z(\beta)\supset \quad\begin{tikzpicture}[baseline={([yshift=-.5ex]current bounding box.center)}, scale=0.4]
 \pgftext{\includegraphics[scale=0.85]{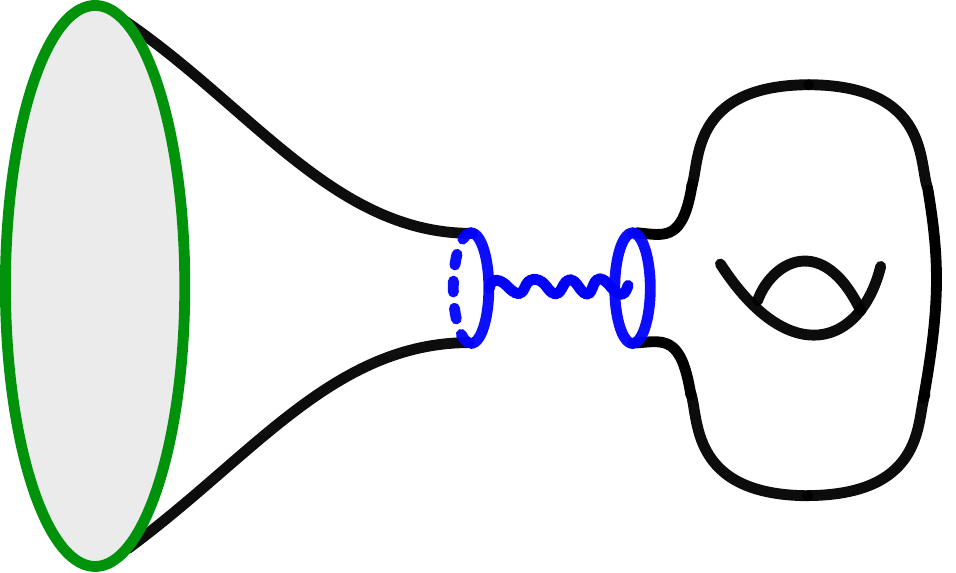}} at (0,0);
  \end{tikzpicture} .
\end{equation}
Throughout section \ref{sect:factor} we do not include degenerate cylindrical surfaces stretching between two geodesic boundaries. This makes sense, just because re-introducing those cylinders just renormalizes the brane correlators $Z_\text{brane}(b_1\dots b_n)$, as we discuss in section \ref{sect:calculation}, where we will need the renormalized correlators.

More generally, taking into account all possible brane correlators, the full expansion for $Z(\beta)$ becomes
\begin{equation}
Z(\b) = Z_{\rm disk}(\b) + \sum_{g=1}^{\infty} e^{\S(1-2g)} \int_0^{\infty} \d b\, b \, V_{g,1}(b)\, Z_{\rm trumpet}(\b,b) + \int_0^\infty \d b\, b\, Z_{\rm trumpet}(\b,b)\,X(b)\,,
\label{eq:brane-model-expansion}
\end{equation}
where $X(b)$ encodes all corrections from the spacetime brane correlators\footnote{We implicitly exclude the degenerate cylinders $g^c=0$, $n^c=2$ from the sums on the second line, as mentioned above.}
\begin{align}
\label{eq:X(b)}
X(b) &=\sum_{g=0}^\infty \sum_{n=1}^\infty\frac{1}{n!}\, e^{\S (1-2g-n)}\int_0^\infty \d b_1 b_1 \dots \int_0^\infty \d b_n b_n\,V_{g,n+1}(b,b_1,\dots, b_n)\,Z_\text{brane}(b_1,\dots, b_n)\\ &+  \nn \sum_{g=0}^\infty \sum_{n=1}^\infty\sum_{g^c=0}^\infty \sum_{n^c=1}^\infty\frac{1}{n!}\frac{1}{n^c!}\, e^{\S (1-2g-2g^c-n-n^c)}\, \int_0^\infty \d b_1 b_1 \dots \int_0^\infty\d b_n b_n\, V_{g, n+1}(b,b_1,\dots, b_n)   \\\nn   &\hspace{4cm}\int_0^\infty\d b_1^c b_1^c \dots \int_0^\infty\d b_n^c b_n^c\, V_{g^c,n^c}(b_1^c,\dots, b_n^c) \,Z_\text{brane}(b_1,\dots, b_n,b_1^c,\dots, b_{n^c}^c) \nn \\ &+ \text{branes connecting to multiple closed universes}
\end{align}
The first line are all cases where the brane correlators do not connect to closed universes, on the second line we have all contributions connected to one closed universe (hence the superscript $c$), etcetera. Therefore, the brane correlator $Z_\text{brane}(b_1,\dots, b_n,b_1^c,\dots, b_{n^c}^c)$ should be read as having at least one connected contribution between a brane on the surface with the asymptotic boundary and a brane on the extra closed universe.

\subsection{Geometric setup}

To summarize, the rules for constructing all geometries such that we account for the contribution of the correlated branes are as follows:
\begin{enumerate}
    \item Besides the asymptotic boundary, the partition function includes a sum over an arbitrary number of geodesic boundaries, for example one more contributing geometry is
    \be 
  \text{Example of new geometry with multiple one-point functions:}\quad \begin{tikzpicture}[baseline={([yshift=-.5ex]current bounding box.center)}, scale=0.4]
 \pgftext{\includegraphics[scale=0.75]{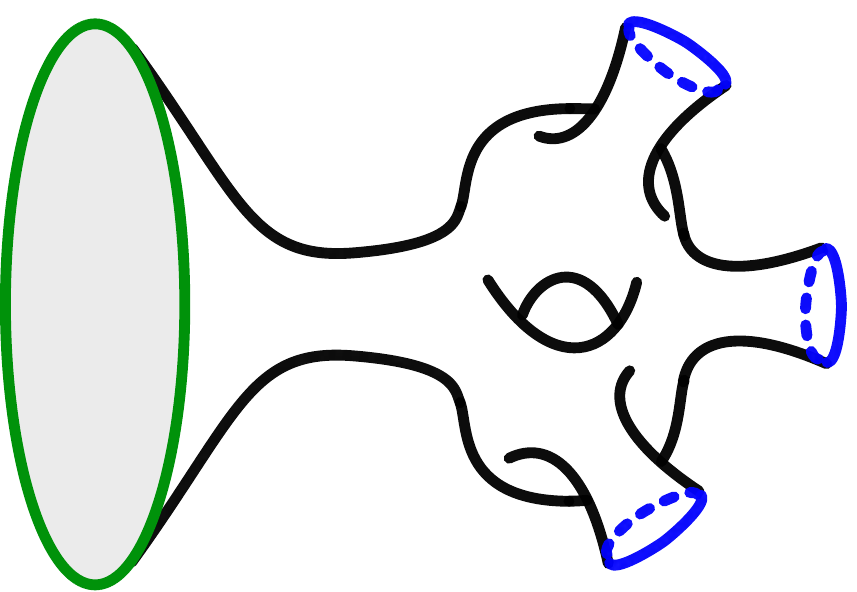}} at (0,0);
  \end{tikzpicture}\,\,.
  \label{eq:example1}
    \ee
    \item The extra brane boundaries are smeared with wavefunctions, and the joint connected wavefunction for $n$ boundaries is $Z_\text{brane}(b_1,\dots, b_n)$. We denote this connected component of the brane-correlator, as in for example \eqref{3} and \eqref{eq:example-connection-to-closed-universe}.
   \be 
   \text{Notation for connected brane correlators:}\quad Z_\text{brane}(b_1,\dots, b_n) = \begin{tikzpicture}[baseline={([yshift=0ex]current bounding box.center)}, scale=0.6]
  \pgftext{\includegraphics[scale=0.30]{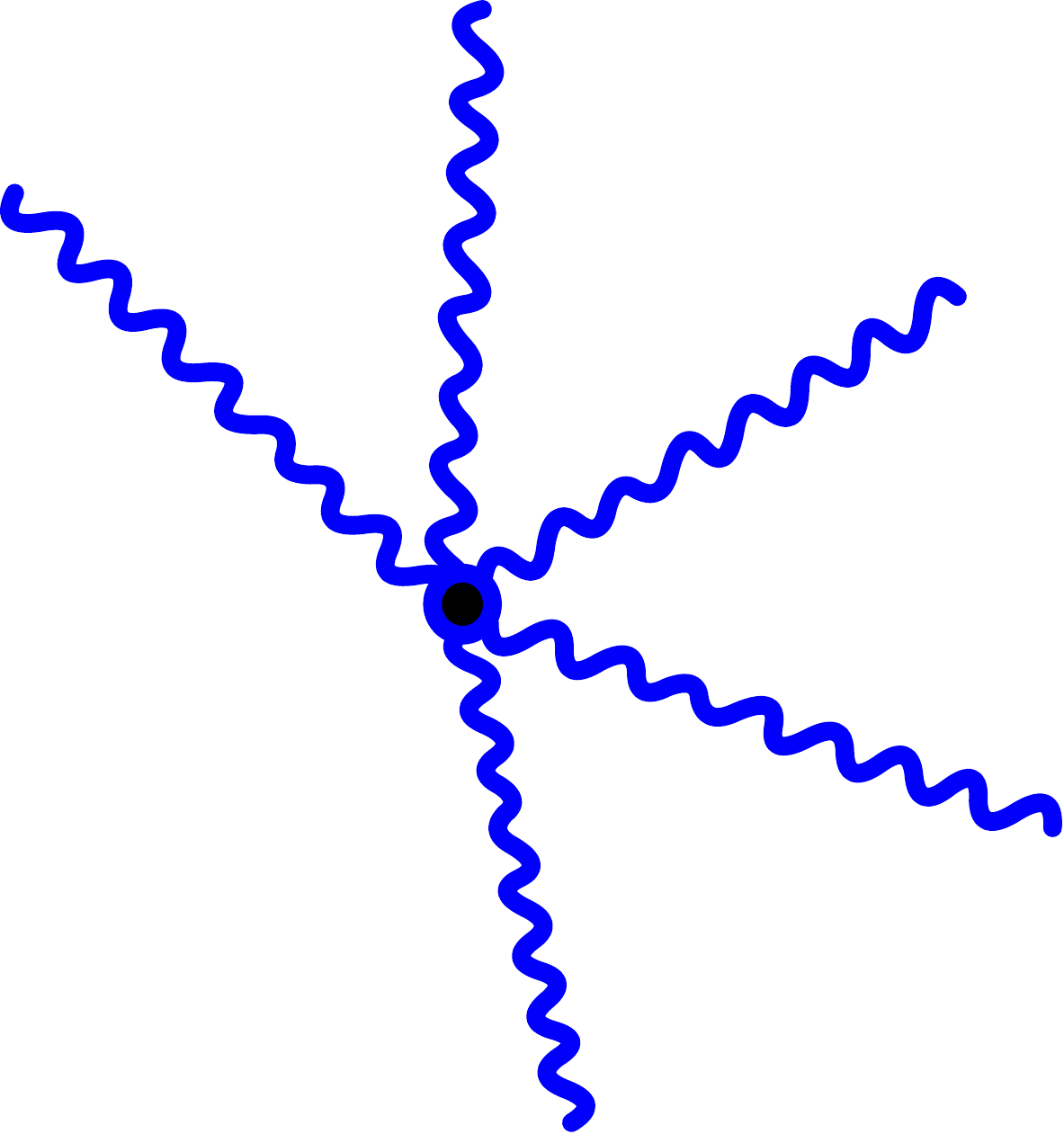}} at (0,0);
  \end{tikzpicture}
    \ee
   These wavefunctions are integrated over $b_1,\dots, b_n$ using the Weil-Peterson measure.

    \item Brane correlators can connect spacetime components with asymptotic boundaries (open universes) to closed Euclidean spacetime components (closed universes), which would otherwise factor out of boundary observables. In addition to the closed universe connected through the brane correlator to an open universe as in \eqref{eq:example-connection-to-closed-universe}, this also generates extra connected contributions, where closed universes connect to multiple open universes (and potentially also to each other).
    \be 
    \text{Examples how closed universes contribute: } \quad \begin{tikzpicture}[baseline={([yshift=-.5ex]current bounding box.center)}, scale=0.4]
 \pgftext{\includegraphics[scale=0.75]{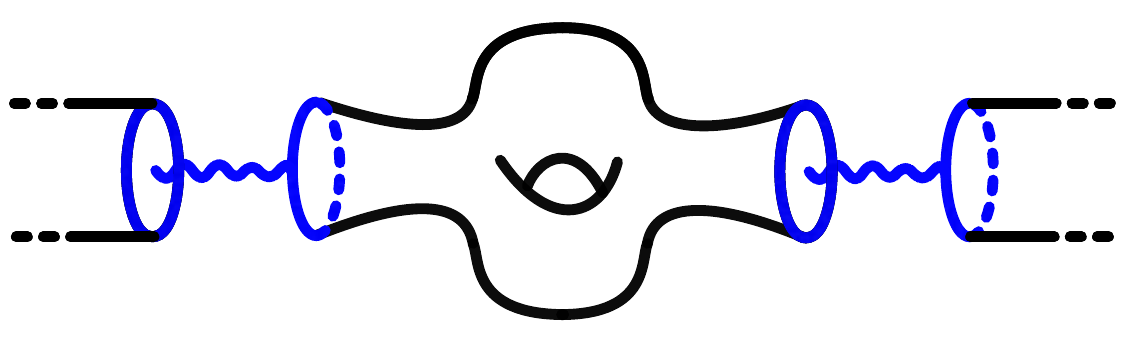}} at (0,0);
  \end{tikzpicture}\,\,.
    \ee
    The closed universes can alternatively be absorbed in a renormalization of the brane correlations, see section \ref{sec:additional-comments}. 
\end{enumerate}

\section{Factorization and discreteness from correlated spacetime branes}\label{sect:factor}
The purpose of this section is the present the bulk gravitational description of one quantum mechanical system with Hamiltonian $\Hh$.\footnote{Earlier work in this direction includes \cite{Blommaert:2019wfy,Blommaert:2020seb,Blommaert:2021gha,Blommaert:2021etf,Saad:2021rcu,Saad:2021uzi,Mukhametzhanov:2021hdi}.} The dual gravitational theory has a discrete spectrum (the eigenvalues $E_1,\dots, E_L$ of $\Hh$) and factorizes
\begin{equation}
    Z(\beta_1,\dots,\beta_n)=Z(\beta_1)\dots Z(\beta_n)\,.
\end{equation}
We claim that the sought-after gravitational theory is nothing but JT gravity with correlated spacetime branes, where the connected spacetime brane correlators take the specific values
\begin{align}
    &\text{Factorization} &&\Leftrightarrow\quad Z_\text{brane}(b_1,b_2)=-\frac{1}{b_1}\delta(b_1-b_2)\,,\,\,\text{ and }\,\,\, Z_\text{brane}(b_1\dots b_n)=0\text{ for $n>2$} \nn\\&\text{Discreteness}&&\Leftrightarrow\quad  Z_\text{brane}(b)=\sum_{i=1}^L  \frac{2}b \cos(b E_i^{1/2})-\int_0^\infty d E\,\rho_0(E)\,\frac{2}{b}\cos( b E^{1/2})\,.\label{solution}
\end{align}
The first line follows from demanding factorization, and the second line from the requirement that the theory has the discrete spectrum $E_i$ of $\Hh$.

We will first prove that \eqref{solution} is a solution to the factorization puzzle, after which, in section \ref{sect:unique}, we prove it is the unique solution. In section \ref{sect:onepoint}, we derive the formula for the one-point function that is needed to observe a discrete spectrum, and we quantify the magnitude of this one-point function in section \ref{sect:recovering}. In section \ref{sect:matrix-integral-localization} we further stack this up by proving that these brane correlators also have the desired effect in the dual matrix integral description non-perturbatively in $e^{-\S}$. 
\subsection{All orders factorization}\label{sect:allorders}

To understand why \eqref{solution} resolves factorization, notice first that in $Z(\beta_1,\beta_2)$ the genus zero wormhole is precisely canceled by two trumpets connected by a brane two-point function
\be 
\label{eq:leading-wormhole-cancelling}
\begin{tikzpicture}[baseline={([yshift=-.5ex]current bounding box.center)}, scale=0.4]
 \pgftext{\includegraphics[scale=0.85]{Factorization_leading_cancellation.pdf}} at (0,0);
          \draw (10,0) node  {$ = 0 $}; 
  \end{tikzpicture} ,
\ee
or explicitly in formulas (the trumpet partition function is in \eqref{ExpansionZcorr} and the wormhole partition function is equation (135) in \cite{Saad:2019lba})
\begin{equation}
   Z_\text{wormhole}(\beta_1,\beta_2)+\int_0^\infty \d b_1 b_1\int_0^\infty \d b_2 b_2\, Z_{\rm trumpet}(\beta_1, b_1) Z_{\rm trumpet}  (\beta_2, b_2) \,Z_\text{brane}(b_1,b_2)=0\,.
\end{equation}
This indeed vanishes when we insert \eqref{solution} and compute the integrals.

We can prove that \eqref{solution} is a solution to the factorization puzzle to all orders in the genus expansion, for two boundaries, by noticing that can group together geometries in such a way that partition functions that include branes on cycles which are homotopic to the asymptotic boundary, cancel the contributions of geometries that have no brane on these same cycles.

Explicitly, using  \eqref{solution} (and gluing branes to the asymptotic boundary using the Weil-Peterson measure) one can remark that, for any surface $\Sigma$, which may or may not have extra branes, and which may or may not connect the two asymptotic boundaries (either via geometries or via extra brane correlators on geometries), we have the cancellation
\begin{align}
 &   \begin{tikzpicture}[baseline={([yshift=-.5ex]current bounding box.center)}, scale=0.7]
  \pgftext{\includegraphics[scale=0.5]{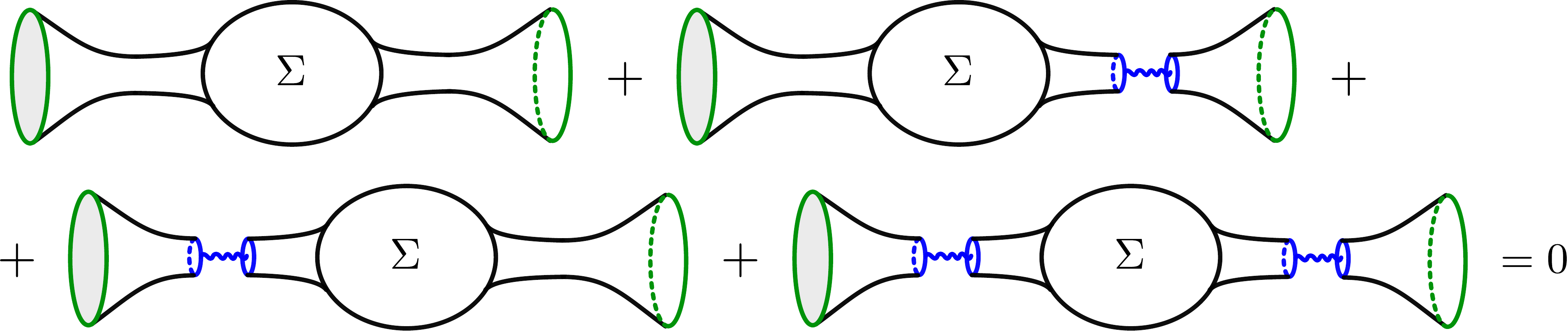}} at (0,0);
   \end{tikzpicture} 
  \label{eq:cancelllation-2-boundaries}
\end{align}
All surfaces having two asymptotic boundaries are among those listed in \eqref{eq:cancelllation-2-boundaries}, and therefore $Z(\beta_1,\beta_2)=Z(\beta_1)Z(\beta_2)$ indeed.

To show that the gravitational path integral factorizes when considering $n$-asymptotic boundaries, we can generalize \eqref{eq:cancelllation-2-boundaries} by grouping geometries in essentially the same way. When the boundaries have temperatures $\beta_1$, $\dots$, $\beta_n$, then, for any geometry $\Sigma$ which might include handles or an arbitrary number of brane correlators, we have the cancellation
\begin{align}
    \sum_{k=0}^n  \sum_{\s_k^n}
    \;\begin{tikzpicture}[baseline={([yshift=-.5ex]current bounding box.center)}, scale=0.70]
 \pgftext{\includegraphics[scale=0.38]{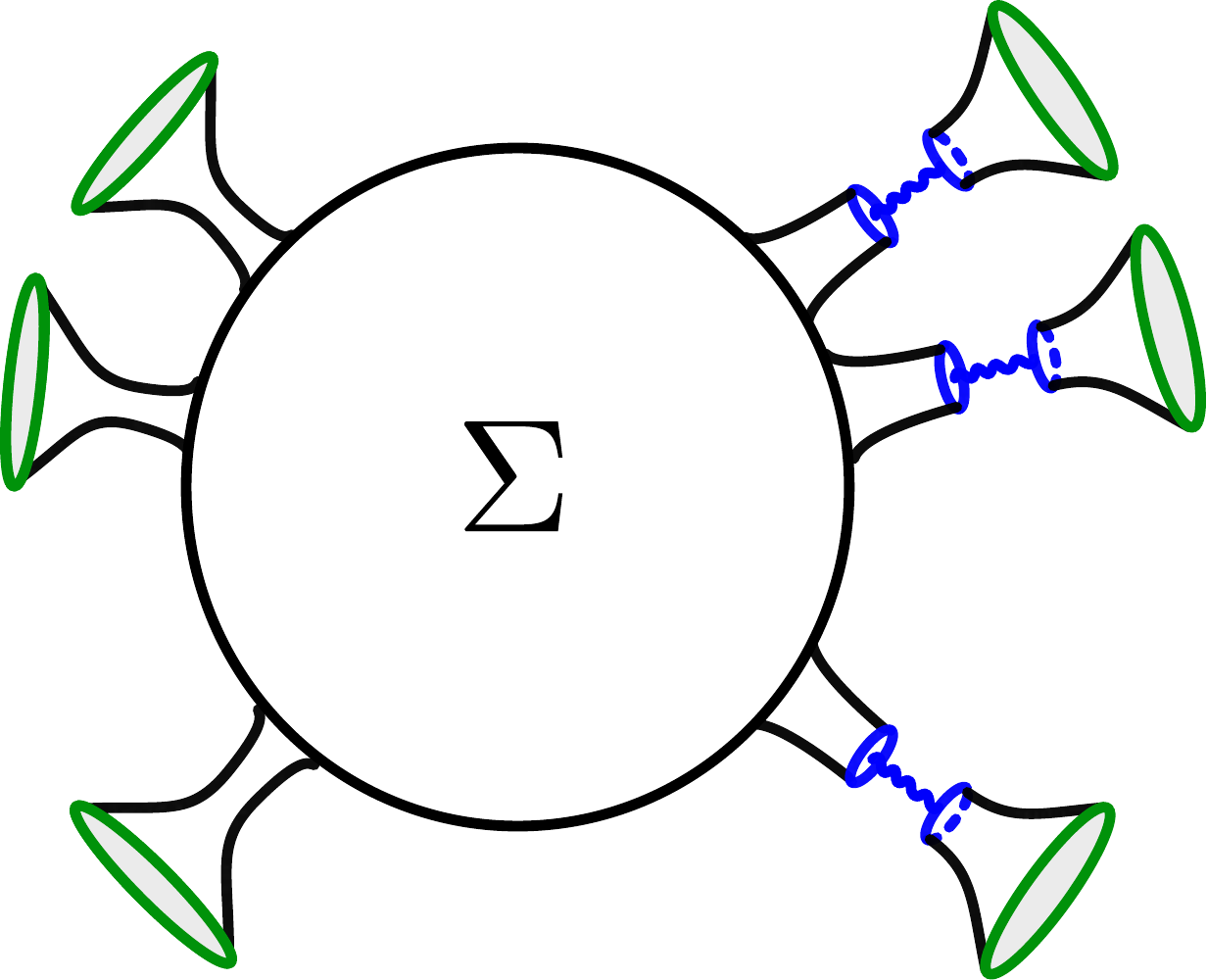}} at (0,0);
  \end{tikzpicture} \hspace{-0.3cm} 
  \quad\;\; = \quad \begin{tikzpicture}[baseline={([yshift=-.5ex]current bounding box.center)}, scale=0.70]
 \pgftext{\includegraphics[scale=0.38]{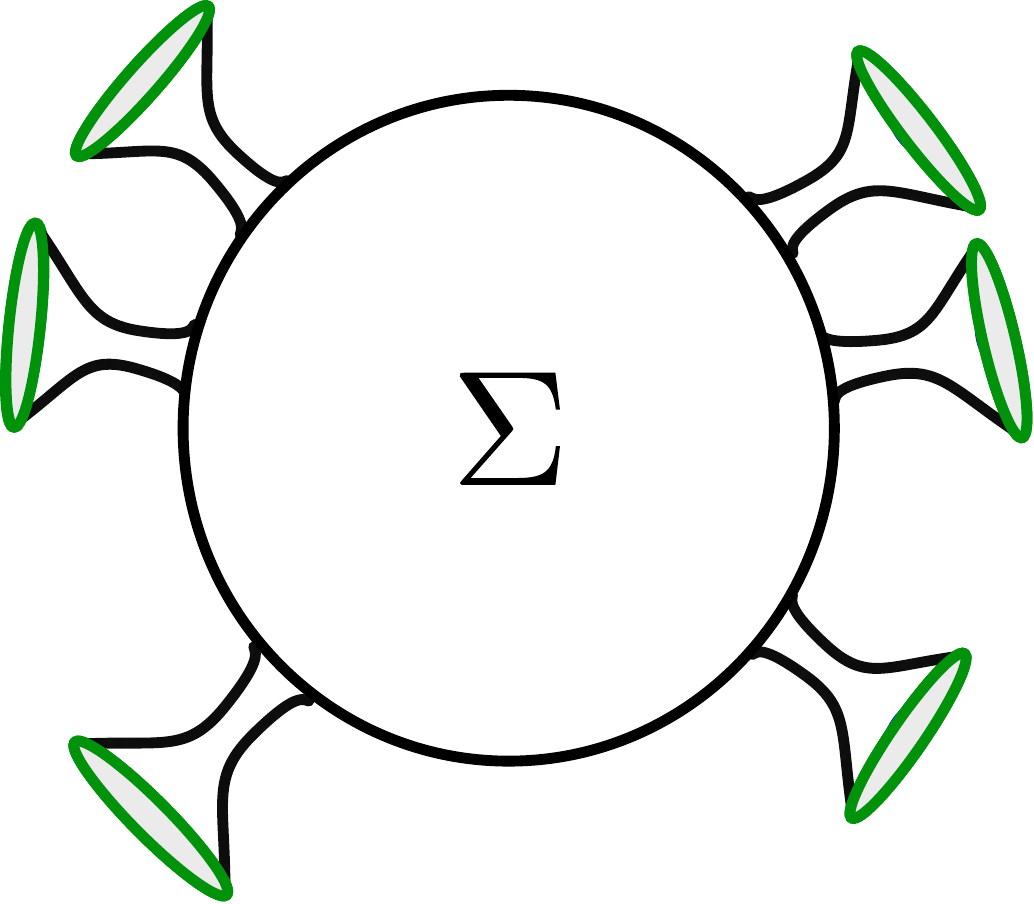}} at (0,0);
  \end{tikzpicture} \; \sum_{k=0}^n (-1)^k{n \choose k}=0\,.
\label{eq:general-identity-that-makes-geometries-vanish}
\end{align}
Above, the sum over $\sigma_k^n$ is over all combinations diving the set of $n$ boundaries of $\Sigma$ into $k$ boundaries that end on an asymptotic boundary and $n-k$ boundaries ending on a brane. Generically all boundaries can have different inverse temperatures $\beta_i$. The first equation follows by inserting $Z_\text{brane}(b_1,b_2)$ as given in \eqref{solution} for all the blue brane correlators, and gluing the branes using the Weil-Petersson measure. One finds that each term in the sum is proportional to the same gravitational partition function, and the prefactors add up to zero. Notice that the case $n=1$ implies that all higher genus corrections to $Z(\beta)$ are canceled.

The only exceptions to these enormous cancellations are the genus zero disk, and the disk with a brane one point function \eqref{eq:Z(beta)-after-simplification}, which do not have geodesic (that are not a boundary) homotopic to the asymptotic boundary, therefore the above argument does not go through. In summary, when we include an exponential of brane two-point function \eqref{solution} in JT gravity, all wormhole geometries are cancelled to all orders in $e^{-\S}$, and only factorizing contributions remain.
\subsection{Uniqueness of the factorizing solution}
\label{sect:unique}
Let us now prove that \eqref{solution} is the unique solution for the brane correlators, if we \emph{demand} factorization. In this section we put $Z_\text{brane}(b)=0$, as the one-point function does not impact the factorization discussion.

To have a genus expansion that still makes sense, it will be important to think of the brane correlators as having an $e^{-\Ss}$ expansion too
\be \label{TopCorExpansion}
Z_\text{brane}(b_1\dots b_n) = \quad \begin{tikzpicture}[baseline={([yshift=0ex]current bounding box.center)}, scale=0.6]
  \pgftext{\includegraphics[scale=0.30]{nPtCorrelator.pdf}} at (0,0);
  \end{tikzpicture}\quad = \sum_{k = -\infty}^{2-n}\bigg( \begin{tikzpicture}[baseline={([yshift=0ex]current bounding box.center)}, scale=0.6]
  \pgftext{\includegraphics[scale=0.30]{nPtCorrelator.pdf}} at (0,0);
  \end{tikzpicture}\bigg)_k\,,
\ee
where the terms on the right side are assumed to scale as $e^{k\Ss}$. The upper bound $k_\text{max}=2-n$ is fixed by demanding that the connected $n$-point partition function vanishes at order $O(e^{k\Ss})$ when $k>2-n$.\footnote{Since the leading brane-free connected geometry is $O(e^{\S(2-n)})$, we indeed can not have $n$-brane vertices for $k>2-n$.} 

The idea is now to impose factorization, order per order in the $e^{-\Ss}$ expansion, and for any number of boundaries, and use this to find the expansion coefficients in \eqref{TopCorExpansion}. Demanding $Z(\beta_1,\beta_2)=0$ at order $O(1)$ gives the equation \eqref{eq:leading-wormhole-cancelling}, which in our current mindset fixes the leading order propagator to
\begin{equation}
    \bigg(\begin{tikzpicture}[baseline={([yshift=-.5ex]current bounding box.center)}, scale=0.8]
 \pgftext{\includegraphics[scale=0.38]{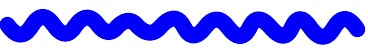}} at (0,0);
  \end{tikzpicture}\bigg)_0 = -\frac{1}{b_1}\delta(b_1 - b_2)\,.
\end{equation}

Continuing to focus on the two-boundary gravitational partition function, at next order $O(e^{-\S})$ we have no contributions other than that coming from subleading contributions to the two-brane correlator, and so factorization of $Z(\beta_1,\beta_2)$ at order $O(e^{-\Ss})$ demands that
\be  \bigg(\begin{tikzpicture}[baseline={([yshift=-.5ex]current bounding box.center)}, scale=0.8]
 \pgftext{\includegraphics[scale=0.38]{propagator.pdf}} at (0,0);
  \end{tikzpicture}\bigg)_{-1} =0\,.
\ee
At order $O(e^{-2\S})$ there are several diagrams that can contribute, so factorization becomes less trivial\footnote{Here we will write symmetry factors as if all asymptotic boundaries have the same temperature, to avoid cluttering. Furthermore, the brane correlators in drawings always denote the leading order ones, unless explicitly specified otherwise.}
\begin{align}
 \begin{tikzpicture}[baseline={([yshift=-.5ex]current bounding box.center)}, scale=0.4]
 \pgftext{\includegraphics[scale=0.75]{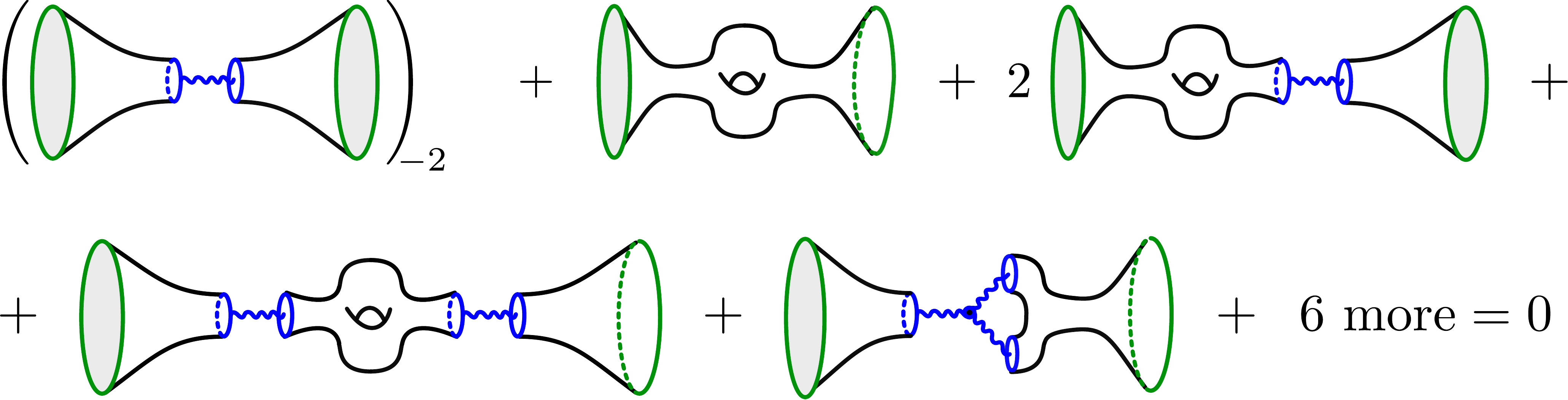}} at (0,0);
  \end{tikzpicture}
  \label{eq:Oe-2S0-2bdies}
\end{align}
The other six diagrams involve only leading order brane two-point correlators. Using \eqref{eq:cancelllation-2-boundaries}, we learn that most diagrams cancel, except the first on the first line and the second on the second line. To determine the $O(e^{-2\S})$  contribution to the two-brane correlator, we must hence first determine the leading order $3$-brane correlator. Setting $Z(\beta_1,\beta_2,\beta_3)=0$ at (first non-trivial) order $O(e^{-\Ss})$ gives
\begin{align}
\begin{tikzpicture}[baseline={([yshift=-.5ex]current bounding box.center)}, scale=0.4]
 \pgftext{\includegraphics[scale=0.75]{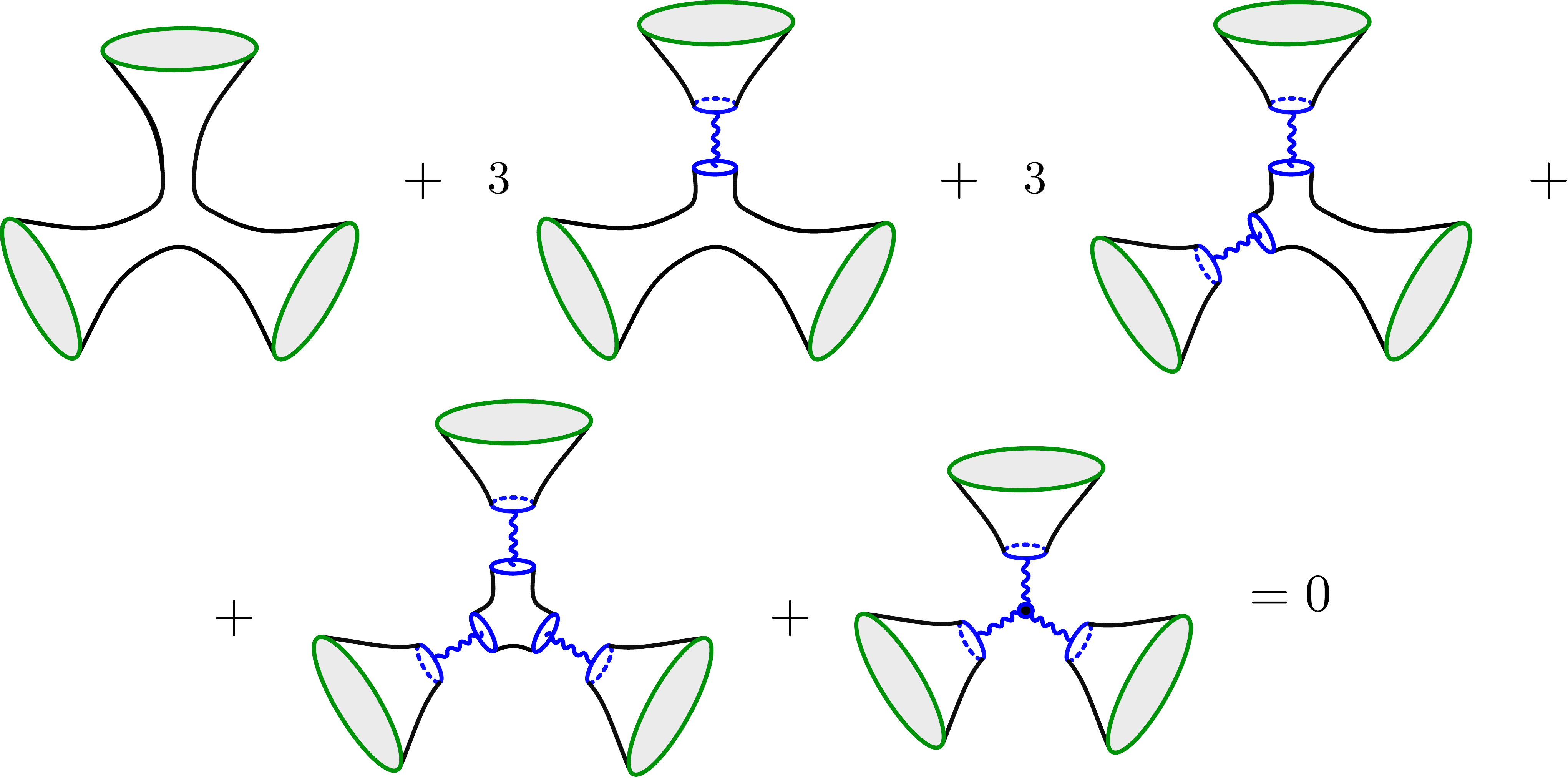}} at (0,0);
  \end{tikzpicture}
\end{align}
Using \eqref{eq:general-identity-that-makes-geometries-vanish} for $n=3$, this equation imposes that the leading order brane $3$-point vertex vanishes, which, in turn, can be inserted in \eqref{eq:Oe-2S0-2bdies} to prove that the second-subleading order propagator also vanishes
\be 
\bigg(\begin{tikzpicture}[baseline={([yshift=-.5ex]current bounding box.center)}, scale=0.16]
 \pgftext{\includegraphics[scale=0.45]{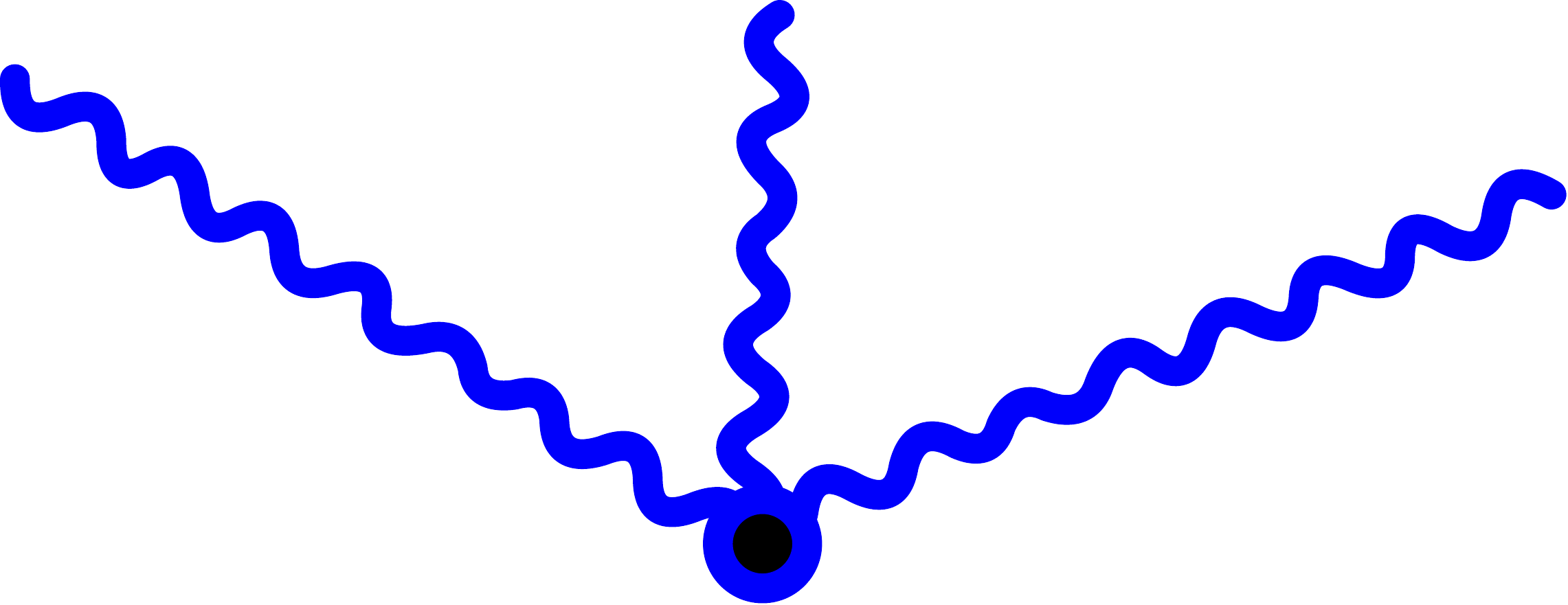}} at (0,0);
  \end{tikzpicture} \bigg)_{-1} = 0\,,\quad \bigg(\begin{tikzpicture}[baseline={([yshift=-.5ex]current bounding box.center)}, scale=0.8]
 \pgftext{\includegraphics[scale=0.38]{propagator.pdf}} at (0,0);
  \end{tikzpicture}\bigg)_{-2} =0\,.
\ee

Similarly, demanding that $Z(\beta_1\dots\beta_n)=0$ at leading non-trivial order, and using \eqref{eq:general-identity-that-makes-geometries-vanish}, we prove (by induction) that the leading non-trivial order $n$-point brane vertices must all vanish in a factorizing theory
\begin{equation}
     \bigg(\begin{tikzpicture}[baseline={([yshift=-.5ex]current bounding box.center)}, scale=0.55]
 \pgftext{\includegraphics[scale=0.30]{nPtCorrelator.pdf}} at (0,0);
  \end{tikzpicture}\bigg)_{2-n} = 0\quad \text{for }n>2\,.
\end{equation}
This can in turn be used (again in combination with \eqref{eq:general-identity-that-makes-geometries-vanish}) to prove inductively that subleading lower-point brane vertices must all vanish. In summary, we have proven that demanding factorization uniquely fixes the brane correlators to be Gaussian \eqref{solution}, all higher point brane correlators must vanish
\be
\begin{tikzpicture}[baseline={([yshift=-.5ex]current bounding box.center)}, scale=0.8]
 \pgftext{\includegraphics[scale=0.38]{propagator.pdf}} at (0,0);
  \end{tikzpicture} = -\frac{1}{b_1}\delta(b_1-b_2)\,,\quad\begin{tikzpicture}[baseline={([yshift=-.5ex]current bounding box.center)}, scale=0.55]
 \pgftext{\includegraphics[scale=0.30]{nPtCorrelator.pdf}} at (0,0);
  \end{tikzpicture} = 0\quad \text{for }n>2\,.
  \label{eq:solution-to-factorization}
\ee

\subsection{Discreteness and exact half-wormholes}\label{sect:onepoint}
We now turn on some non-zero brane one-point function $Z_\text{brane}(b)$. For most geometries with one-point functions we can still use equation \eqref{eq:general-identity-that-makes-geometries-vanish}, the only new exception is the geometry with only one $Z_\text{brane}(b)$ inserted.\footnote{Remember from section \ref{sect:why} that we do not allow degenerate cylinders. Including them does not affect this argument, it only renormalizes the brane correlators, see section \ref{sect:calculation}.} Therefore our theory remains extremely simple, the partition function becomes \emph{exactly}
\begin{align}
\label{eq:Z(beta)-after-simplification}
    Z(\beta)=\,Z_\text{disk}(\beta)+\int_0^\infty \d b\,b\,Z_\text{trumpet}(\beta,b)\,Z_\text{brane}(b) = \begin{tikzpicture}[baseline={([yshift=1.7ex]current bounding box.center)}, scale=0.7 ]
 \pgftext{\includegraphics[scale=0.45]{Halfwormhole_contribution.pdf}} at (0,0);
  \draw (-1.9, -2) node {black hole};
         \draw (2.3, -2) node {one-point function};
  \end{tikzpicture}\,.
\end{align}
The disk captures the contribution of the black hole saddle. The only other remaining contribution can be identified as the half-wormhole discussed in Saad, Shenker, Stanford and Yao \cite{Saad:2021rcu,Saad:2021uzi} with the important distinction that here we proved that the partition function receives no other perturbative contributions in $e^{-\S}$. We will prove in section \ref{sect:matrix-integral-localization} that this remains exactly true non-perturbatively in $e^{-\S}$.

In this sense what we have done is prove that in this setup the half-wormhole approximation is exact.

We now want to check that the expression for $Z_\text{brane}$ in \eqref{solution} results indeed in a theory with discrete spectrum $E_1,\dots, E_L$. Using the Gaussian integral
\begin{equation}
    \int_0^\infty \d b\, b\,Z_\text{trumpet}(\beta,b)\,\frac{2}{b}\,\cos(b E^{1/2})=\frac{1}{2\pi^{1/2}\beta^{1/2}}\int_{-\infty}^{+\infty} \d b\,\exp(-\frac{b^2}{4\beta}+i b E^{1/2})  =e^{-\beta E}\,,\label{gaussian}
\end{equation}
we find that the half-wormhole geometry in \eqref{eq:Z(beta)-after-simplification} contributes
\begin{equation}
    \int_0^\infty \d b\, b\,Z_\text{trumpet}(\beta,b)\,Z_\text{disk}(b)=\sum_{i=1}^Le^{\beta E_i}-\int_0^\infty d E\,\rho_0(E)\,e^{-\beta E}=\Tr e^{-\beta \Hh}-Z_\text{disk}(\beta)\,,
\end{equation}
and hence this gravitational theory indeed has a discrete spectrum determined by the eigenvalues of $\Hh$
\begin{equation}
    Z(\beta)=\Tr e^{-\beta \Hh} \,.
\end{equation}
Multi-boundary partition functions become just products of this, because all the wormholes still cancel.

Note that using \eqref{gaussian}, we can alternatively write the full partition function of the gravity theory as
\begin{align}
\label{fullXb}
Z(\beta) &= \int_0^\infty \d b\, b\,Z_\text{trumpet}(\beta,b) \sum_{i=1}^L  \frac{2}b \cos(b E_i^{1/2})\nn\\
&= - \int_0^\infty \d b\, b\,Z_\text{trumpet}(\beta,b) \sum_{i=1}^L\bigg(M_{\i E_i^{1/2}}(b)+M_{-\i E_i^{1/2}}(b)\bigg) \,.
\end{align}
In particular we recognize (both sheets of) the wavefunction of FZZT boundaries \eqref{210} in the allowed region. Hence, the full partition function can be viewed as a sum of cylinder geometries ending on FZZT boundaries, and the wavefunction $M_{\i E_i^{1/2}}(b)+M_{-\i E_i^{1/2}}(b)$ can be interpreted as preparing the microstate with energy $E_i$. Pictorially
  \begin{equation}
  \label{eq:H0-fixed}
       \Tr e^{-\beta \Hh} = -\sum_{i=1}^L\quad \int_0^\infty \d b \,b \,\bigg( \,\,\,\begin{tikzpicture}[baseline={([yshift=-.5ex]current bounding box.center)}, scale=0.55]
 \pgftext{\includegraphics[scale=0.25]{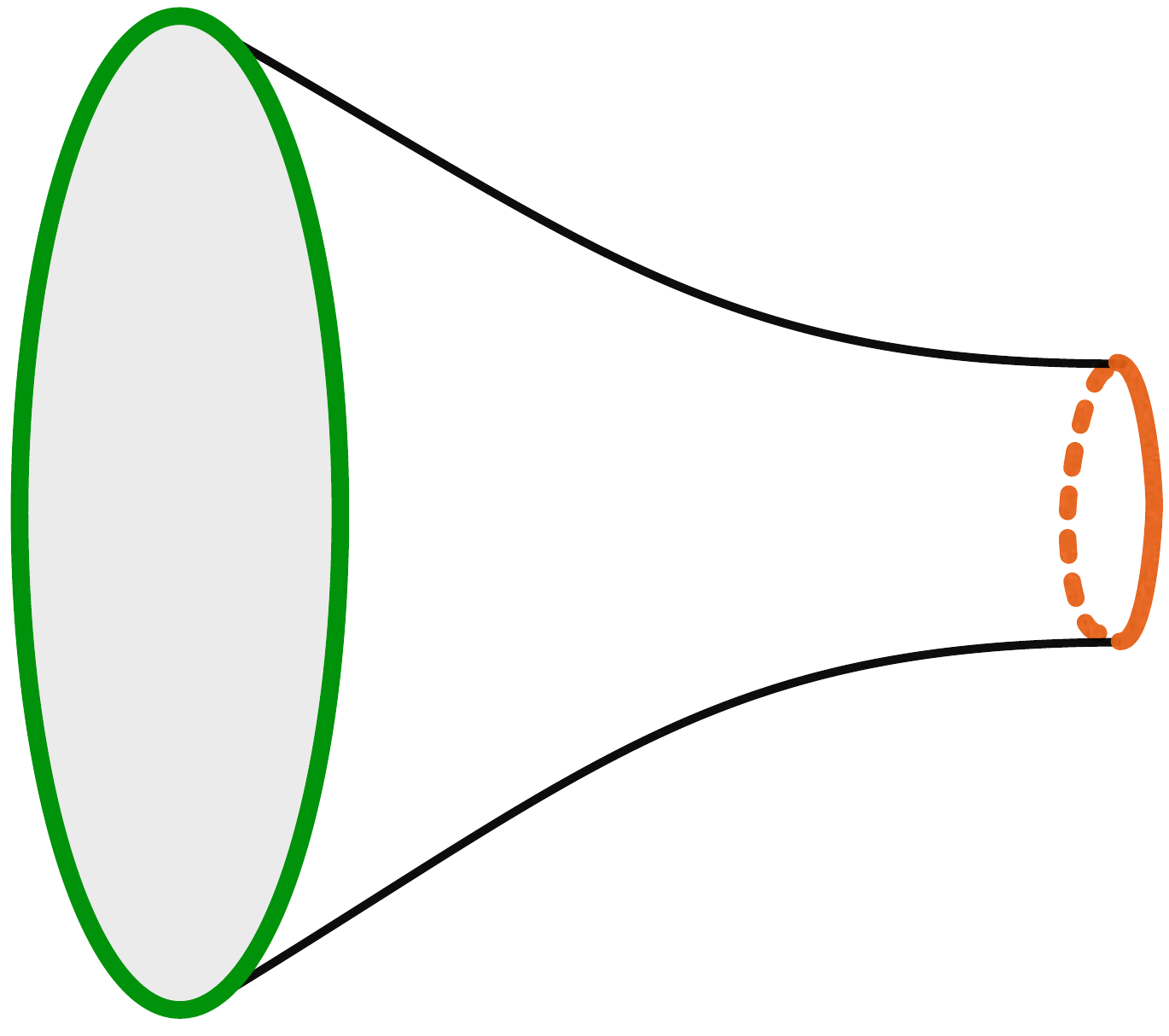}} at (0,0);
   \draw (1.85,-1.2) node  {$M_{\i E_i^{1/2}}(b)$}; 
  \end{tikzpicture}   +\quad\begin{tikzpicture}[baseline={([yshift=-.5ex]current bounding box.center)}, scale=0.55]
 \pgftext{\includegraphics[scale=0.25]{Half_wormhole_fixed_energy.pdf}} at (0,0);
   \draw (2,-1.2) node  {$M_{-\i E_i^{1/2}}(b)$}; 
  \end{tikzpicture}\bigg)\quad,
    \end{equation}
This has similar flavor to the eigenbranes picture of \cite{Blommaert:2019wfy}, with the important distinction that here we have a semiclassical interpretation without complicated sum over geometries. We have thus found that for each Hamiltonian $\Hh$ there is a correspondence
\be 
\Hh\quad  \Longleftrightarrow \quad Z_\text{brane}(b)\,,
\ee
which, in turn, can be understood as a linear combination of FZZT wavefunctions.

\subsection{Magnitude of UV corrections}\label{sect:recovering}
We want to understand when the half-wormhole or brane one-point function in \eqref{eq:Z(beta)-after-simplification} is a small correction to the leading black hole (as expected of UV corrections). We can distinguish two situations, depending on whether $\Hh$ is a typical (realistic) draw of the original JT gravity ensemble, or an atypical draw.

For typical draws we can accurately estimate the importance of the half-wormhole or brane one-point function by computing ensemble averaged over $\Hh$, using the same measure as in the original JT gravity ensemble. We denoted these averages by $\rL\dots  \rR$, to distinguish from the original matrix ensemble. Notice now that $Z_\text{brane}(b)$ as given by \eqref{solution} plays exactly the same role in the ensemble average over $\Hh$, as the role played by $\mo_{\rm G}(b)$ in \eqref{obdef} in the original ensemble average. In particular, ensemble averaging over $\Hh$ one recovers the exact analogue of \eqref{212}
\begin{equation}
    \rL Z_\text{brane}(b)\rR=\sum_{g=1}^\infty e^{\Ss(1-2g)}\,V_{g,1}(b)\,,\quad \rL Z_\text{brane}(b_1)\dots Z_\text{brane}(b_2) \rR_\text{\color{red}conn}=\sum_{g=0}^\infty e^{\Ss(2-2g-n)}\,V_{g,n}(b_1\dots b_n)\,.\label{316}
\end{equation}
The last equation can graphically be represented as
\be 
\label{eq:red-average}
\begin{tikzpicture}[baseline={([yshift=-.5ex]current bounding box.center)}, scale=0.6]
 \pgftext{\includegraphics[scale=0.7]{red_average_half_wormholes.pdf}} at (0,0);
     \draw (3.8,-2.5) node  {$\color{red}{\text{conn}}$};
  \end{tikzpicture} \quad = \quad 
  \begin{tikzpicture}[baseline={([yshift=-.5ex]current bounding box.center)}, scale=0.6]
 \pgftext{\includegraphics[scale=0.7]{n-wormhole.pdf}} at (0,0);
  \end{tikzpicture}  + \quad \text{higher genus}\,.
\ee 

Thus, even though the wormholes have canceled in the geometric expansion of the partition function $Z(\beta)$, they are still encoded in the statistics of the brane one-point function (or half-wormhole). From \eqref{eq:red-average} we see that all the moments of the one-point function are suppressed by powers of $e^{-\S}$ when we compute the partition function, 
so we conclude that for any typical draws $\Hh$ the UV corrections to the partition function are indeed suppressed as compared to the black hole geometry $Z(\beta) - Z_\text{disk}(\beta) \sim O(1)$, which is nontrivial since both terms on the left are $\sim O(e^{\Ss})$.

For atypical draws $\Hh$ there is no such cancellation occurring, and  $Z(\beta) - Z_\text{Disk}(\beta) \sim O(e^{\Ss})$. This implies that the brane one-point function scales as $\sim O(e^{\S})$, thus the half-wormhole contribution is not parametrically smaller than the disk contribution. This is no surprise, atypical draws are simply poorly described by JT gravity, so for those matrices $\Hh$, this construction makes little sense from the get-go. In summary,
\begin{align}
   \Hh\text{ typical member of the JT ensemble}&\quad\Longleftrightarrow\quad\text{brane corrections suppressed by $\sim O(e^{-\Ss})$} \nn \\  \Hh\text{ atypical member of the JT ensemble}&\quad\Longleftrightarrow\quad\text{brane corrections never suppressed}\nn  \,.
\end{align}

We note that for typical $\Hh$ and self-averaging quantities, the red average in \eqref{eq:red-average} can essentially be dropped. The wormhole computation for such observables approximates the answer in each member of the ensemble, which itself is captured by only the black hole and the half-wormhole geometries \eqref{eq:Z(beta)-after-simplification}. As a concrete example, consider the spectral form factor \cite{Cotler:2016fpe}. At exponentially late times $t>t_\text{ramp}$, the spectral form factor is not self-averaging, but a smeared version where we time average over an interval $\Delta T\gg 1$ \emph{is} self-averaging.\footnote{This is because at times on the ramp both the standard deviation of the spectral form, and the spectral form factor itself, scale as $\sim t/\beta$. After time averaging, the standard deviation scales as $\sim t/(\beta \Delta T)$, and the averaged signal as $\sim t/\beta$.} This implies the following approximate equality\footnote{The figure above involves an analytic continuation from Euclidean to Lorentzian geometry which here we consider occurs along a geodesic in the geometry. For the contribution of the half-wormhole we take this geodesic to be located along a geodesic that separates the spacetime into a region that contains a  brane and one that does  not. Thus, with our choice, the brane is located entirely in the Euclidean preparation region of the bra or the ket states.  This choice however does not appear to be unique and it would be interesting to explore other possible continuations to Lorentzian signature.  }
\begin{align}
\frac{1}{\Delta T}\int_{t-\Delta T}^{t+\Delta T} \d\tau\,  Z(\beta - \i\tau)Z(\beta + \i \tau)
&= \frac{1}{\Delta T}\int_{t-\Delta T}^{t+\Delta T} \d \tau\,\, \bigg| \begin{tikzpicture}[baseline={([yshift=-.5ex]current bounding box.center)}, scale=0.3]
 \pgftext{\includegraphics[scale=0.35]{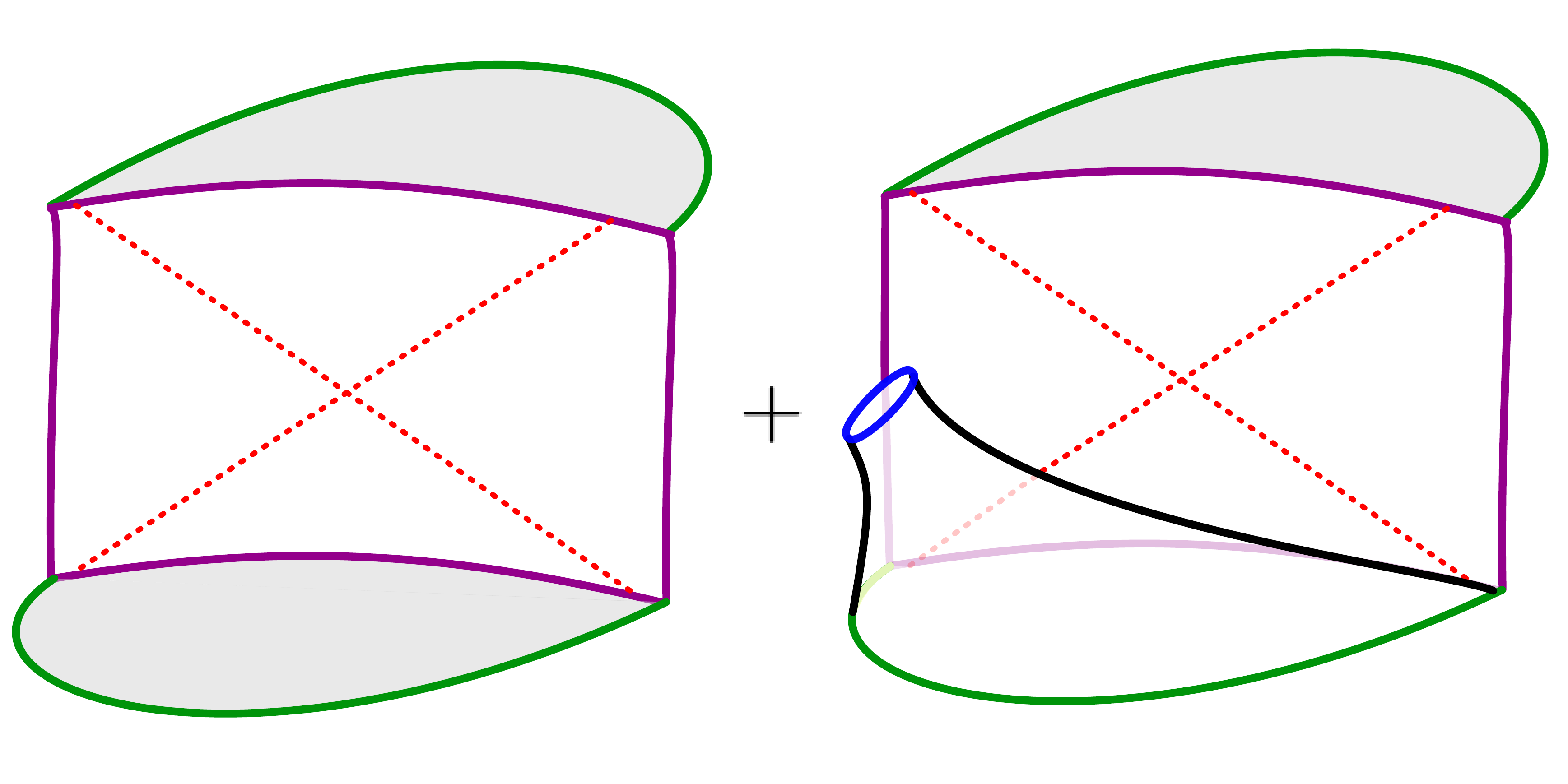}} at (-1,0);
  \end{tikzpicture}\bigg|^2\,\,  
 \\ 
 & {\approx}\,\,\begin{cases} 
 \,\, \bigg|\begin{tikzpicture}[baseline={([yshift=-.5ex]current bounding box.center)}, scale=0.21]
 \pgftext{\includegraphics[scale=0.35]{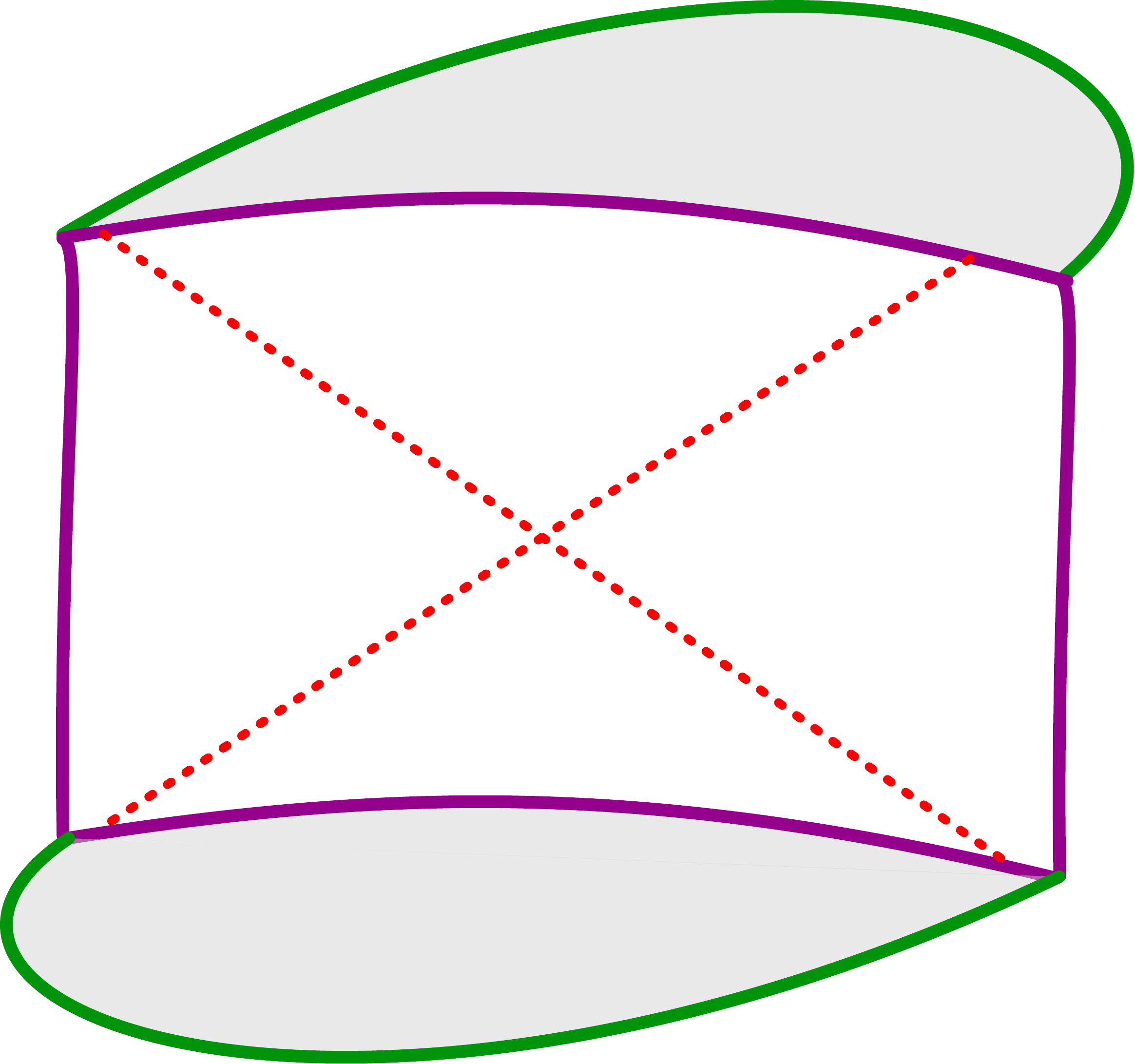}} at (-1,0);
  \end{tikzpicture} \bigg|^2 &\sim  \frac{e^{\S}}{t^3} e^{\frac{4\pi \beta}{t^2}}\,\text{ for }t < t_\text{ramp}\\
  \,\, \begin{tikzpicture}[baseline={([yshift=-.5ex]current bounding box.center)}, scale=0.25]
 \pgftext{\includegraphics[scale=0.35]{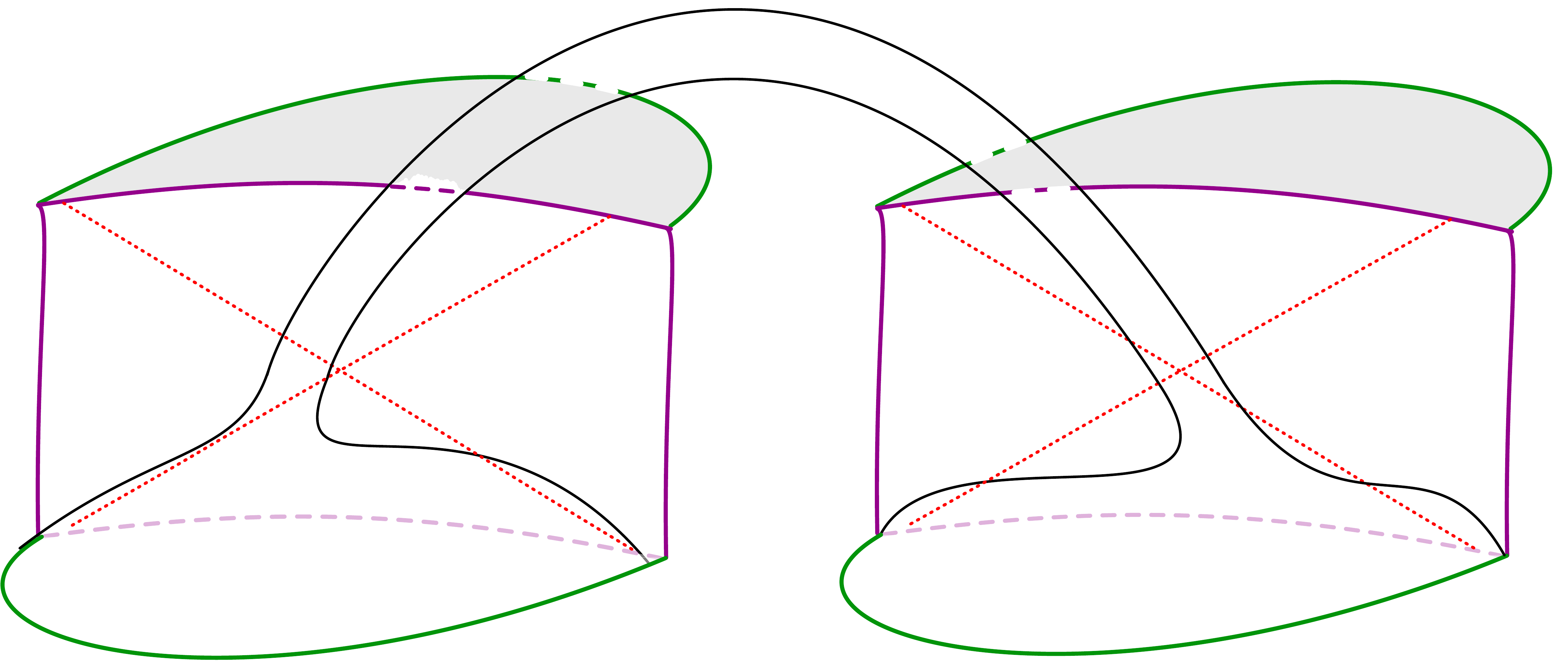}} at (-1,0);
  \end{tikzpicture}&\sim \frac{t}{\beta}\,\hspace{1cm}\text{ for }t_\text{ramp}<t<t_\text{plateau}\nn
  \end{cases}
\end{align}
In going from the second to the third line, we used the fact that the time averaged spectral form factor in quantum systems with (typical) Hamiltonians $\Hh$ is self-averaging, such that one can effectively apply \eqref{eq:red-average}. So, at early times one sees that both the half-wormhole (in our model) and  geometric wormhole (in JT gravity) are subleading but, when $t>t_\text{ramp}$, the half-wormhole contribution starts dominating in the time-averaged spectral form factor, and crucially the result is well approximated by the JT gravity wormhole. This is why simple gravity models, with wormholes, but without branes, are oftentimes good approximations.


\subsection{Additional comments}
\label{sec:additional-comments}

A few comments are in order.
\begin{enumerate}
    \item An important ingredient that has led to the simple expression for the brane couplings \eqref{eq:solution-to-factorization} comes from allowing correlations between branes on closed universes and branes on geometries that include an asymptotic boundary. One could attempt to absorb the closed universes in dressed brane correlators (black) that only attach to asymptotic universes, which would be non-Gaussian
    \be 
    \begin{tikzpicture}[baseline={([yshift=-.5ex]current bounding box.center)}, scale=0.55]
 \pgftext{\includegraphics[scale=0.60]{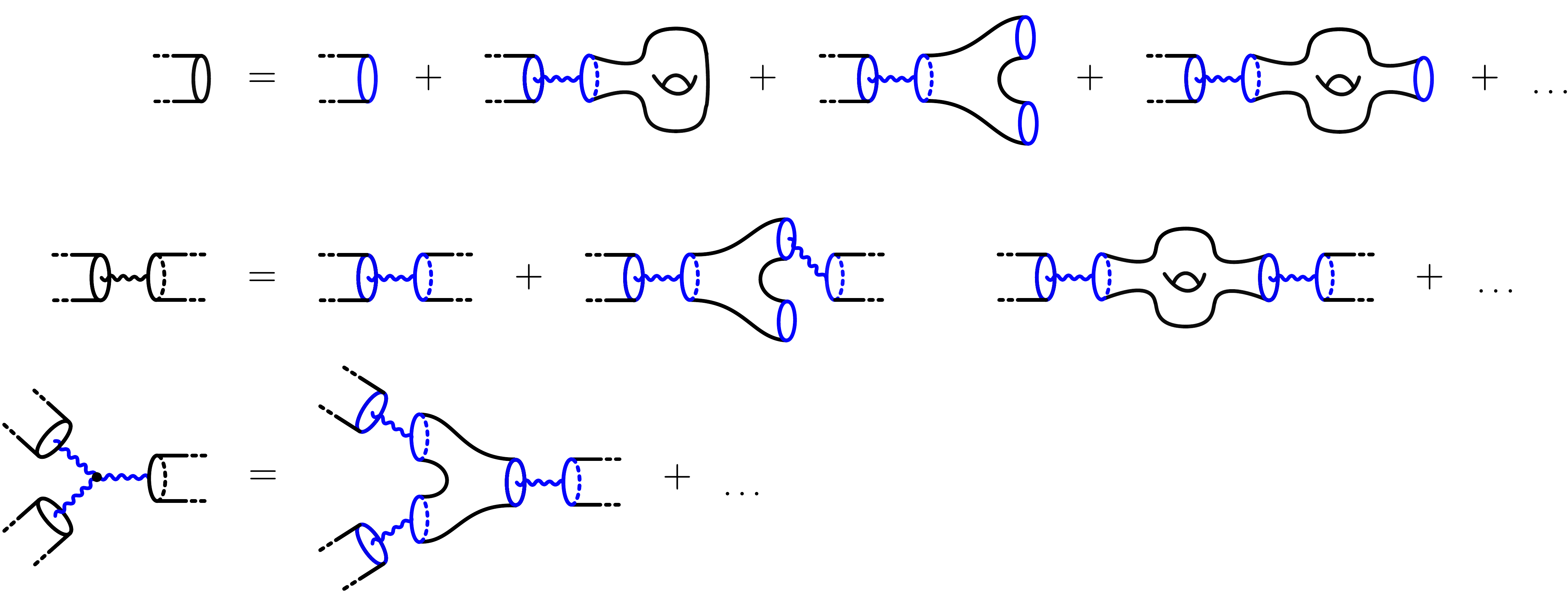}} at (0,0);
  \end{tikzpicture}
    \ee

    However, some of the terms in this expansion are divergent, which suggests that the more natural expansion is the original one, which does allow branes connecting closed and asymptotic universes. We believe that the closed universes are an important conceptual ingredient in this story. From the perspective of dilaton gravity with nonlocal interactions \eqref{nonlocex} it also makes most sense to think about things in terms of there being closed universes as well.
    
    \item Tracing the presence of the brane correlators back to the non-local deformations of the action, we see that the cancellation of all wormholes is due to the perturbative expansion of these nonlocal deformations around the leading black hole saddle.  For example, the cancellation of the leading two-boundary wormhole is obtained by expanding the non-local deformation around the two copies of the black hole saddle-point \eqref{eq:leading-wormhole-cancelling}. This re-emphasizes the importance of tracking such non-local terms in the low-energy effective theory of gravity. We will further discuss the possible meaning of such nonlocal terms in the discussion in section \ref{sect:discussion}.
    
    \item It is essential for the arguments of section \ref{sect:allorders} to work, that there is an exponential of two-brane correlators in our theory which results in a sum over all possible brane insertions. In these points,  we substantially from \cite{Saad:2021uzi}, and it is why we obtain factorization to all orders in the $e^{-S_0}$ expansion.
    \item The simplicity of the result is a consequence of the fact that the integration measure and integration range when gluing trumpets to any surface $\Sigma$ in \eqref{eq:general-identity-that-makes-geometries-vanish}, is independent of the topology and moduli of $\Sigma$. This is because gluing a trumpet to $\Sigma$ changes the mapping class group just by introducing an extra factor $\text{MCG}_\Sigma \times \mathbb Z$.

    \item The brane correlators are independent of the parameters that determine the leading black hole geometry, indeed \eqref{eq:solution-to-factorization} is independent of $e^{\S}$ and $\beta$. This suggests a possible non-geometric origin for the brane correlators, consistent with our intuition about the stringy realization of these UV corrections in the section \ref{sec:intro} and section \ref{sect:discussion}.
    \item \textit{Factorization implies null-states in the baby-universe Hilbert space.} One might worry that the negativity of the connected brane correlator $Z_{\rm brane}(b_1, b_2)$ might lead to a lack of reflection positivity in the gravitational Hilbert space of the theory. To show that this is not the case we will use the baby universe Hilbert space formalism of \cite{Marolf:2020xie}. In this Hilbert space, states are obtained from boundary creation operator, which we can take to create asymptotic boundaries with boundary length $\beta$ and fixed dilaton. The norm of any such state can be expressed as\footnote{We thank D.~Stanford and Z.~Yang for useful comments in this direction.}  
  \be 
    \<\Psi| \Psi\> =  \alpha^\dagger \mathcal Z \alpha \,,\qquad \text{ where } \qquad |\Psi\> = \sum_{i=1}^{n_\text{bdies}} \alpha_i\, |Z(\beta)^i\>\,,
    \ee
    where $n_\text{bdies}$ is the maximum number of boundaries involved in the state $|\Psi\>$. Factorization implies that the inner product matrix $\mathcal{Z}_{ij} = Z(\b)^{i+j}$ (whose dimension is $n_\text{bdies}\times n_\text{bdies}$) has 
    rank $1$ with its one non-zero eigenvalue being positive and equal to $\sum_{i=1}^{n_\text{bdies}} Z(\beta)^{2i}$. This implies that $  \<\Psi| \Psi\> \geq 0$ for all $\alpha$, and thus reflection positivity is not violated in this baby universe Hilbert space. Furthermore, the fact that  $\mathcal Z$ is rank $1$ is a signature of the existence of null-states which are associated to imposing factorization within in our model. In the usual GNS construction these are then modded out and the final Hilbert space is one-dimensional, consistent with being in a single $\alpha$-state. The geodesic boundaries can be thought of as an additional set of universe creating operators $\hat{\mathcal{O}_{\rm G}}(b)$, with correlations \eqref{solution}. These correlations are only accessible to observers with access to different universes, and are inaccessible to observers within a single universe. Observers within a single universe can only measures the total correlation (which vanishes), see section \ref{sect:discussion}.
    
    \item  \textit{The $b$ integral in \eqref{eq:H0-fixed} has saddlepoints. }The integrand of \eqref{eq:H0-fixed} is even, so we can extend the integral to the entire real axis and deform the contour through the saddlepoint at $b_i = 2\b \i E^{1/2}_i$ \eqref{gaussian}. Thus for $E_i>0$ (which is the set of energies we focus on) this saddle is at \emph{imaginary} $b$. This means we can actually view it (on-shell) as a defect geometry with defect angle 
    \be 
    \a_i = 2\pi + \i b_i = 2\pi - 2 \b E_i^{1/2}. 
    \ee
    It would be interesting to understand the precise meaning of the emergence of defect geometries here and in particular its fate upon analytic continuation to Lorentzian signature. In particular how we cut the Euclidean geometry and glue it to a Lorentzian section. 
\end{enumerate}
All statements thus for hold to all orders in $e^{-\S}$ perturbation theory. We now prove that actually \eqref{solution} yields \eqref{fullXb} even non-perturbatively, the correlated branes truly fix the eigenvalues of the theory to $E_i$.
\section{Factorization and discreteness from matrix integral localization}
\label{sect:matrix-integral-localization}

Here we explain from the matrix integral description of two dimensional dilaton gravity why introducing extra bilocal interactions in the spacetime action is sufficient to obtain factorizing gravity systems, and how the brane one point functions discretize the spectrum.

The partition functions of the relevant matrix ensembles are defined as \cite{mehta2004random,Saad:2019lba}
\begin{align}
    \mathcal{Z}&=\prod_{i=1}^L\int_\mathcal{\cont}d\lambda_i\exp\bigg(-L\sum_{i=1}^L V(\l_i)\bigg)\prod_{i<j}^L(\lambda_i-\l_j)^2=\prod_{i=1}^L\int_\mathcal{\cont}d\lambda_i\exp\bigg(-L\sum_{i=1}^L V(\l_i)+\sum_{i\neq j}^L\log(\l_i-\l_j)\bigg)\,,\label{matint}
\end{align}
where the potential $V(E)$ determines which dilaton gravity we are studying. Typical observables in this matrix integral are products of spectral densities
\begin{equation}
    \rho(E)=\sum_{i=1}^L\delta(E-\lambda_i).\label{speccor}
\end{equation}
This corresponds in gravity with computing amplitudes with fixed energy boundaries, which are inverse Laplace transforms of the typical fixed length partition functions $Z(\beta)$. The equations of motion from varying the action in \eqref{matint} with respect to $\l_i$ relate the potential with the large $L$ leading order spectrum $\rho_0(E)$ \cite{brezin1993planar}
\begin{equation}
   \quad L V(\l)=2\fint_\G d E\,\log(\l-E)\,\rho_0(E)+\text{constant} \quad \Rightarrow \quad L V'(\l)=2\fint_\G d E\,\frac{1}{\l-E}\,\rho_0(E)\,,\label{eom}
\end{equation}
where $\G$ is the cut along which $\rho_0(E)$ has non-zero support. This principal value integral can be inverted to determine $\rho_0(E)$ in terms of the potential \cite{migdal1983loop}. The constant is the Lagrange multiplier which is used to fix the normalization of $\rho_0(E)$ in the collective field formulation; its integral over the region $\G$ is $L$. Subleading corrections and higher point correlators of $\rho(E)$ can be calculated by expanding around the saddle $\rho_0(E)$. The Vandermonde encodes non-trivial correlation between different eigenvalues \cite{Cotler:2016fpe}. 

The gravitational dual of correlated eigenvalues are connected geometries (i.e. wormholes) between different asymptotic regions. For example the leading order variance of the spectral density is computed in gravity by computing the wormhole between two different asymptotic regions \cite{Saad:2019lba}
\begin{equation}
    \rho(E_1,E_2)_\text{conn}=-\frac{1}{4\pi^2}\frac{E_1+E_2}{E_1^{1/2}E_2^{1/2}}\frac{1}{(E_1-E_2)^2}+\text{subleading higher genus corrections.}
\end{equation}
Converting this to fixed $\b$, one simply sums over the wormhole geometry, with any number of handles.
\subsection{Localization concept}
In section \ref{sect:factor} we saw that one can obtain a factorizing gravitational theory by including branes with specific correlation functions. In particular, we saw that bilocal interactions between different branes can conspire to precisely cancel the geometric wormhole correlation between different asymptotic boundaries. It is surprising that bilocal interactions are sufficient to attain factorization, since a priori there seems little gravitational reason to expect that generic multi-local interactions would not feature. 

However from the matrix model side this has a simple explanation. Introducing bilocal interactions in gravity, corresponds in the matrix integral with introducing a double-trace deformation,
\begin{equation}
    \exp\bigg( \sum_{m,n=0}^\infty f_2^{m n}\,\Tr H^m \,\Tr H^n \bigg) = \exp\bigg(\sum_{i,j=1}^L f_2(\l_i,\l_j)\bigg)\,.
\end{equation}
This modifies the repulsive force between eigenvalues. The basic intuition how a discrete spectrum then arises is that we will tune the repulsive interactions and the potential in such a way that the eigenvalues are frozen at the locations of the eigenvalues of a matrix $\Hh$, drawn from the JT ensemble.
Since the undeformed matrix integral itself is comprised of just single-and double trace terms, it is not necessary to include higher-trace deformations to achieve this.

More concretely, we will prove that the deformation corresponding to the brane insertions discussed in section \ref{sect:factor}, result in a deformed matrix integral of the form
\begin{align}
    \mathcal{Z}(\Hh)=\lim_{q\to\infty}\prod_{i=1}^L\int_\cont d\l_i\,\exp\bigg( q\sum_{i,j=1}^L I(\l_i, E_j)\bigg) \,,\label{finmatint}
\end{align}
where $q$ is a parameter, larger than any other quantity in the matrix integral, whose origin we explain shortly. Because $q\to\infty$ the integrals over $\l_i$ localize onto the stationary points of the ``action'' $I(\l_i,E_j)$, given in \eqref{426}. We will prove that, with the brane deformations of section \ref{sect:factor}, the stationary points are\footnote{There can be other saddles, but we will prove below that these are the dominant ones, to which the integral localizes.}
\be 
\label{eq:saddle-point-equation}
\frac{\partial}{\partial \l_i}\sum_{i,j=1}^L I(\l_i ,E_j)= 0\quad\Leftrightarrow\quad \l_1\dots\lambda_L = \text{permutations of }E_1\dots E_L\,.
\ee
Therefore, indeed, the set of random eigenvalues $\lambda_i$ are frozen to the energy spectrum $E_i$ of $\Hh$.\footnote{There are different ways of achieving this localization, here we use the single-and double-trace deformations of section \ref{sect:factor}. This differs from the matrix integral action used in \cite{Blommaert:2021gha}, which involved a deformation that broke the $U(L)$ invariance.} Thus, for all purposes, the matrix integral \eqref{finmatint} reduces for $q\to\infty$ to
\begin{equation}
    \mathcal{Z}(\Hh)=\prod_{i=1}^L\int_\cont d\l_i\,\prod_{j=1}^L \sum_{i_j=1}^L\delta(\l_{i_j}-E_j)\,.\label{final}
\end{equation}
This same result was obtained with the eigenbrane picture \cite{Blommaert:2019wfy,Blommaert:2020seb} and the proposal in \cite{Blommaert:2021gha}, but notably the gravitational description we presented in section \ref{sect:factor} is much simpler; we never lose control of gravity as the full theory \eqref{simple} has a good semiclassical interpretations.

We note that already in \cite{Blommaert:2021gha}, there were signs that nonlocal interactions should become important in a factorizing theory of gravity, as pointed out for example in the discussion there. For large $1/\s^2$, which is analogous to large $q$ here, multi-trace deformations in the Harish-Chandra integral become important. We believe that a more careful analysis of the Harish-Chandra integral in the double scaling limit, should reveal the same Gaussian double-trace deformation as the one which we shall discuss in the next section. It would be interesting to make that precise.

We now derive the form of $I$ that follows from inserting the correlated branes of section \ref{sect:factor}, and prove that it indeed satisfies \eqref{eq:saddle-point-equation}.
\subsection{Argument}\label{sect:calculation}
Based on the gravity setup of section \ref{sect:branes}, we are led to the following deformation of the JT matrix integral
\begin{align}
\exp\bigg(\int_0^\infty \d b\, b\,\mo_{\rm G}(b)\,z_\text{brane}(b)+\frac{1}{2}\int_0^\infty \d b_1 b_1 \int_0^\infty \d b_2 b_2\,\mo_{\rm G}(b_1) \mo_{\rm G}(b_2)\,z_\text{brane}(b_1,b_2) \bigg)\,.\label{bb}
\end{align}
Recall from section \ref{sect:branes} that FZZT boundaries corresponds in the matrix integral to insertions of 
\begin{equation}
    \mo_\text{FZZT}(z)=\sum_{i=1}^L \log(\lambda_i+z^2)-\frac{L}{2} V(-z^2)\,,\label{dic}
\end{equation}
where $z^2=-E$. Notice also the potential term for FZZT branes.
Furthermore, as discussed in section \ref{sect:branes}, we can write a smeared FZZT brane as an operator that creates geodesic boundaries in the spacetime\footnote{To see that this is indeed equivalent to \eqref{obdef}, note the relation of the potential with the genus zero spectrum \eqref{eom}.}
\begin{align}
\label{eq:regulated-O(b)}
    \mo_{\rm G}(b)=-\frac{1}{2\pi i}\int_{-i\infty}^{+i\infty} \d z\, e^{b z}\,\mo_{\rm FZZT}(z)=\sum_{i=1}^L\frac{2}{b}\,e^{-b\varepsilon}\cos(b \lambda_i^{1/2}) - \frac{L}{2}W(b)\,,
\end{align}
where we introduced the inverse Laplace transformed potential
\be 
W(b) = \frac{1}{2\pi i}\int_{-i\infty}^{+i\infty} \d z\, e^{b z}\,V(-z^2).\label{Wb}
\ee
Here, the regulator $\varepsilon$ appears because FZZT branes are actually located an $\varepsilon$ left of the imaginary axis, resolvents have small imaginary parts in their energy arguments \cite{Saad:2019lba}. This regulator was not important in section \ref{sect:factor}, but here we should properly keep track of it.

One important point is that, in equation \eqref{bb}, we should be inserting \emph{renormalized} brane correlators
\be 
\label{eq:Z-brane-correlator}
\lim_{q\to\infty}\frac{1}{1+q}\,z_\text{brane}(b) = \sum_{j=1}^L\frac{2}{b}\cos( b E_j^{1/2})-\frac{L}{2}W(b)\,, \quad \lim_{q\to\infty}\frac{1}{q}\,z_{\brane}(b_1,b_2) = -\frac{1}{b_1}\delta(b_1 - b_2)
\ee
This expression for the propagator $z_\text{brane}(b_1,b_2)$ follows from the geometric propagator $Z_{\brane}(b_1,b_2)$ of section \ref{sect:factor}. To understand how this renormalization comes by, notice that in the undeformed JT matrix integral one finds
\be \label{degenerateCylinder}
\average{\mo_{\rm G}(b_1)\mo_{\rm G}(b_2)} =\frac{1}{b_1}\delta(b_1-b_2) + O(e^{-\Ss})\,.
\ee
Geometrically this term represents the connected degenerate cylinder, which we explicitly excluded from our geometric model. However, in the matrix integral we have no choice, these amplitudes do exist. The full two-point function of $\mo_{\rm G}(b)$ in the model with the double trace interaction is then a Dyson series,
because we can insert pairs of $\mo_{\rm G}(b)$ which are correlated via the brane two-point function $z_\text{brane}(b_1,b_2)$. One should view $z_\text{brane}(b_1,b_2)$ as a \emph{bare} brane propagator, and $Z_{\brane}(b_1,b_2)$ as the \emph{dressed} propagator, obtained by resumming a Dyson series of $z_{\brane}$ propagators connected through degenerate cylinders
\begin{equation}
\begin{tikzpicture}[baseline={([yshift=-.5ex]current bounding box.center)}, scale=0.6]
 \pgftext{\includegraphics[scale=0.80]{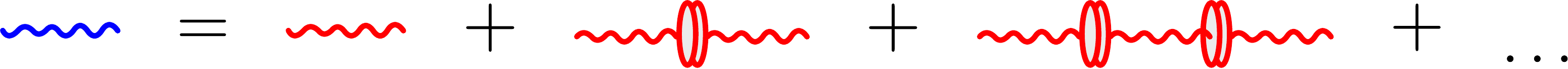}} at (0,0);
  \end{tikzpicture}\; .
  \label{eq:geometric-relation-blue-to-red-prop}
\end{equation}
Note that no other closed universe geometries appear in \eqref{eq:geometric-relation-blue-to-red-prop}; they are already included in the topological expansion defined in section \ref{sect:factor}, which was expressed in terms of blue correlators.  In \eqref{eq:geometric-relation-blue-to-red-prop}, all  diagrams contribute at the same order in $e^{\Ss}$. This is similar to how in QFT one would treat a mass term perturbatively by resumming diagrams. In formulas, the above Dyson equation reads
\begin{align}
Z_{\brane}(b_1,b_2) &= z_{\brane}(b_1,b_2) + \int_0^{\infty} \d b_3 b_3\, z_{\brane}(b_1,b_3)z_{\brane}(b_3,b_2)\nn\\
&\quad + \int_0^{\infty} \d b_3 b_3 \int_0^{\infty}\d b_4 b_4\, z_{\brane}(b_1,b_3)z_{\brane}(b_3,b_4)z_{\brane}(b_4,b_2) + \dots  \label{tosolve}
\end{align}
As $Z_\text{brane}(b_1,b_2)=-\delta(b_1-b_2)/b_1$, the solution is $z_{\brane}(b_1,b_2) = -q \delta(b_1-b_2)/b_1$, where $q$ must satisfy
\be 
\label{eq:naively-divergent}
-1 = -q + q^2 - q^3 + \dots = -\frac{q}{1+q}.
\ee
This has two solutions $q\to \pm \infty$, but since we want the matrix integral including the deformation \eqref{bb} to localize we choose $q \to +\infty$. One might additionally be concerned that \eqref{eq:naively-divergent} is only convergent for $q \to +\infty$ upon analytic continuation in $q$. However, \eqref{eq:naively-divergent} is solely used to determine which operator deformation in the matrix integral is equivalent to the brane insertions discussed in section \ref{sect:factor}; in the resulting matrix integral, no such convergence issues are encountered. 

The renormalization of the bare propagator by a factor $q$ also affects the first term in \eqref{bb}, i.e. the full one-point function of $\mo_{\rm G}(b)$ in the deformed matrix model. Comparing this again with our finding in section \ref{sect:factor}, we see that including degenerate cylinders results in the series
\be 
\begin{tikzpicture}[baseline={([yshift=-.5ex]current bounding box.center)}, scale=0.6]
 \pgftext{\includegraphics[scale=0.80]{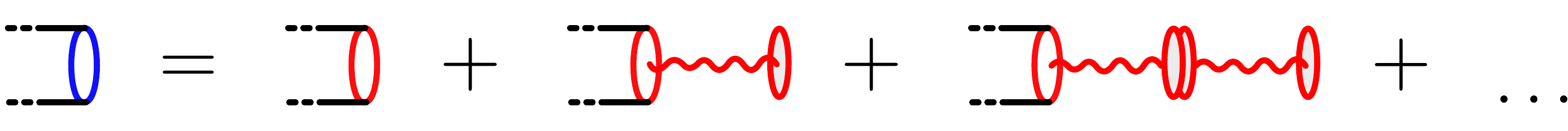}} at (0,0);
  \end{tikzpicture}\; .
\ee
which, concretely, can be rewritten as
\begin{align}
Z_{\brane}(b) &= z_{\brane}(b) + \int_0^{\infty} \d b_1 \, b_1 z_{\brane}(b,b_1)z_{\brane}(b_1) \nn\\
&\quad + \int_0^{\infty} \d b_1 \, b_1 \d b_2 \, b_2 z_{\brane}(b,b_1)z_{\brane}(b_1,b_2)z_{\brane}(b_2) + \dots \,.
\end{align} 
Plugging in the result for $z_{\brane}(b_1,b_2)$ that we just derived, this equation becomes 
\be 
Z_{\brane}(b) = \lim_{q\to \infty} \frac{1}{1+q}\, z_{\brane}(b)\,,
\ee
and so we need $z_{\brane}(b) = (q+1)Z_{\brane}(b)$ modulo subleading terms in $q$ which will not be of importance. We emphasize that requiring consistency with results from section \ref{sect:factor}, which demanded factorization and discreteness, forces $q\to\infty$. Finite $q$ does not result in a factorizing or discrete system, not only because the matrix integral does not localize, but this also simply would not match with our geometric derivation. To summarize, the factors of $q$ come from taking into account the degenerate cylinders.

We now prove that \eqref{bb} indeed gives the localization described around \eqref{eq:saddle-point-equation}. The workhorse formula which we use to compute all integrals in \eqref{bb} is
\be
    -4\int_\delta^\infty \frac{\d b}{b}e^{-b\varepsilon}\cos(b E_1^{1/2})\cos(b E_2^{1/2}) = P(E_1,E_2) + 4 \log \delta\,,
\ee
Here $\delta$ regulates the small $b$ divergence of this integral, which will drop out of all observables, and $P(E_1,E_2)$ is given by
\begin{align}
\label{eq:def-P}
   P(E_1,E_2) &= \log\Big(\left(E_1^{1/2} - E_2^{1/2}\right)^2 + \varepsilon^2\Big) + \log\Big(\left(E_1^{1/2} + E_2^{1/2}\right)^2 + \varepsilon^2\Big)\\&= 
   \left\{\begin{array}{ll}
      2\log|E_1 -E_2|  & \text{ for }E_1\neq E_2 \\
      \log \varepsilon^2 + \log (4|E_1|)  & \text{ for } E_1 = E_2
   \end{array}\right.\,.\label{eq:def-P-regulated}
\end{align}
The quadratic term in $\mo_{\rm G}(b)$ in \eqref{bb} can now be immediately evaluated
\begin{align}
\label{eq:regularized-integral}
    -\frac{q}{2}\int_{\delta}^\infty \d b\, b\,\mo_{\rm G}(b)\mo_{\rm G}(b) &= \frac{q}{2} \sum_{i,j=1}^L P(\l_i,\l_j) + q L \sum_{i=1}^L \int_{\delta}^{\infty} \d b\, W(b)\,e^{-b\varepsilon}\cos(b\l_i^{1/2}) \nn\\
    &\qquad - \frac{q}{2}\frac{L^2}{4}\int_{\delta}^{\infty} \d b\, b\, W(b)W(b) + 2 q L^2 \log \delta\nn\\&=\frac{q}{2} \sum_{i,j=1}^L P(\l_i,\l_j)-q L\sum_{i=1}^L V(\l_i)+\text{constant.}
\end{align}
The constant on the second line is independent of $\l_i$, and thus cancel when normalizing our matrix integral. Subsequently we drop such terms. On the last line we used the definition of the inverse Laplace transformed potential $W(b)$ in \eqref{Wb}, and did the integrals over $b$ and $z$ explicitly. Similarly, the linear in $\mo(b)$ part in \eqref{bb} evaluates to
\begin{align}
    \int_0^\infty db b\, \mo_{\rm G}(b) z_{\brane}(b) &= -(1+q)\sum_{i,j=1}^L P(E_i,\l_j) + (1+q) L \sum_{i=1}^L V(\l_i)\,.
\end{align}
Combining the these two expressions, we find that \eqref{bb} results in the insertion of
\begin{align}
\exp\bigg(\frac{q}{2} \sum_{i,j=1}^L P(\l_i,\l_j) - (1+q)\sum_{i,j=1}^L P(E_i,\l_j)+L\sum_{i=1}^L V(\l_i)\bigg)\,.
\end{align}
The last term cancels the potential in the original matrix integral, so we end up with the matrix integral
\begin{align} 
\label{eq:matrix-int-partition-function}
\mathcal{Z}(\Hh)&=\lim_{q\to\infty}\prod_{i=1}^L\int_\cont \d\l_i\,\exp\bigg(\sum_{i\neq j}^L\log(\l_i-\l_j) + \frac{q}{2} \sum_{i,j=1}^L P(\l_i,\l_j)-(q+1)\sum_{i,j=1}^L P(E_i,\lambda_j)\bigg)\\&=\lim_{q\to\infty}\prod_{i=1}^L\int_\cont \d\l_i\,\exp\bigg(\frac{q+1}{2}\sum_{i\neq j}^L\log(\l_i-\l_j) + \frac{q}{2} \sum_{i=1}^L P(\l_i,\l_i)-(q+1)\sum_{i,j=1}^L P(E_i,\lambda_j)\bigg)\,.\nn
\end{align}
Here we have used \eqref{eq:def-P-regulated} to identify the off-diagonal terms as changing the power of the Vandermonde. The second and third term on the second line can be viewed as the new potential of the matrix integral, which now depends on the spectrum of $\Hh$. The action in \eqref{finmatint} for $q\to\infty$ becomes
\be 
\sum_{i,j=1}^L I(\l_i,E_j) = \frac{1}{2} \sum_{i,j=1}^L P(\l_i,\l_j)-\sum_{i,j=1}^L P(E_i,\lambda_j)+O(1/q)\,.\label{426}
\ee 
The $O(1/q)$ terms just get evaluated on-shell in the $q\to\infty$ limit.

We can now check that the equations of motion \eqref{eq:saddle-point-equation} indeed have the solution $\l_1\dots \l_L=$ permutations of $E_1,\dots, E_L$, by noticing that for $\varepsilon \to 0$ (one should only take $\varepsilon \to 0$ at the very end)
\be 
\frac{\partial}{\partial\l_i} P(\lambda_i, \lambda_i)=\frac{1}{\lambda_i}\,, \quad 
\frac{\partial}{\partial\l_i} P(E_j, \lambda_i) = \frac{1}{2E_j }\text{ when }\l_i=E_j\,.
\ee
Note that the solution is independent of the regulator $\varepsilon$ needed in order to regulate the integral in \eqref{eq:regulated-O(b)}. The on-shell action of this solution is large and scales with both $-\log(\varepsilon)$ and $q$, making it the dominant solution, and the variance is small, therefore the matrix integrals localizes on these solutions for $q\to\infty$.

To compute observables such as $Z(\beta_1\dots\beta_n)$ in this deformed matrix integral, we have to normalize the answer by the matrix integral partition function \eqref{finmatint}, to find 
\begin{align}
   Z(\beta_1\dots\beta_n)&=
   \frac{1}{\mathcal{Z}(\Hh )}\lim_{q\to\infty}\prod_{i=1}^L\int_\cont d\l_i\,\exp\bigg( q\sum_{i,j=1}^L I(\l_i, E_j)\bigg)\,\sum_{i_1=1}^Le^{-\beta_1\l_{i_1}}\dots \sum_{i_n=1}^Le^{-\beta_n\l_{i_n}}\nn \\
     & = \sum_{i_1=1}^L e^{-\beta_1 E_{i_1}}\dots \sum_{i_n=1}^L e^{-\beta_n E_{i_n}}=  \Tr e^{-\beta_1 \Hh} \dots  \Tr e^{-\beta_n \Hh}\,.
\end{align}
We emphasize that the $q$-dependent one-loop determinants, the $L!$ factor coming from the total number of solutions, and constants such as that in \eqref{eq:regularized-integral} drop out between the numerator and the denominator. Thus, our matrix integral computation confirms that, even non-perturbatively, our model (that involves correlated spacetime branes) has resolved the factorization puzzle and yields a discrete spectrum. 

\subsection{Additional comments}

Several comments are in place:
\begin{enumerate}

\item \textit{Factorization and discreteness are intimately linked. } It is interesting to see what becomes of the theory with only the bilocal deformation, so with $z_{\brane}(b) = 0$ and $Z_{\brane}(b) = 0$. According to the discussion of section \ref{sect:factor}, this should be sufficient to obtain a factorizing theory of gravity. For this we consider the matrix integral with only the deformation \eqref{eq:regularized-integral}. One sees that the localization argument still applies, but now the saddle point equations feature the original potential $V(E)$
\begin{equation}
\frac{\partial}{\partial \l_i}\sum_{i,j=1}^L I(\l_i ,E_j)= 0\quad\Leftrightarrow\quad \frac{1}{\lambda_i} + \sum_{j\neq i} \frac{1}{\l_j - \l_i}  = L V'(\lambda_i)\,.\label{eomeom}
\end{equation}
These are the equations of motion for some $(\upalpha,\upbeta)$ ensemble with $\upalpha = 1+2\upbeta$ with $\upbeta\to\infty$ \cite{Stanford:2020wkf,Dijkgraaf:2009pc}. In the limit $q \to \infty$ and in particular $q \gg L$ the usual techniques for solving matrix models break down. In particular, one can no longer make the approximation that the spectrum is continuous, and the loop equations collapse in this regime to the above discrete set of equations.\footnote{In the standard large $L$ saddle-point, the equations of motion are imposed for all $\l$ in some to-be-determined spectral cut. One can interpret the discrete set here as the matrix integral having $L$ cuts around the eigenvalues, with width going to zero for $q\to\infty$. This behavior of nearly-factorized matrix integrals was also observed using different techniques in \cite{Blommaert:2021gha}.} The limit $q\to\infty$ \emph{freezes} the eigenvalues to one classical solution of \eqref{eomeom} (up to permutations). Even though we fixed $Z_{\brane}(b) = 0$, the eigenvalues are fixed non-perturbatively, away from the continuous leading density of states $\rho_0$, to the solutions of the saddle-point equation \eqref{eomeom} (once again, well-approximated by $\rho_0$).  A similar localization in the context of the $\upbeta\to\infty$ limit of the $\upbeta$ ensembles (with a Gaussian potential) was discussed in \cite{dumitriu2005eigenvalues}, where the matrix integral localized on the zeros of the $L$-th Hermite polynomial.\footnote{The matrix model action in that case looks like $\upbeta \sum_{i\neq j} \log(\l_i - \l_j) - \upbeta L \sum_{i}\l_i$ and so at large $\upbeta$ it localizes on the solutions of $\sum_{i\neq j} (\l_i - \l_j)^{-1} = 2L\l_j$, which are the equations for the zeros of the Hermite polynomial $H_L(\sqrt{L}x)$. The solutions for large $\upbeta$ in our case $(\upalpha,\upbeta) = (1+2\upbeta,\upbeta)$ would then again be the roots of the relevant orthogonal polynomial.} In summary, turning on only the bilocal deformation (which is sufficient to obtain factorization), already discretizes the spectrum non-perturbatively. The parameter $q$ did not make any immediate appearance in section \ref{sect:factor}, but here we see that treating the limit $q\to\infty$ non-perturbatively, gives an intimate relation between factorization and discreteness.

\item \textit{Large deformations can lead to small changes.} Consider $\Hh$ to be a typical matrix in the JT matrix integral ensemble.  While the deformations performed in the matrix integral are very large (since we were forced to take the $q\to \infty$ limit), typical observables in the matrix integral are only affected at subleading order in $e^{-\S}$. This is so because the leading order density of states $\rho_0(E)$  remains unaffected after turning on the deformation. This is equivalent to our observation in section \ref{sect:recovering}  that the half-wormhole correction is subleading to the black hole solution.

\item Notice that \eqref{final} is consistent with the fact that observables in the gravity theory dual to the fixed quantum mechanics, reproduce the observables in the original dilaton gravity upon ensemble averaging again over the spectrum $E_i$ of the fixed Hamiltonian $\Hh$
\begin{align}
    \frac{1}{\mathcal{Z}}\prod_{i=1}^L\int_\mathcal{\cont}dE_i&\exp\bigg(-L\sum_{i=1}^L V(E_i)\bigg)\prod_{i<j}^L(E_i-E_j)^2\nn\\&\frac{L!}{\mathcal{Z}(\Hh)}\prod_{i=1}^L\int_\cont d\l_i\,\prod_{j=1}^L \delta(\l_j-E_j)\,\mo(H)=\rL{\mo(\Hh)}\rR=\average{\mo(H)}\,.
\end{align}
This is trivial and boring here, but has interesting gravity consequences, namely that, as mentioned before in section \ref{sect:recovering}, more complicated geometries like wormholes are effectively encoded in the statistical properties of the brane one-point functions, the coupling constants in the spacetime action. This is the whole reason why the original dilaton gravity theory which includes wormhole serves an extremely good approximation to the gravity dual of one quantum system drawn from the ensemble.

\end{enumerate}

\section{Including bulk matter}\label{sect:matter}
So far, we have focused on effective field theories that did not include any matter fields in the $2$d bulk. The inclusion of such fields introduces two complications. First, we don't know an exact matrix model that is dual to JT gravity plus matter fields. Second, there are divergences due to Casimir energies associated to spacetime wormholes with narrow necks. In this section we will address the latter problem, and show that even with matter one can obtain a discrete and factorizing theory. 
\subsection{Factorization
}
In our model with additional (correlated) boundaries, the inclusion of matter requires us to put certain boundary conditions for the matter fields on those additional boundaries (on the asymptotic boundaries we use the usual Dirichlet boundary condition). At this point we will be agnostic about what boundary conditions we impose at the branes, and label some basis of matter boundary conditions by $\mathcal{B}$.

From the discussion in section \ref{sect:factor} it is then clear that we can essentially run the argument again with matter fields present, one finds that the only non-zero brane correlator is (because, by definition, the set of all boundary conditions $\mathcal{B}$ form an orthonormal basis)
\be
\label{eq:correlated-brane-w-matter}
\text{Factorization} \quad\Leftrightarrow\quad Z_\text{brane matter}((b_1, \mathcal{B}_1), (b_2, \mathcal{B}_2)) = -\frac{1}{b_1} \delta(b_1 - b_2)\,\delta( \mathcal{B}_1 -  \mathcal{B}_2)\,.
\ee
Following the same arguments that resulted in \eqref{eq:Z(beta)-after-simplification}, the total partition function reduces to
\be 
Z(\beta) = Z_\text{disk}(\beta) + \sum_{\mathcal{B}} \int_0^{\infty} \d b\, b\, Z_\text{trumpet matter}(\beta, b, \mathcal{B})\, Z_\text{brane matter}(b, \mathcal{B})\,.\label{52}
\ee
Here the trumpet partition function is that of matter-coupled JT gravity. Knowing this trumpet partition function, and the eigenvalues $E_i$ of the Hamiltonian $\Hh$ of the putative exact quantum mechanical dual, one could in principle determine $Z_\text{brane matter}(b, \mathcal{B})$ explicitly, which will again depend on $E_i$. Ensemble averaging over $\Hh$ with the appropriate measure (which as mentioned above, for matter coupled JT gravity, is not currently known) gives back the genus expansion of wormholes plus matter. 

This immediately raises the question what happens in our model to the $e^{a/b}$ (with a constant $a>0$) divergence from the Casimir energy of the matter fields, which features in the trumpet partition function $Z_\text{trumpet matter}(\beta, b, \mathcal{B})$. In a full UV complete model of JT gravity with matter, these divergence should not be there, some (unknown) matrix model description should presumably resolve it. From our perspective, we are imposing that \eqref{52} is finite (with some given discrete spectrum). This means that this would-be divergence will by construction be cancelled by a factor $e^{-a/b}$ in $Z_\text{brane matter}(b, \mathcal{B})$. This is independent of the precise set of typical eigenvalues that one chooses for $\Hh$ and so upon averaging this divergence will remain absent, just as one would expect from a UV complete description of JT plus matter. Additionally, we expect that when averaging over $\Hh$ the contribution of geometries which have all moduli sizes sufficiently large is however well approximated by simply coupling JT gravity to matter fields. It would be interesting to make this precise, should a dual of JT plus matter ever become available.

\subsection{Probe matter}
\label{sect:matter-correlator}

To be more concrete, let us consider probe matter, so we ignore any backreaction on the matter fields but we entirely include the backreaction of such probes on the metric. Consider a free bulk matter field. In the boundary this corresponds to some operator $\mathcal{O}$ with dimension $\D$. Since this is probe matter we can use the techniques developed in \cite{Yang:2018gdb, Saad:2019pqd,Mertens:2017mtv,Mertens:2018fds,Kitaev:2018wpr,Blommaert:2018iqz,Iliesiu:2019xuh,Blommaert:2018oro} and we can use the fact that the free propagator on a hyperbolic disk is $e^{-\D \ell}$ with $\ell$ the (regularized) geodesic length between the two boundary points. 

Let us assume for now that the extra boundaries due to the branes are not allowed to cut the geodesic between boundary operators. This means we don't include contributions coming from geodesics starting on the asymptotic boundary, and ending on the extra brane boundaries. Our claim is that the two-point function is then given by the sum of four geometries (we consider the un-normalized two-point function)
\be 
\< \mathcal{O}(x_1) \mathcal{O}(x_2)\> \quad =\quad  \begin{tikzpicture}[baseline={([yshift=-.5ex]current bounding box.center)}, scale=0.43]
 \pgftext{\includegraphics[scale=0.35]{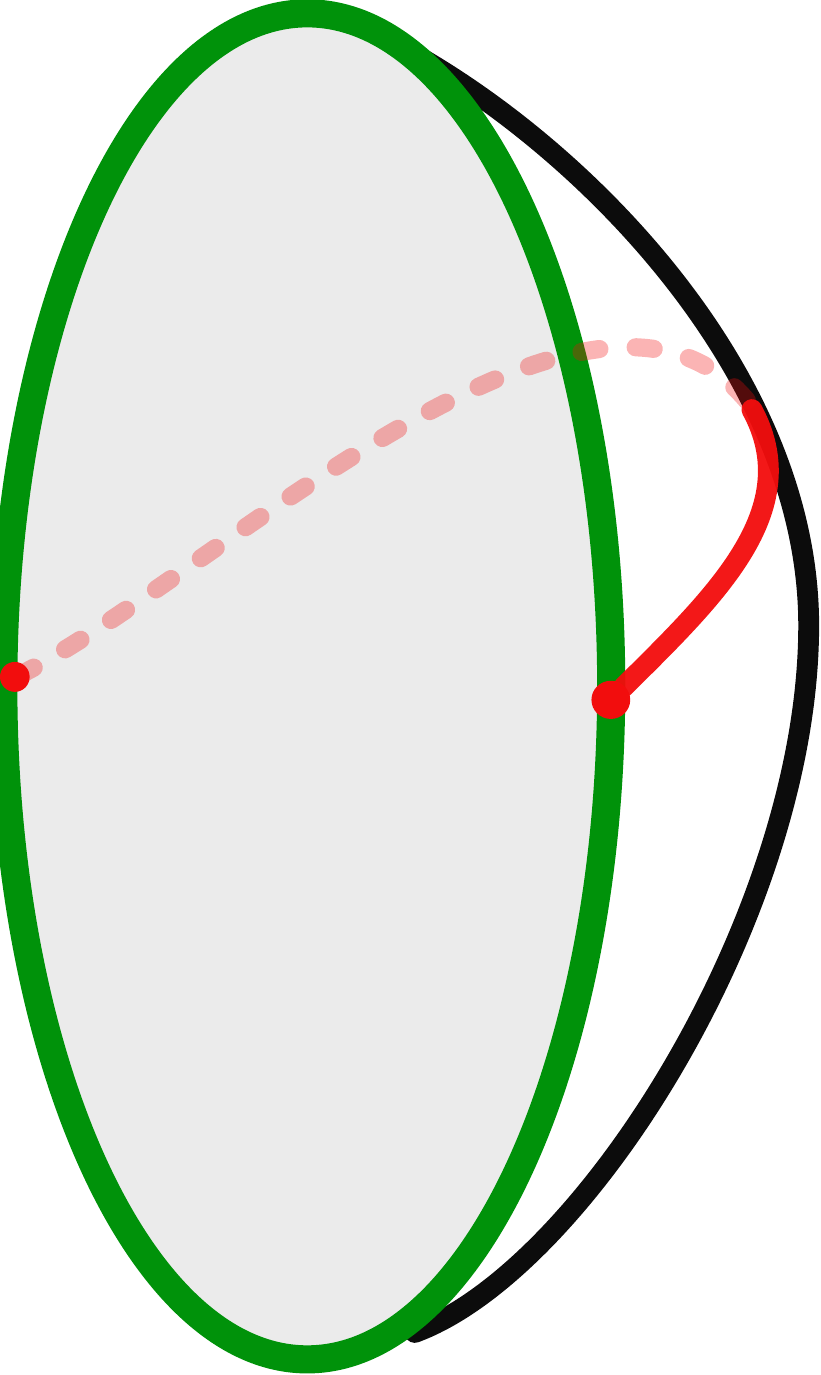}} at (-1,0);
  \end{tikzpicture}\quad + \quad \begin{tikzpicture}[baseline={([yshift=-.5ex]current bounding box.center)}, scale=0.43]
 \pgftext{\includegraphics[scale=0.35]{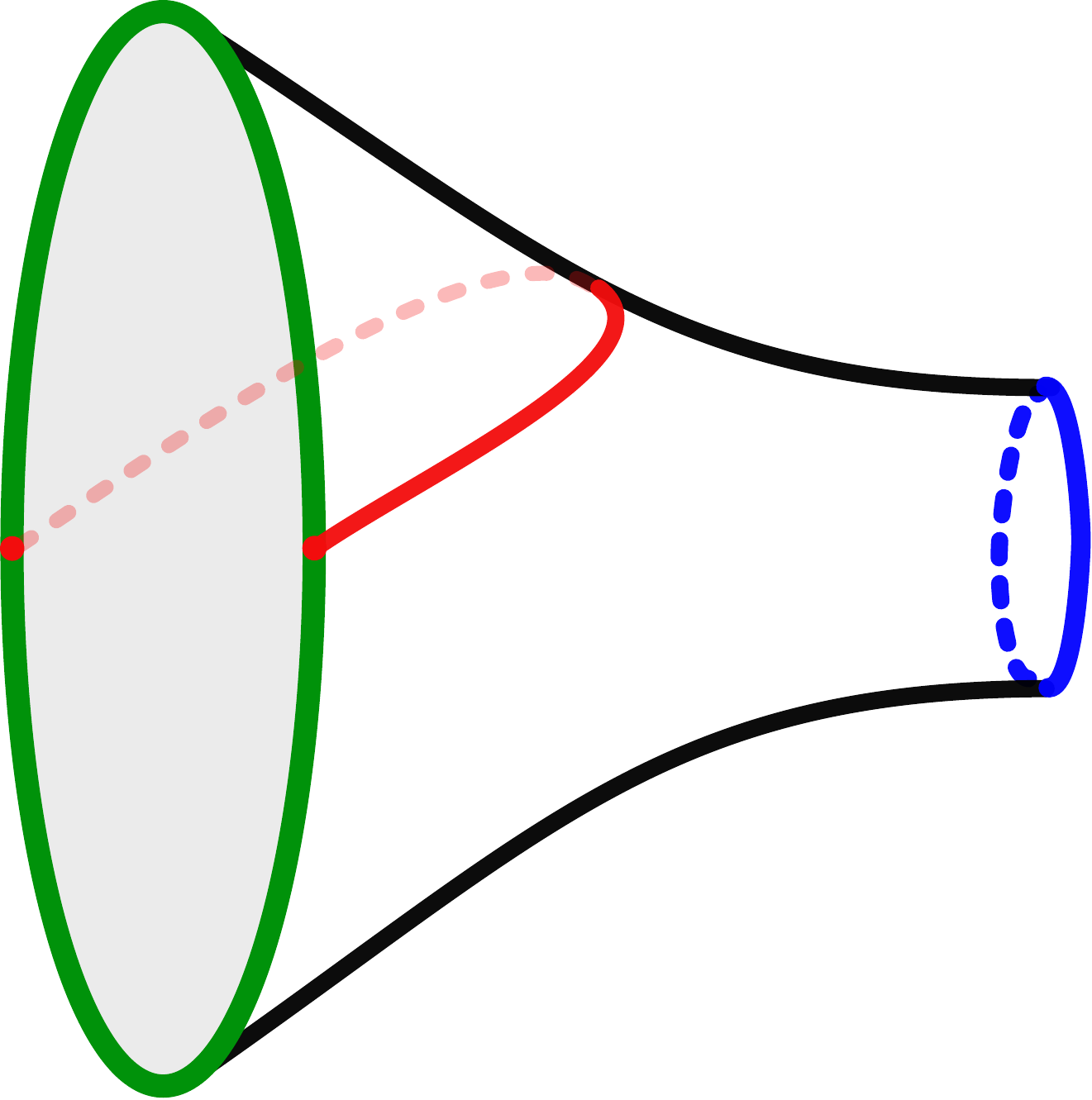}} at (-1,0);
  \end{tikzpicture}\quad + \quad \begin{tikzpicture}[baseline={([yshift=-.5ex]current bounding box.center)}, scale=0.43]
 \pgftext{\includegraphics[scale=0.35]{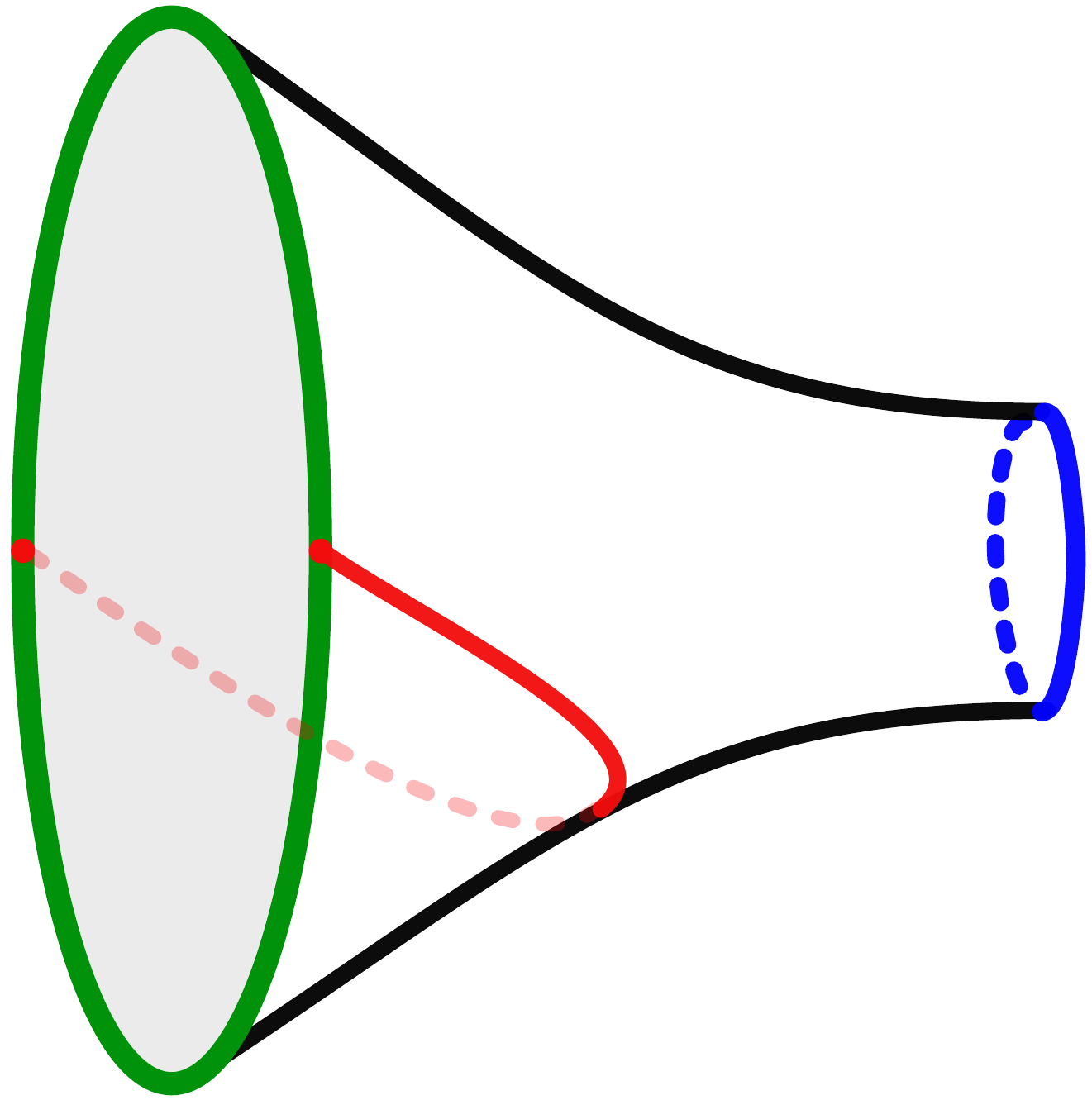}} at (-1,0);
  \end{tikzpicture}\quad + \quad 
  \begin{tikzpicture}[baseline={([yshift=-.5ex]current bounding box.center)}, scale=0.43]
 \pgftext{\includegraphics[scale=0.35]{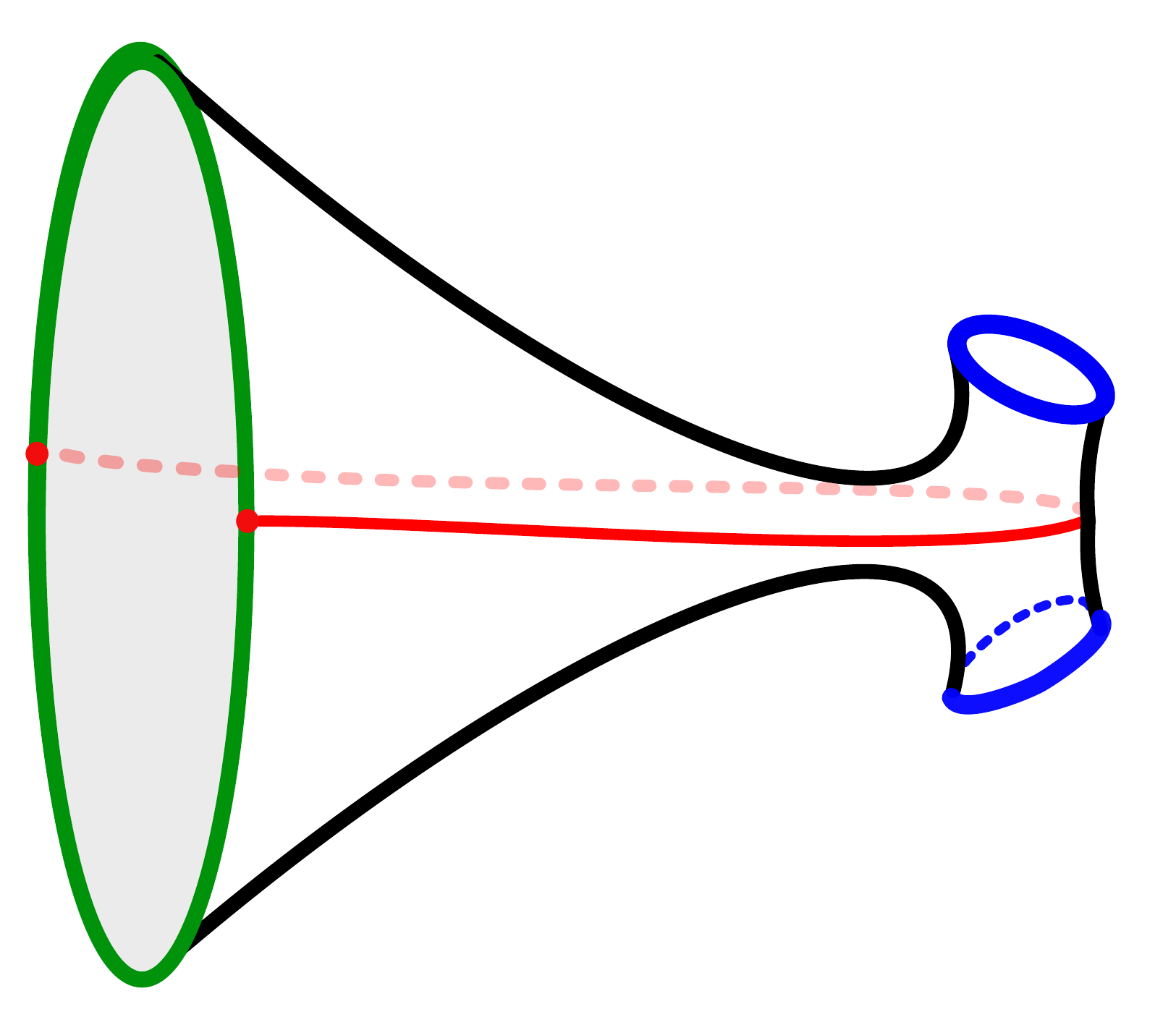}} at (-1,0);
  \end{tikzpicture}.\label{54}
\ee
From the discussion in section \ref{sect:factor} it is clear that the first three geometries contribute, these are just the disk and half-wormhole contributions, but now with a geodesic one them. The fourth geometry can not be cancelled by other contributions, and thus remains as well. All other geometries cancel, for example
\eqref{eq:leading-wormhole-cancelling} implies that
\be 
\label{eq:geodesic-insertion-wormhole-still-cancels}
 \begin{tikzpicture}[baseline={([yshift=-.5ex]current bounding box.center)}, scale=0.45]
 \pgftext{\includegraphics[scale=0.35]{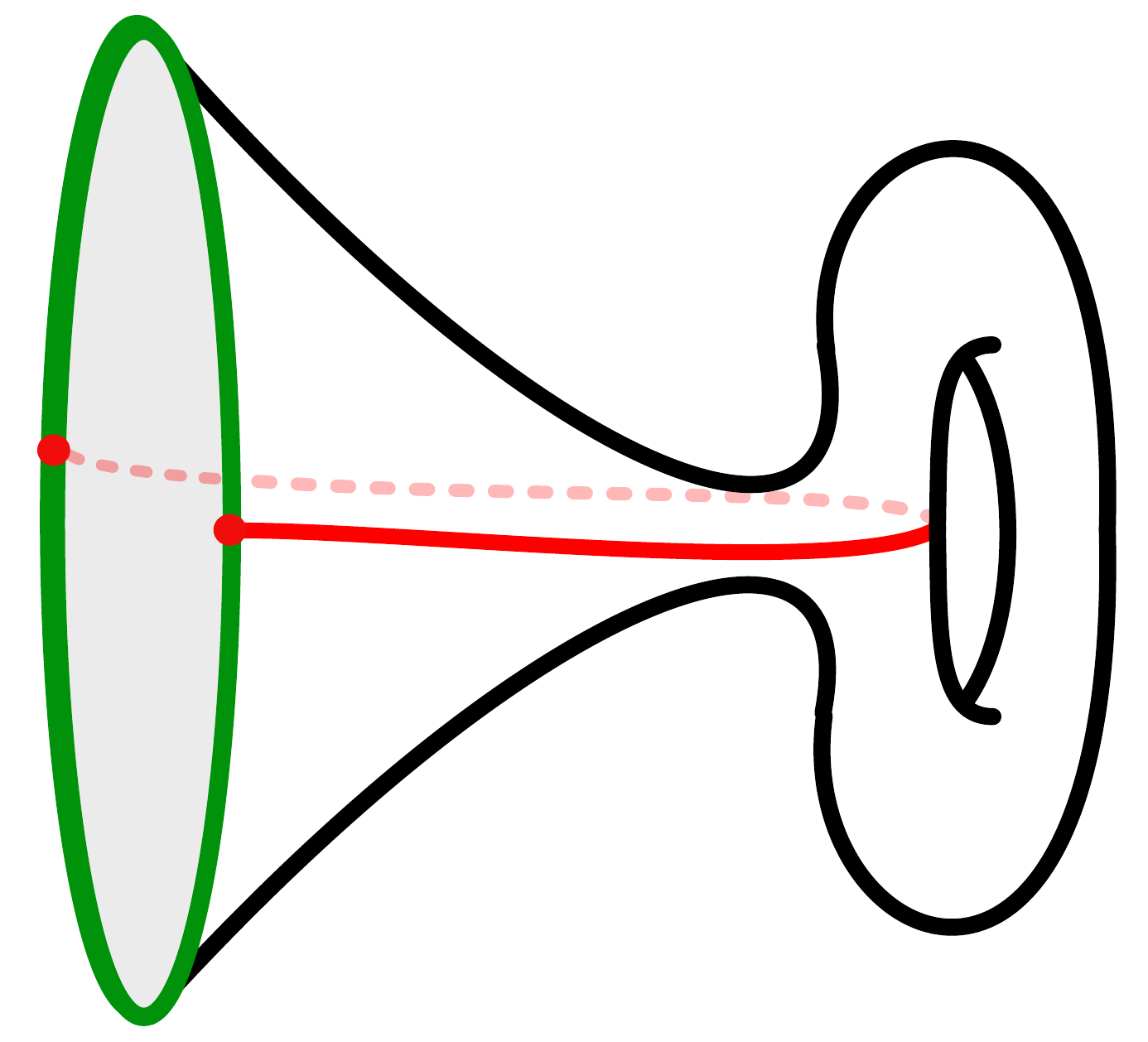}} at (-1,0);
  \end{tikzpicture}
  \quad + \quad
  \begin{tikzpicture}[baseline={([yshift=-.5ex]current bounding box.center)}, scale=0.45]
 \pgftext{\includegraphics[scale=0.35]{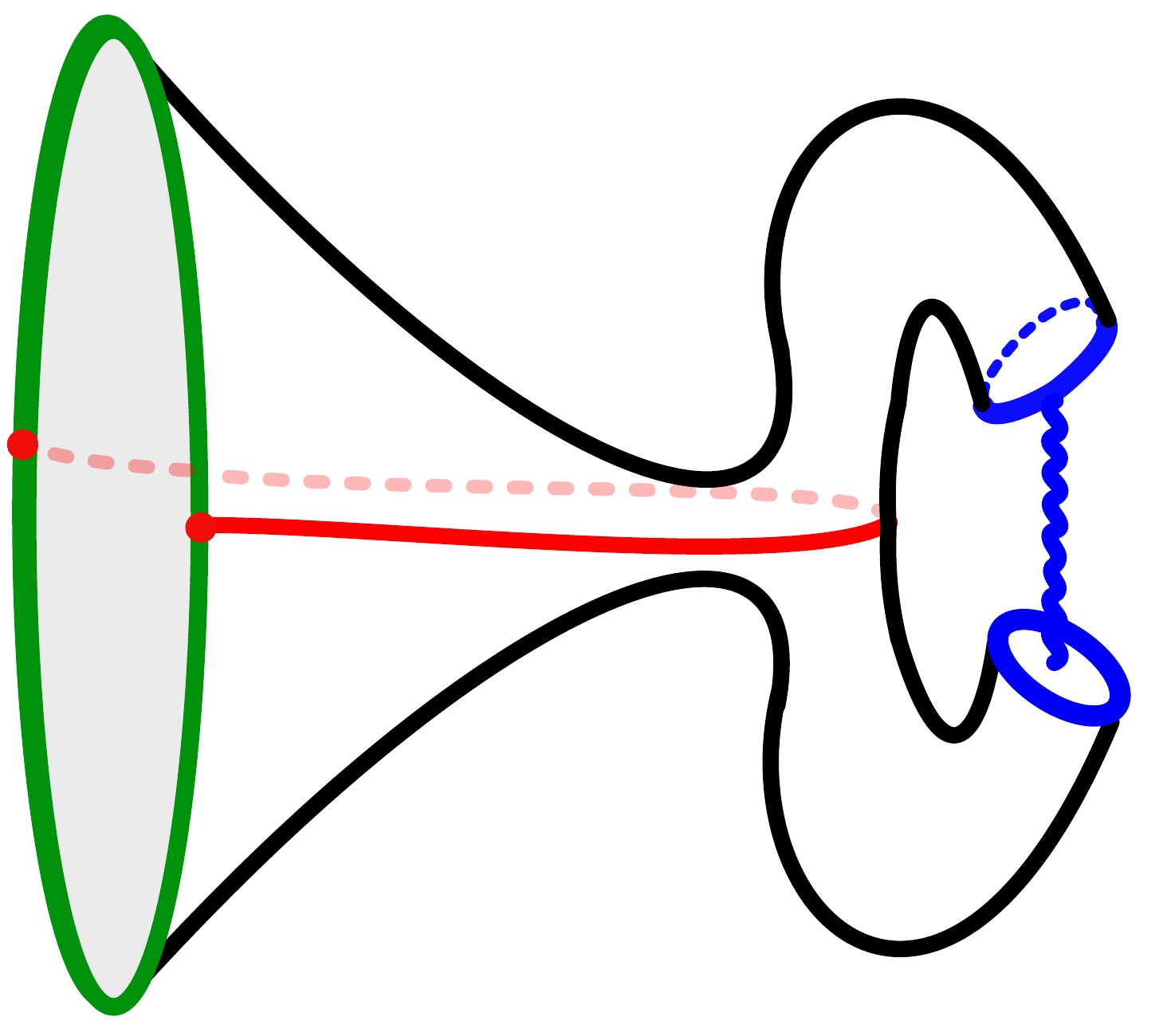}} at (-1,0);
  \end{tikzpicture} = 0\,.
\ee
This is because the presence of the geodesic effectively reduces the mapping class group of the disk with a handle to that of the cylinder \cite{Blommaert:2019hjr,Saad:2019pqd,Blommaert:2020seb}. Similarly, the mapping class group of a surface of genus $g$ and one boundary is effectively reduced by the presence of the geodesic to either that of a connected surface with two boundaries and genus $g-1$, or that of two surfaces, each with one  boundary, whose genera sum to $g$ \cite{Blommaert:2019hjr,Saad:2019pqd,Blommaert:2020seb,Iliesiu:2021ari}. One can then use the same arguments as in section \ref{sect:factor} to prove that only \eqref{54} remains. 

In terms of formulas, the disk geometry in \eqref{54} contributes \cite{Yang:2018gdb, Saad:2019pqd,Mertens:2017mtv,Mertens:2018fds,Kitaev:2018wpr,Blommaert:2018iqz,Iliesiu:2019xuh,Blommaert:2018oro}
\begin{align} 
\average{\mathcal{O}(x_1)\mathcal{O}(x_2)}_\text{disk} &=\int_{-\infty}^{+\infty} \d \ell e^\ell\, \psi_\text{disk}^*(\ell,\beta-x_2+x_1)\,\psi_\text{disk}(\ell,x_2-x_1)\, e^{-\D \ell}\nn\\&=
\int_0^\infty d E_1\,\rho_0(E)\,e^{-(\beta-x_2+x_1) E_1}\int_0^\infty d E_2\,\rho_0(E_2)\,e^{-(x_2-x_1) E_2}\,\abs{\mo_{E_1E_2}}^2\label{DiskM}\,,
\end{align}
with 
\be 
\psi_{\rm disk}(\ell,z) = \int_0^{\infty} \d E \rho_0(E) e^{-\frac{z}{2} E} \left(4 e^{-\ell/2}K_{\i \sqrt{8E}}(4 e^{-\ell/2})\right)\,,\quad \rho_0(E) = e^{\S} \frac{\sinh 2\pi\sqrt{E}}{4\pi^2}\, ,
\ee
and $|\mathcal{O}_{E_1 E_2}|^2$ given by
\be 
|\mathcal{O}_{E_1 E_2}|^2 = \frac{|\Gamma(\D + \i(\sqrt{E_1} + \sqrt{E_2}))\Gamma(\D - \i (\sqrt{E_1} - \sqrt{E_2}))|^2}{2^{2\D + 1}\Gamma(2\D)}\,.
\ee
The two half-wormhole diagrams contribute \cite{Saad:2019pqd}
\begin{align}
&\average{\mathcal{O}(x_1)\mathcal{O}(x_2)}_\text{half-wormhole} = \int_0^{\infty} \d b\, b\, Z_\text{brane}(b) \int_{-\infty}^{+\infty} \d \ell e^\ell\, \psi_\text{disk}^*(\ell,\beta-x_2+x_1)\,\psi_\text{trumpet}(\ell,b,x_2-x_1)\, e^{-\D \ell}\nn\\&\hspace{2cm}+  \int_0^{\infty} \d b\, b\, Z_\text{brane}(b) \int_{-\infty}^{+\infty} \d \ell e^\ell\, \psi_\text{trumpet}^*(\ell,b,\beta-x_2+x_1)\,\psi_\text{disk}(\ell,x_2-x_1)\, e^{-\D \ell}\,,\label{HWM}
\end{align}
where 
\be 
\psi_{\rm trumpet}(\ell,b,z) = \int_0^{\infty} \d E \rho(E,b) e^{-\frac{z}{2} E} \left(4 e^{-\ell/2}K_{\i \sqrt{8E}}(4 e^{-\ell/2})\right)\,,\quad \rho(E,b) = \frac{\cos b\sqrt{E}}{\pi \sqrt{E}}
\ee
and the fourth diagram (which we will call the nose geometry) contributes\footnote{There is a subtlety that now one has to quotient the path-integral over $g_{\mu \nu}$ by the mapping class group of the three-holed sphere, which is non-trivial. However, because in the two-point ${\average{\mathcal{O}(x_1)\mathcal{O}(x_2)}}$ we are implicitly summing over all geodesics going from $x_1$ to $x_2$ on the three-holed spherical geometry, including geodesics that wind around the branes, the quotient by the group can be traded-off with the sum over all such geodesics as pointed out in \cite{Saad:2019pqd, Blommaert:2020seb, Iliesiu:2021ari}. }
\begin{align}
\label{eq:nose-geometry}
    &\average{\mathcal{O}(x_1)\mathcal{O}(x_2)}_\text{nose}=\int_0^{\infty} \d b_1 b_1\, Z_\text{brane}(b_1)\int_0^{\infty} \d b_2 b_2\, Z_\text{brane}(b_2)\nn\\ &\hspace{4cm}\int_{-\infty}^{+\infty} \d \ell e^\ell\, \psi_\text{trumpet}^*(\ell,b_1,\beta-x_2+x_1)\,\psi_\text{trumpet}(\ell,b_2,x_2-x_1)\, e^{-\D \ell}\,.
\end{align}
Adding up these contributions, and writing everything in terms of energy variables as in \eqref{DiskM}, one finds
\be 
\average{\mathcal{O}(x_1)\mathcal{O}(x_2)} = 
\int_0^\infty d E_1\,\rho(E)\,e^{-(\beta-x_2+x_1) E_1}\int_0^\infty d E_2\,\rho(E_2)\,e^{-(x_2-x_1) E_2}\,\abs{\mo_{E_1E_2}}^2 \,,
\ee
where the combined spectral density is that of the discretized system (using the explicit expression \eqref{solution} for $Z_\text{brane}(b)$ in the second equality)
\begin{equation}
    \rho(E)=\rho_0(E)+\int_0^\infty \d b\, b\, Z_\text{brane}(b)\,\rho_\text{trumpet}(E,b)=\sum_{i=1}^L\delta(E-E_i)
\end{equation}
Therefore, the full probe matter two-point function becomes a sum over energies, instead of an integral
\be 
\label{eq:2-pt-function-exact}
\average{\mathcal{O}(x_1)\mathcal{O}(x_2)} = \sum_{i,j=1}^Le^{-(\b - x_2+x_1)E_i}e^{-(x_2-x_1)E_j} \abs{\mo_{E_i E_j}}^2 = \Tr( e^{-(\b - x_2+x_1)\Hh}\,\mathcal{O}\,e^{-(x_2-x_1)\Hh}\,\mathcal{O})\,.
\ee
This is precisely what one would expect from a system with a discrete spectrum, and it is a non-trivial check to see that our model also works in the case of probe matter. Explicitly, if we again consider an ensemble average as in section \ref{sect:recovering}, we find that after averaging the two-point function \eqref{eq:2-pt-function-exact}, the result is identical to the two-point function of probe matter fields in JT gravity to all orders in $e^{-\S}$ \cite{Blommaert:2019hjr,Saad:2019pqd,Blommaert:2020seb, Iliesiu:2021ari}, and also non-perturbatively in $e^{-\S}$ (since this is the operator one would insert in the matrix ensemble). One comment here is that our assumption about the geodesic boundaries not intersecting the probe geodesic wordlines seems to be justified, because it gives the correct final result \eqref{eq:2-pt-function-exact}.

The nose geometry can also be interpreted in a slightly different way. Specifically, we can think of it as giving an additional contribution to the brane one-point function $Z_\text{brane matter}(b,\mathcal{B})$ as shown below
\be
  \begin{tikzpicture}[baseline={([yshift=-.5ex]current bounding box.center)}, scale=0.5]
 \pgftext{\includegraphics[scale=0.75]{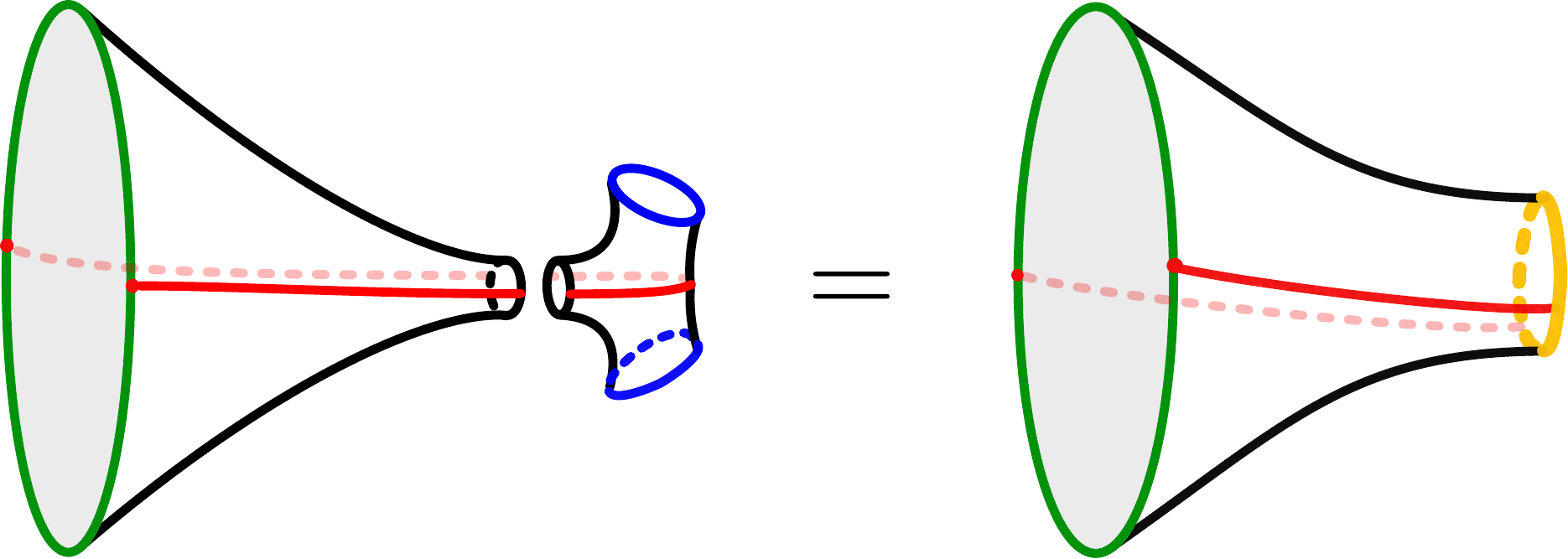}} at (-1,0);
  \end{tikzpicture}
  \label{eq:rewriting-of-geodesic-ending-on-brane}
\ee
Here we have cut the nose geometry along its waist, and the detached three-holed sphere part now encodes the coupling between the matter and the degrees of freedom on the brane boundary. What we mean with that is that, because in the right figure we now have geodesics ending on the brane boundary, the detached three-holed sphere now is an operator insertion on the yellow boundary. Or in other words, this yields a non-trivial boundary state of the matter on the geodesic boundary. Going back to the case of fully dynamical matter, we expect this non-trivial state on the geodesic boundary to become the state $\mathcal{B}$ we mentioned above. Moreover, the full two-point function in that case will be given by a sum over geodesics that not only start and end on the same (asymptotic) boundary (like in the first three figures in \eqref{54}) but also a sum over (pairs) of geodesics that start on the asymptotic boundary and end on the geodesic one (like the figure on right of \eqref{eq:rewriting-of-geodesic-ending-on-brane}).

\subsection{Why wormholes are oftentimes good approximations
}

As in section \ref{sect:recovering}, let us explain why wormholes are good approximations for self-averaging observables, here including matter or other geometric probes, even in the regime when they dominate over the (typically leading) black hole geometries.

\subsection*{The time-averaged probe matter two-point function } 

Just like the spectral form factor discussed in section \ref{sect:recovering}, the probe matter two-point function \eqref{eq:2-pt-function-exact} is not by itself self-averaging, but its time-averaged version over the time-interval $\Delta T\gg 1$ \emph{is}. For early times $t<t_\text{ramp}$, the time-averaged correlator is dominated  by the black hole, the first diagram in \eqref{54}. This quantity decreases with time forever, which poses a problem for black hole unitarity \cite{Maldacena:2001kr}. At later times $t>t_\text{ramp}$, the nose geometry in \eqref{54} dominates. Crucially, because the variance of the time-averaged two-point function is small, it is well approximated by evaluating the correlator on a disk with a handle, the left diagram in \eqref{eq:geodesic-insertion-wormhole-still-cancels}. As shown in \cite{Saad:2019pqd}, this quantity \emph{grows} with time. Furthermore, while in JT gravity the linear growth of the probe two-point function is exact, in our model the exact two-point function (without a time-average) has additional erratic noise, characteristic of discrete quantum systems. So, the contribution of the half-wormhole addresses the apparent lack of unitarity in the black hole geometry \cite{Maldacena:2001kr}.  

\subsection*{The volume of the black hole interior } 

A second self-averaging quantity which we can discuss is the volume of the two-sided black hole interior \cite{Iliesiu:2021ari}. This can be evaluated by inserting $\ell$, instead of $e^{-\Delta \ell}$ in \eqref{DiskM} to \eqref{eq:nose-geometry}. Up to unimportant multiplicative factors, one finds that the length of the Einstein-Rosen bridge 
\be
\<\ell(t) - \ell(0)\> \sim {e^{-\S}}\sum_{i< j}^L \frac{e^{-\frac{\beta}4(E_i+E_j)} \left(\cos\left(\frac{ t (E_i-E_j)}2\right)-1 \right)}{(E_i-E_j)\left(\cosh(2\pi \sqrt{E_i}) - \cosh(2\pi \sqrt{E_j})\right)}\,,
\ee
which agrees with the length operator computed in \cite{Iliesiu:2021ari}, after taking an ensemble average over $\Hh$.

As opposed to the spectral-form factor, this volume is self-averaging up to very-long times $t \sim e^{2\S}$, so no time-average needs to be considered up to very long times. The black hole dominates when computing the volume up to $t \sim e^{\S}$, a time until which the volume experiences a period of linear growth. Past this time, the contribution of the half-wormholes in \eqref{54} (or the doubly non-perturbative contributions in the spectral correlator $\<\rho \rho\>$ in the ensemble average) becomes comparable to the black hole saddle. As observed in \cite{Iliesiu:2021ari}, this precisely cancels the linear growth in the black hole saddle, and results in a plateau $\sim e^{\S}$. Thus, the plateau for the volume of the black hole interior, which is typically attributed to a breakdown of GR \cite{Susskind:2018pmk}, can be concretely attributed in our model to the brane corrections in \eqref{54}. 

\subsection*{The Page curve}

Finally we can discuss the computation of the Page curve in our model. For concreteness we will focus on the setup of \cite{Almheiri:2019qdq}, which couples a patch of spacetime where gravity is dynamical to a bath where the metric is fixed; within the patch where gravity is dynamical we will consider our two-dimensional model of gravity with the two-point correlator between branes given by \eqref{eq:correlated-brane-w-matter}. The coupling between the two region is realized by introducing a common set of matter fields that live in both parts of the spacetime. The goal is to compute the entanglement entropy of these matter fields along a segment located in the bath region, as a function of Lorentzian time $t$.

While we will not go through all the details of computing the Page curve in our model due to the similarity to \cite{Almheiri:2019qdq}, we will focus on two points which are necessary in our setup: (i) understanding why the Page curve is self-averaging in the model of JT gravity coupled to matter which includes wormholes, and (ii)  understanding why the backreaction from matter fields make half-wormholes geometries dominate over configurations that involve the standard black hole saddle past the Page time, in the same way in which in the setup discussed in \cite{Almheiri:2019qdq} replica wormholes become the dominant saddle. To this end one considers $n$-replicas of the system (sewed in the proper manner) and computes the $n$-th Renyi entropy $S_n = \tr \rho_R^n$ for the density matrix $\rho_R$ of the matter fields along the segment in the reservoir; finally, one takes the analytic continuation $n \to 1$ to obtain the entanglement entropy.    

(i) The Page curve is self-averaging if $\sigma_{S_n} \ll  \rL S_n \rR $. To check this inequality we consider two copies of the $n$-replicated system, with each of the two replicated sets of baths glued in the proper manner needed to compute the Renyi entropy, in order to compute $\rL(S_n)^2\rR$. Importantly, note that while each of the $n$-replicas are connected along the cuts in the reservoir regions, the two distinct copies of the $n$-replicas are not connected through the cuts (this is thus different than considering the $2n$-replicas of the system that compute $S_{2n}$).  At early times ($t<t_{\rm Page}$) the Renyi entropy is dominated solely by the black hole saddle and the two copies of the  $n$-replicated system are completely separated. At later times, there are replica wormhole saddles that dominate. In contrast to the case of pure JT gravity where no wormhole saddles can be found (since there is nothing that fixes the modulus $b$ of each wormhole), these saddles exist due to the backreaction of the matter fields. This backreaction can be understood from the contribution of closed worldlines for the matter fields that propagate through the wormhole in between different replicas of the reservoir, finally forming a closed loop by traveling through the cuts in the reservoir. Denoting the length of such loops by $\ell_L$, it is the balance between $\sum_L e^{-\Delta \ell_L}$ (where the sum over $L$ involves a sum over all such possible loops including those that wind multiple times in each one of the $n$-replicas) and the JT gravity action which stabilizes the moduli of the wormhole. Since the two distinct copies of the $n$-replicated system are not glued along any of the cuts in the reservoir, then there is no new saddle that involves a replica wormhole geometry that connects the two distinct $n$-replicated copies. Therefore, we have that 
\begin{align}
\rL (S_n)^2 \rR =  \rL S_n \rR^2 + O(e^{-\S}) = \begin{cases}
  \bigg( \begin{tikzpicture}[baseline={([yshift=-.5ex]current bounding box.center)}, scale=0.7]
 \pgftext{\includegraphics[scale=0.35]{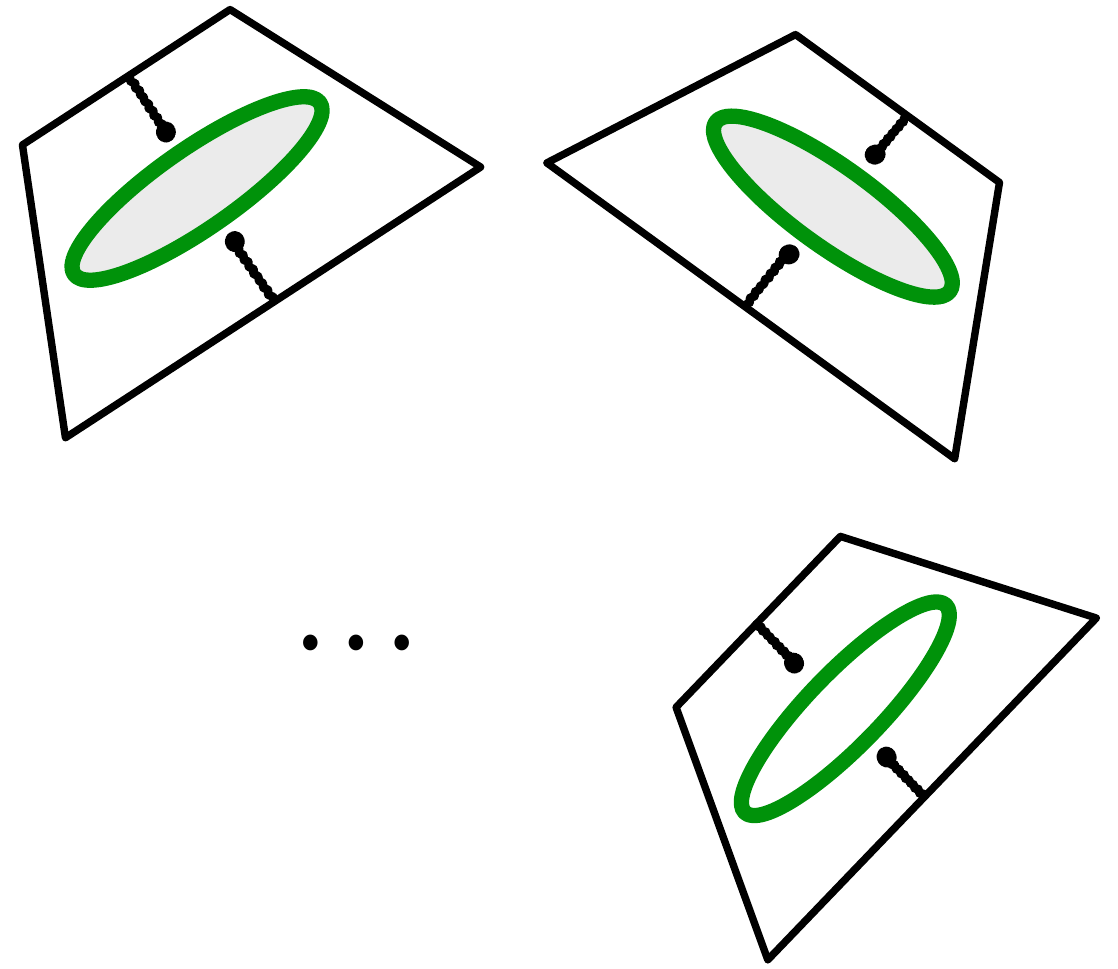}} at (-1,0);
  \end{tikzpicture} \bigg)^2 + O(e^{-\S})  \quad \quad  \text{ for } t<t_\text{Page}\,,\\
   \bigg(\begin{tikzpicture}[baseline={([yshift=-.5ex]current bounding box.center)}, scale=0.7]
 \pgftext{\includegraphics[scale=0.35]{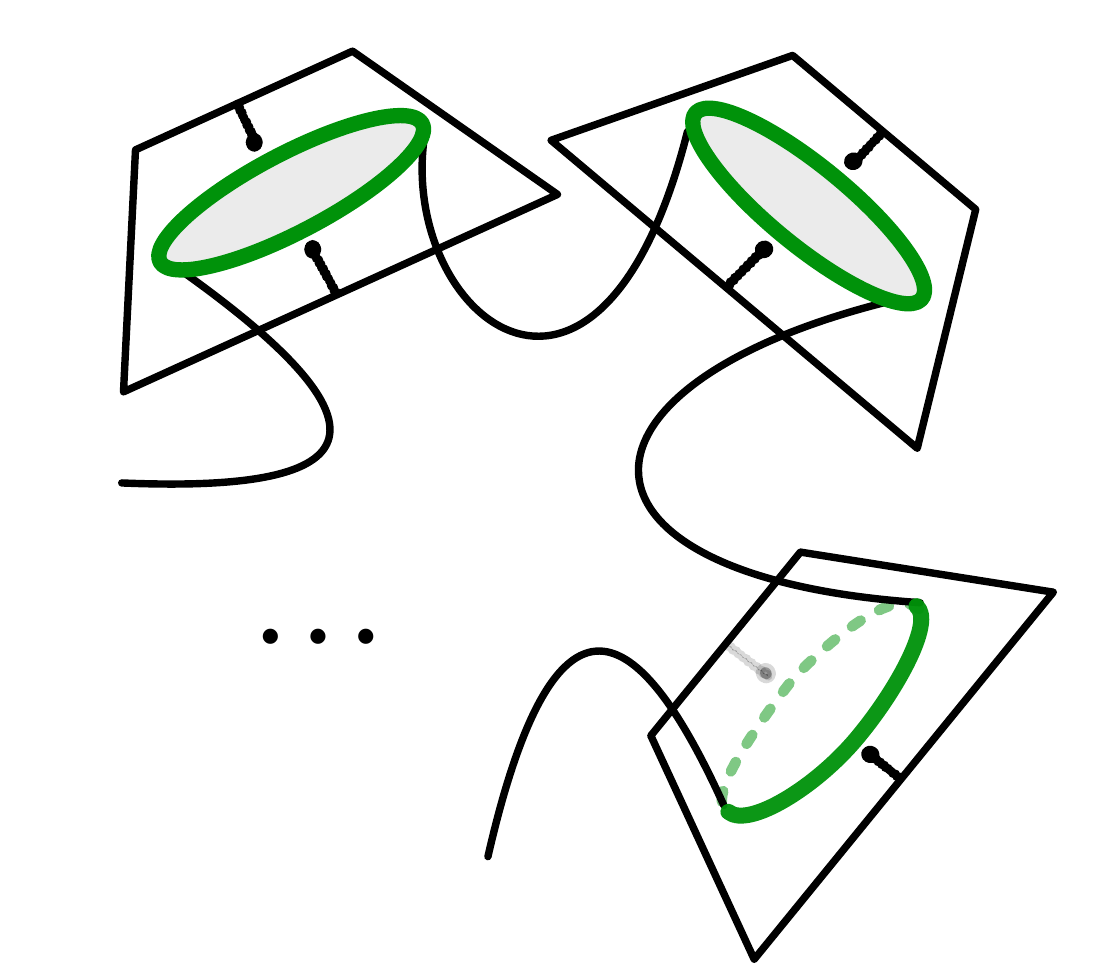}} at (-1,0);
  \end{tikzpicture} \bigg)^2 + O(e^{-\S}) \quad \quad \text{ for } t>t_\text{Page}\,.
  \end{cases}
\end{align}
where the $e^{-\S}$ corrections comes from configurations that are not necessarily associated to a saddle-point, such as geometries that connect the two $n$-replicated copies.  Consequently, we find that
\be 
\label{eq:variance-Renyi-entropies}
\sigma_{S_n} \ll S_n\,, \quad \text{for all } t\,,
\ee
which implies that the Renyi entropies are self-averaging quantities. 

(ii) As discussed below \eqref{eq:rewriting-of-geodesic-ending-on-brane}, the matter states $\mathcal{B}$ on the brane can be recast through the insertion of operators (formed out of the matter fields) on the brane. Thus, rewriting the integral over fields as a sum over worldlines there will be worldlines propagating between the branes of the half-wormholes living on different replicas. \eqref{eq:rewriting-of-geodesic-ending-on-brane} can then be extended to rewrite the interaction between the worldline of the particles in the theory and the brane in terms of geometries where the worldline never intersect a brane boundary. For instance, for the $n=2$ replicated system, we can take the average over $\Hh$ for the correlator of a pair of operators inserted on the two branes, to obtain 
\be 
\begin{tikzpicture}[baseline={([yshift=-.5ex]current bounding box.center)}, scale=0.6]
 \pgftext{\includegraphics[scale=0.35]{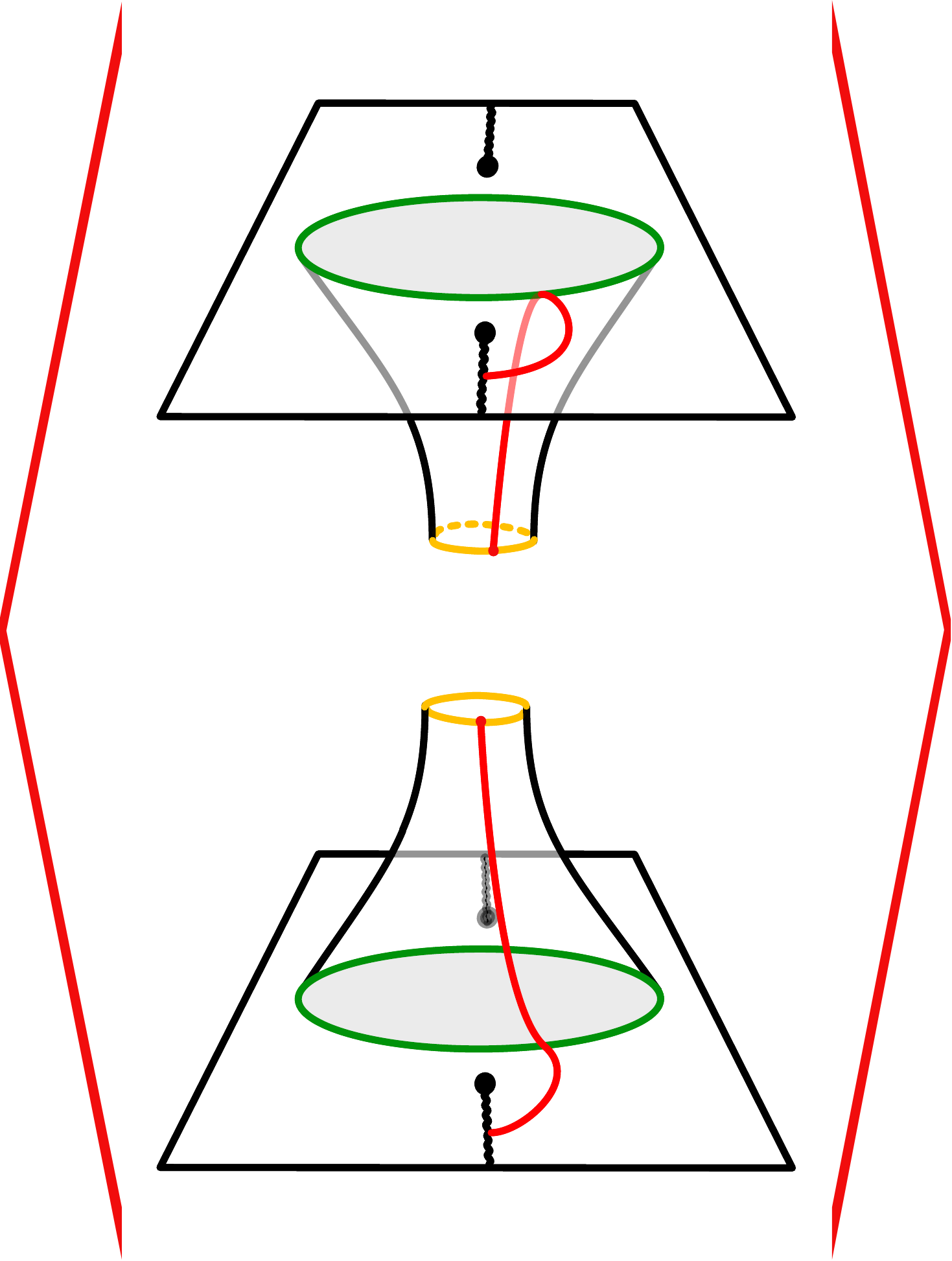}} at (0,0);
     \draw (3.0,-2.8) node  {$\color{red}{\text{conn}}$};
  \end{tikzpicture} \quad = \quad 
  \begin{tikzpicture}[baseline={([yshift=-.5ex]current bounding box.center)}, scale=0.6]
 \pgftext{\includegraphics[scale=0.35]{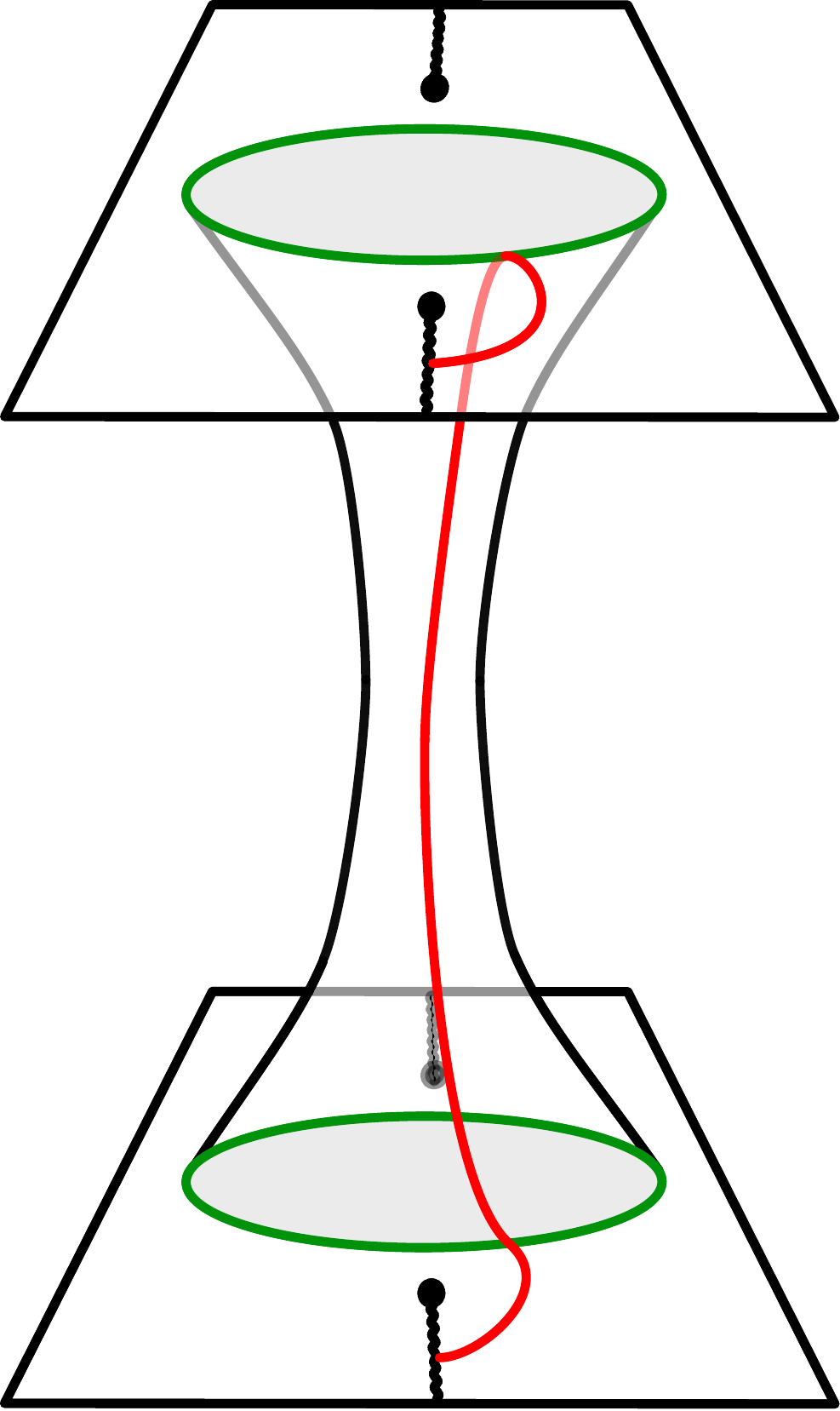}} at (0,0);
  \end{tikzpicture}  \,,
\ee
where the right-hand side represents the closed particle worldline which stabilizes the modulus $b$ of the wormhole. As discussed above, such closed worldlines are necessary to obtain the replica wormhole saddle-point for $n=2$. For higher values of $n$ one can similarly evaluate the correlation functions of matter operators inserted on the branes. Once again, after averaging over $\Hh$ one can rewrite this correlator in terms of the closed worldlines that are necessary to stabilize the moduli of the replica wormhole with $n$ asymptotic boundaries.  

In our model, we thus have that the $n$-th Renyi entropy is computed by
\begin{align}
\label{eq:Page-curve-self-averaging}
S_n &=\,  \begin{tikzpicture}[baseline={([yshift=-.5ex]current bounding box.center)}, scale=0.7]
 \pgftext{\includegraphics[scale=0.45]{Page_curve1.pdf}} at (-1,0);
  \end{tikzpicture} + \,\dots\,+
 \begin{tikzpicture}[baseline={([yshift=-.5ex]current bounding box.center)}, scale=0.7]
 \pgftext{\includegraphics[scale=0.45]{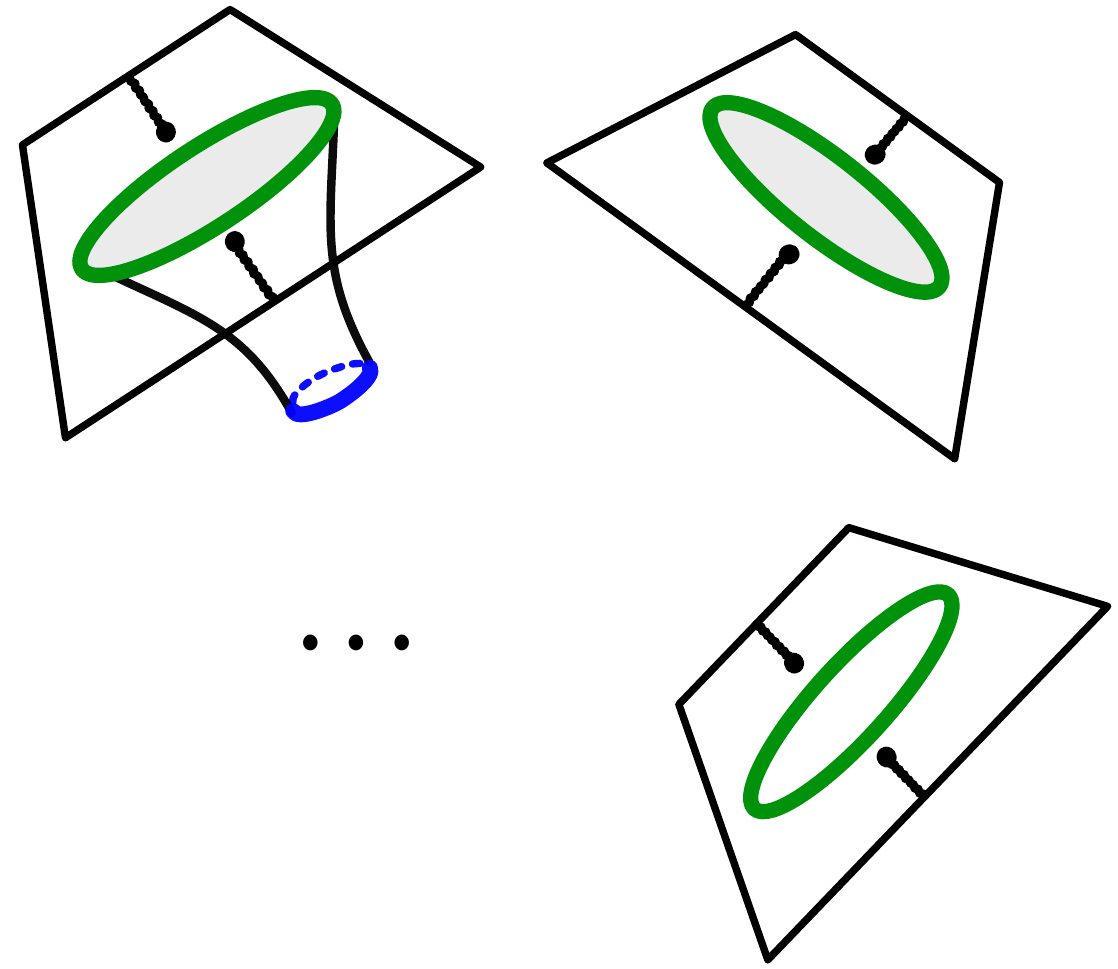}} at (-1,0);
  \end{tikzpicture} + \,\dots\,+
   \begin{tikzpicture}[baseline={([yshift=-.5ex]current bounding box.center)}, scale=0.7]
 \pgftext{\includegraphics[scale=0.45]{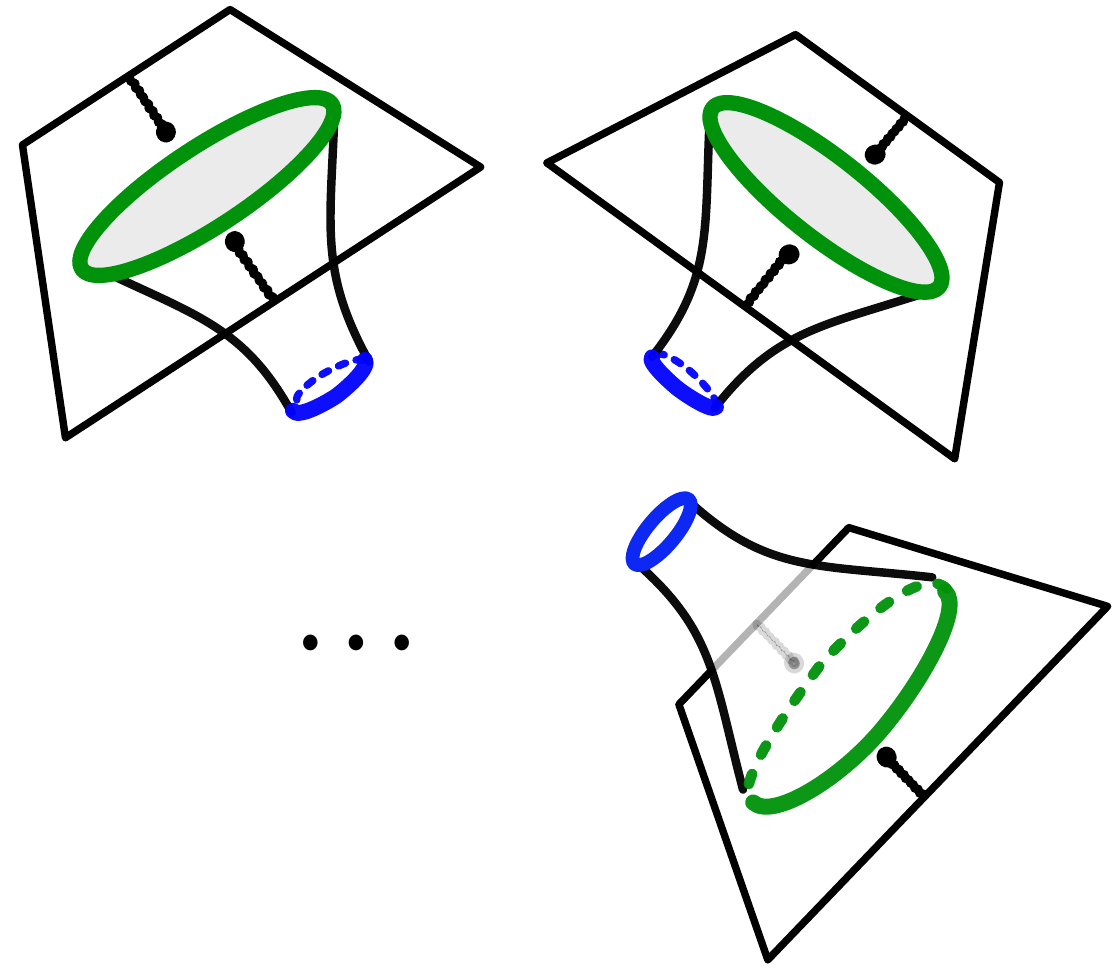}} at (-1,0);
  \end{tikzpicture} \nn \\ &\quad \quad \quad \quad \approx 
  \begin{cases}
  \begin{tikzpicture}[baseline={([yshift=-.5ex]current bounding box.center)}, scale=0.7]
 \pgftext{\includegraphics[scale=0.45]{Page_curve1.pdf}} at (-1,0);
  \end{tikzpicture} + O(e^{-\S})  \quad \quad  \text{ for } t<t_\text{Page}\,,\\
   \begin{tikzpicture}[baseline={([yshift=-.5ex]current bounding box.center)}, scale=0.7]
 \pgftext{\includegraphics[scale=0.45]{Page_curve_approximate.pdf}} at (-1,0);
  \end{tikzpicture}  + O(e^{-\S}) \quad \quad \text{ for } t>t_\text{Page}\,,
  \end{cases}
\end{align}
where in going between the first and second line we have taken the average over $\Hh$, used the fact that the matter backreaction is the same in the half-wormhole as in the replica wormhole,  and then used \eqref{eq:variance-Renyi-entropies} to study each member of the ensemble.

To summarize, at times $t<t_{\rm Page}$ the black hole saddles  dominate and the entanglement entropy along the segments in \eqref{eq:Page-curve-self-averaging} grows with time. At times $t>t_{\rm Page}$, the backreaction of matter fields cause the half-wormhole configuration (the last diagram on the first line in \eqref{eq:Page-curve-self-averaging}) to become dominant. Due to self-averaging, for a typical choice of $\Hh$, such a half-wormhole configuration is well approximated (up to $e^{-S_0}$ corrections) by the replica wormhole configuration in \eqref{eq:Page-curve-self-averaging}. Analytically continuing in $n$ to $n \to 1$, such a replica wormhole  saddle was shown to correctly reproduce the expected time-dependence for the entanglement entropy in the Page curve.  Consequently, the same analytic continuation in the half-wormhole geometry reproduces the Page curve.

One might wonder whether, when it comes to the density matrix of the Hawking radiation $\rho_R$, our model has a computational advantage compared to the models which include replica wormholes. The main issue with the replica wormhole computation is that it is difficult to imagine how to compute individual elements of the radiation density matrix $(\rho_R)_{ij}$, in place of the Renyi entropies $\Tr (\rho_R^n)$. In other words, it is unclear whether it is possible to compute the entanglement entropy in the JT gravitational theory coupled to matter, without performing the replica trick and taking the analytic continuation to $n\to 1$. In our model however, the computation of the Renyi entropy is much closer to that in a standard QFT; there is no unexpected topology changes when one considers the $n$-replica geometry as in the first line of  \eqref{eq:Page-curve-self-averaging}. Thus, there is, at least in principle, no obstruction to computing $(\rho_R)_{ij}$ for a given choice of $Z_{\text{brane}}(b, \,\mathcal B)$. While it is difficult to do such a computation when our model is coupled to matter fields since we don't know the precise matrix integral dual from which typical energies $E_i$ are picked from, such a computation might be greatly simplified by incorporating the correlated branes discussed in this paper within the EOW brane model of \cite{Penington:2019kki,Blommaert:2021etf}. We leave such an analysis to future work. 

Additionally, it would be interesting to explore the effect of brane insertions for other self-averaging observables. For instance it would be interesting to explain how the presence of the brane should explicitly break any global symmetry present for matter fields in the bulk (as proposed in \cite{Hsin:2020mfa}), analyze how correlated branes play a role in the state reconstruction for the  black hole interior \cite{Penington:2019kki} or how they can drastically reduce the dimension of the matter Hilbert space \cite{Hsin:2020mfa}.

\section{Discussion}\label{sect:discussion}

In this work we considered an effective geometric model with correlated spacetime branes. Those, as we have argued, would arise from integrating our UV degrees of freedom. If the original UV theory factorizes, this heavily constrains the resulting low-energy effective theory, which by construction should factorize too. We showed that imposing factorization completely constrains all correlations between the branes, except for the brane one-point function. This one-point function is instead completely determined by the discrete spectrum of the putative dual quantum mechanics. We have shown that these statements uphold non-perturbatively, by studying the dual of the spacetime brane insertions in the JT gravity matrix integral. 

Our brane correlator $Z_\text{brane}(b_1,b_2)$ is similar in spirit to the two-boundary component of the baby-universe wavefunction in \cite{Saad:2021uzi}, and to the linked half-wormhole \cite{Saad:2021rcu}, but for one key difference. Namely that we have an \emph{exponential} of this correlator \eqref{nonlocex}. The exponential enables us to resolve factorization and discreteness to all orders in the genus expansion, and non-perturbatively, where the mechanisms proposed in \cite{Saad:2021rcu,Saad:2021uzi} were designed to resolve factorization to leading order.

In the remainder of this section, we discuss multiple interpretations for these brane correlators, propose a possible higher-dimensional extensions of the mechanism found to restore factorization and analyze the  UV implications of our findings.

\subsection*{Fight ensemble with ensemble}

The mechanism for factorization described here boils down to having the correct Gaussian correlation between different disconnected components of the spacetime. To emphasize that this is the key ingredient, rather than describing our model in terms of branes per se, we discuss several alternative perspectives. 

Combining \eqref{map} and \eqref{nonlocex}, and inserting the appropriate value \eqref{solution} for $Z_\text{brane}(b_1,b_2)$, one can rewrite our bilocal deformation of the dilaton gravity action as a local deformation, by introducing an auxiliary Hubbard-Stratanovich field $Q(b)$ (a normalization constant cancels in all observables)
\begin{align}
\label{eq:HS-deformation}
    &\exp\bigg(\int_0^\infty \d b\, b\,\mo_{\rm G}(b)\,Z_\text{brane}(b)-\frac{1}{2}\int_0^\infty \d b\, b\,\mo_{\rm G}(b)\mo_{\rm G}(b)\bigg)\nn\\
    &\hspace{3cm}=\int \mathcal{D}Q(b)\, \exp\bigg(\frac{1}{2}\int_0^{\infty} \d b\,(Q(b)-Z_\text{brane}(b))^2+\int_0^\infty \d b\, b\,Q(b)\,\mo_{\rm G}(b)\bigg)\nn\\&\hspace{3cm}\Leftrightarrow\average{\exp\bigg(\frac{1}{2}\int_\Sigma \d^2 x\sqrt{g}\,U_{\rm local}(\Phi,Q(b))\bigg)}_{\text{couplings}}\,,
\end{align}
where we also used the dictionary \eqref{map} between $\mo_{\rm G}(b)$ and a local deformation in the JT path integral. The average in the last equality is over $Q(b)$, using the Gaussian measure from the second line. The field $Q(b)$ thus introduces random coupling constants, since it does not depend on the spacetime coordinates $x$.\footnote{The field $Q(b)$ lives in the target space in string theory language. The relevance of target space in this setup is unclear.}  We see therefore that we can interpret our model as having a dilaton potential picked from an \emph{ensemble average}. In other words, an ensemble average in the bulk theory can potentially lead to a factorizable answer in the boundary theory. To see how this is possible, we need to remember that when computing the gravitational partition function (with let's say one asymptotic boundary) one should also in principle include a sum over all possible disconnected spacetimes. Typically, if the bulk interactions are fixed, this multiplicative factor is removed when normalizing the gravitational partition function with one asymptotic boundary by the gravitational path integral, this time, with no asymptotic boundaries. However, the ensemble average over $Q(b)$ implies that the contribution of disconnected closed universes is no longer removed by factorization; this is simply due to the typical observation that the average of a product in an ensemble is different than the product of averages. 

This trick of writing nonlocal interactions as random local interactions is similar to how Giddings-Strominger and Coleman thought about wormholes \cite{giddings1988loss,COLEMAN1988867}. Here though, we are introducing random coupling constants that cancel wormholes. The point is that all random couplings can be viewed as generating some type of connection between far-away regions of spacetime, but only in certain cases does that connection resemble a semiclassical wormhole. Other random couplings create non-geometric connections, which we believe encode details about the UV completion, examples are the tiny wormholes of \cite{Das:1989fq}.\footnote{See the discussion of \cite{Blommaert:2021gha}.} We have shown that those non-geometric connections can completely undo the the contribution of connected geometries (real wormholes) to the gravitational path integral.

Alternatively, as advertised in section \ref{sect:branes}, one can view our model as having explicit nonlocal spacetime interactions (the potential can be obtained explicitly from \eqref{nonlocex} and \eqref{solution} by doing the integrals over $b_1$ and $b_2$)
\begin{align}
\label{eq:bilocal-potential}
    -I_\text{nonlocal} &= e^{-2\S}
    \sum_{i,j=1}^n\int_{\Sigma_i} d^2 x_1\sqrt{g(x_1)}\int_{\Sigma_j} d^2 x_2  \sqrt{g(x_2)}\,U_\text{nonlocal}(\Phi(x_1),\Phi(x_2))\,,
\end{align}
where $\Sigma=\Sigma_1\cup \dots \cup \Sigma_n$, and the disconnected components can be closed universes or surfaces with an asymptotic boundary. Thanks to the sum over $i,j$ there are once again interactions between closed universes and the asymptotic boundary, and between different asymptotic boundaries: in this case the interaction will be manifest by correlating the profile of the dilaton on these different components. As emphasized before, these correlations were key for factorization.\footnote{It would be interesting to understand this nonlocal theory classically (not treating \eqref{eq:bilocal-potential} as a deformation that we expand). The one-point function \eqref{map} with \eqref{solution} seems to introduce a type of raggedness (or discretization) in the classical metric, but the classical role of the nonlocal interactions is more obscure.}

What are the experiences of some observer in this theory?  Even though the bilocal potential \eqref{eq:bilocal-potential} seems to induce interactions that might violate causality, the equivalence to the Hubbard-Stratanovich formulation (describing the boundary theory with fixed Hamiltonian as an ensemble average over dilaton couplings in the bulk \eqref{eq:HS-deformation}), and to the half-wormhole picture \eqref{eq:Z(beta)-after-simplification} reassure us that, at least as far as we can see, there is no violation of causality. Furthermore, the half-wormhole picture \eqref{eq:Z(beta)-after-simplification} implies that from the perspective of a boundary observer there is no real indication of the correlation between different spacetime components (since all spacetime wormholes precisely cancels the contributions of the correlated closed universes): the observer's only possible ``clues'' are simply the existence of a discrete spectrum and the factorization of all boundary observables.

\subsection*{Implications for higher dimensions}

The fact that factorization boils down to correlating disconnected components of the spacetime, appears to be independent of the number of spacetime dimensions and is not uniquely tied to our model of correlated branes. It is therefore not unreasonable to expect a similar mechanism to also hold in spacetime dimensions $D=d+1>2$. 

In the two dimensional case we studied in this paper, the boundary is always a circle and is the only boundary topology one needs to worry about. The resolution of factorization was then to just include branes such that manifolds with topology $I\times S^1$ are included in the path integral. Spacetimes with other topologies are always canceled with closed universes that are Gaussian correlated with the spacetime on which the asymptotic boundary resides. In higher dimensions, the boundary topology can be much more complicated and we would have a more general manifold $M_{d-1}$ as the spatial manifold. Extending our logic, it would then be enough to factorize topologies of the form $I\times S^1 \times M_{d-1}$, which is always true sufficiently close to the asymptotic boundary. However, since the spectrum depends on the spatial manifold $M_{d-1}$, each $M_{d-1}$ needs to be factorized and discretized seperately. From the gravitational theory we would then include a similar bilocal term to the action, just like \eqref{eq:bilocal-potential}. Again this can be decoupled using a Hubbard-Stratonovich field and would again have the interpretation of averaging over bulk couplings. Notice that this is independent of whatever is going on in the bulk (i.e. there could be fluxes, strings or branes wrapping cycles), just as our arguments were independent of what $\Sigma$ was in \eqref{eq:cancelllation-2-boundaries}. 

\subsection*{Non-geometric wormholes in string theory}

In full-fledged string theory, there should be configurations of strings that correspond with including wormholes on an asymptotic spacetime, because all allowable metric configurations should be included in the (non-perturbative) string description. Here, with wormholes, we mean the pure gravitational configurations, like the ones discussed in this work that are also include in JT gravity. If there are wormholes on an asymptotic spacetimes, there are also wormholes connecting asymptotic regions.\footnote{For example, one could imagine take a pinching limit of some higher genus boundary to obtain disconnected boundaries, as in \cite{Collier:2021rsn,Blommaert:2019hjr}, then some wormholes that were originally attached to the same asymptotic boundary will connect two different asymptotic regions, you can therefore not have one type of wormholes without having the other type too.} Therefore, even in full-fledged string theory, there is a factorization puzzle, as something needs to cancel these wormholes, such that the full answer factorizes. This ``something'' would be the UV degrees of freedom which after being integrated-out results in the nonlocal deformation which we rewrote in terms of correlated branes.

What could those UV degrees of freedom be? We suspect that the analogue of the brane two-point function are very non-geometric string configurations that connect two separated regions in spacetime.\footnote{One piece of evidence in favor of non-geometric configurations is that our brane two-point function is independent of of $\beta$ and $e^{\Ss}$ as discussed in section \ref{sec:additional-comments}.} Indeed, many string configurations have no semiclassical gravity interpretation, but they can still be important. To make progress on this, one would have to find some string theory for which there are some more-or-less classical wormhole configurations (unlike in the model of \cite{Eberhardt:2021jvj}), and find other string configurations that cancel those wormholes. 
For instance, it would be interesting to understand whether the mechanism of tachyon condensation, discussed in \cite{Horowitz:2006mr, Adams:2005rb}, which results in a spacetime with disconnected components, leaves behind non-trivial correlation between those components.

\section*{Acknowledgments}

We are grateful for discussions with Ahmed Almheiri, Christian Copetti, Clifford Johnson, Juan Maldacena, Thomas Mertens, Mark Mezei, Eva Silverstein, Steve Shenker, Douglas Stanford, Joaquin Gustavo Turiaci,  and Zhenbin Yang. AB was supported in part by a BEAF fellowship, by the SITP at Stanford, and by the ERC-COG Grant NP-QFT No. 864583. LVI was supported in part by the US NSF under Grant No. PHY-1820651 and by the Simons Collaboration on Ultra-Quantum Matter, a Simons Foundation Grant with No. 651440. JK is supported by the Simons Foundation.

\appendix 

\section{Branes versus defects}\label{apt:branes}

In this appendix we will discuss in a bit more detail the defect and trumpet geometry, and their analytic relation. Let us consider $2$d hyperbolic metrics in Fefferman-Graham gauge (with constant $B$)
\be 
\d s^2 = \d \rho^2 + (e^{\rho} - B\, e^{-\rho})^2\, \d u^2\,,\quad u\sim u+2\pi
\ee
Depending on the value of $B$ we have different geometries: $B = 1/4$ is the disk, $B$ positive (but unequal to $1/4$) are conical defect geometries and finally negative $B$ are trumpet geometries. The defect geometry has a conical singularity at the origin (we can always choose this to be at the origin)  with deficit angle $2\pi(1-2\sqrt{B})$, therefore it is the solution to\footnote{Here we follow \cite{Witten:2020ert} and normalize the delta function as $\int \d^2 x \sqrt{g}\, \delta^2(x) = 1$} 
\be \label{R2delta}
R + 2 = 4\pi(1-2\sqrt{B})\,\delta^2(x).
\ee

The trumpet geometry with $B<0$ has one asymptotically AdS boundary, and one geodesic boundary of fixed length $b$ and is related to the defect geometry by analytic continuation. To understand this a bit more, notice that for the trumpet geometry $g_{uu}$ will be positive for all $\rho$ and has a minimum at some $\rho = \rho_0$. The trumpet is then obtained by restricting the range of $\rho$ from $\rho_0$ to $\infty$ (or from $-\infty$ to $\rho_0$). For negative $B$ the coefficient of the delta function in \eqref{R2delta} becomes complex
\be \label{complexR2}
R + 2 = 2(2\pi \pm \i b)\delta^2(x)\,,
\ee
where we used that the size of the neck $b$ is related to $B$ as $b = 4\pi \sqrt{-B}$. So we get the familiar relation $b = \pm \i (2\pi - \a)$ \cite{Witten:2020ert} where the different signs correspond with the two choices of branch $\sqrt{-B}=\pm i\sqrt{B}$. The imaginary defect \eqref{complexR2} is located at $\rho_0\pm \i \pi/2$, which is why we see no curvature singularity in the trumpet geometry. When we insert an imaginary defect, we can choose the contour for the $\rho$ coordinate along the real axis from $\infty$ to $\rho_0$, and then along the imaginary axis until we reach the defect, analogous to how an expanding dS transitions to the semisphere. So, imaginary defects are the same as geodesic boundaries, they are only distinguished by where we imagine cutting off the geometry (at $\rho_0$ or $\rho_0\pm \i\pi/2$).

In \eqref{map} we also secretly encountered \eqref{complexR2}. Namely, if we take the dilaton potential corresponding to the insertion of an exponential of geodesic boundaries, we found the dilaton action
\be 
I[g,\Phi] = - \frac{1}{2} \int_{\Sigma} \d^2 x \sqrt{g} \left( \Phi(R+2) + 2e^{-\S}\cos(b\Phi)e^{-2\pi\Phi} \right)
\ee
As with the usual defect calculation \cite{Witten:2020wvy}, we now expand the part of the exponential $e^{-I}$ containing the non-trivial dilaton potential. This gives the correction to the usual JT partition function of the form 
\begin{align}
\delta Z(\beta)&=\frac{e^{-\S}}{2}\int \d^2 y \sqrt{g(y)} \int \frac{\mathcal{D}g\mathcal{D}\Phi}{\text{Vol}} \exp\left( -\frac{1}{2} \int_{\Sigma} \d^2 x \sqrt{g} ( \Phi(R+2) - 2(2\pi - \i b)\, \Phi\, \delta^2(x-y) ) \right)\nn\\&\quad +\frac{e^{-\S}}{2}\int \d^2 y \sqrt{g(y)} \int \frac{\mathcal{D}g\mathcal{D}\Phi}{\text{Vol}} \exp\left( -\frac{1}{2} \int_{\Sigma} \d^2 x \sqrt{g} ( \Phi(R+2) - 2(2\pi + \i b)\, \Phi\, \delta^2(x-y) ) \right)\nn\\
&\quad +\text{multiple defects.}
\end{align}
The equations of motion associated to each branch are precisely those we had found before by analytically continuing the defect geometry \eqref{R2delta}. The deformed dilaton potential takes both defects with $\pm b$ into account with equal weight. This is also what one expects from the BF perspective; the holonomies with $\pm b$ are conjugate by an SO($2)$ rotation (so then all physical observables are an even function of $b$) \cite{Mertens:2019tcm}.

\bibliographystyle{ourbst}
\bibliography{main}

\providecommand{\href}[2]{#2}\begingroup\raggedright\begin{thebibliography}{10}

\bibitem{tHooft:1993dmi}
G.~'t~Hooft, ``{Dimensional reduction in quantum gravity},'' {\em Conf. Proc.
  C} {\bfseries 930308} (1993) 284--296,
  \href{http://arxiv.org/abs/gr-qc/9310026}{{\ttfamily arXiv:gr-qc/9310026}}.

\bibitem{Susskind:1994vu}
L.~Susskind, ``{The World as a hologram},''
  \href{http://dx.doi.org/10.1063/1.531249}{{\em J. Math. Phys.} {\bfseries 36}
  (1995) 6377--6396}, \href{http://arxiv.org/abs/hep-th/9409089}{{\ttfamily
  arXiv:hep-th/9409089}}.

\bibitem{Maldacena:1997re}
J.~M. Maldacena, ``{The Large N limit of superconformal field theories and
  supergravity},'' \href{http://dx.doi.org/10.1023/A:1026654312961}{{\em Adv.
  Theor. Math. Phys.} {\bfseries 2} (1998) 231--252},
  \href{http://arxiv.org/abs/hep-th/9711200}{{\ttfamily arXiv:hep-th/9711200}}.

\bibitem{Page:1993df}
D.~N. Page, ``{Average entropy of a subsystem},''
  \href{http://dx.doi.org/10.1103/PhysRevLett.71.1291}{{\em Phys. Rev. Lett.}
  {\bfseries 71} (1993) 1291--1294},
  \href{http://arxiv.org/abs/gr-qc/9305007}{{\ttfamily arXiv:gr-qc/9305007}}.

\bibitem{Almheiri:2019qdq}
A.~Almheiri, T.~Hartman, J.~Maldacena, E.~Shaghoulian, and A.~Tajdini,
  ``{Replica Wormholes and the Entropy of Hawking Radiation},''
  \href{http://dx.doi.org/10.1007/JHEP05(2020)013}{{\em JHEP} {\bfseries 05}
  (2020) 013}, \href{http://arxiv.org/abs/1911.12333}{{\ttfamily
  arXiv:1911.12333 [hep-th]}}.

\bibitem{Penington:2019kki}
G.~Penington, S.~H. Shenker, D.~Stanford, and Z.~Yang, ``{Replica wormholes and
  the black hole interior},'' \href{http://arxiv.org/abs/1911.11977}{{\ttfamily
  arXiv:1911.11977 [hep-th]}}.

\bibitem{Almheiri:2020cfm}
A.~Almheiri, T.~Hartman, J.~Maldacena, E.~Shaghoulian, and A.~Tajdini, ``{The
  entropy of Hawking radiation},''
  \href{http://dx.doi.org/10.1103/RevModPhys.93.035002}{{\em Rev. Mod. Phys.}
  {\bfseries 93} no.~3, (2021) 035002},
  \href{http://arxiv.org/abs/2006.06872}{{\ttfamily arXiv:2006.06872
  [hep-th]}}.

\bibitem{Maldacena:2004rf}
J.~M. Maldacena and L.~Maoz, ``{Wormholes in AdS},''
  \href{http://dx.doi.org/10.1088/1126-6708/2004/02/053}{{\em JHEP} {\bfseries
  02} (2004) 053}, \href{http://arxiv.org/abs/hep-th/0401024}{{\ttfamily
  arXiv:hep-th/0401024}}.

\bibitem{Iliesiu:2021are}
L.~V. Iliesiu, M.~Kologlu, and G.~J. Turiaci, ``{Supersymmetric indices
  factorize},'' \href{http://arxiv.org/abs/2107.09062}{{\ttfamily
  arXiv:2107.09062 [hep-th]}}.

\bibitem{Maloney:2007ud}
A.~Maloney and E.~Witten, ``{Quantum Gravity Partition Functions in Three
  Dimensions},'' \href{http://dx.doi.org/10.1007/JHEP02(2010)029}{{\em JHEP}
  {\bfseries 02} (2010) 029}, \href{http://arxiv.org/abs/0712.0155}{{\ttfamily
  arXiv:0712.0155 [hep-th]}}.

\bibitem{Cotler:2020ugk}
J.~Cotler and K.~Jensen, ``{AdS$_{3}$ gravity and random CFT},''
  \href{http://dx.doi.org/10.1007/JHEP04(2021)033}{{\em JHEP} {\bfseries 04}
  (2021) 033}, \href{http://arxiv.org/abs/2006.08648}{{\ttfamily
  arXiv:2006.08648 [hep-th]}}.

\bibitem{Maxfield:2020ale}
H.~Maxfield and G.~J. Turiaci, ``{The path integral of 3D gravity near
  extremality; or, JT gravity with defects as a matrix integral},''
  \href{http://dx.doi.org/10.1007/JHEP01(2021)118}{{\em JHEP} {\bfseries 01}
  (2021) 118}, \href{http://arxiv.org/abs/2006.11317}{{\ttfamily
  arXiv:2006.11317 [hep-th]}}.

\bibitem{Maldacena:2001kr}
J.~M. Maldacena, ``{Eternal black holes in anti-de Sitter},''
  \href{http://dx.doi.org/10.1088/1126-6708/2003/04/021}{{\em JHEP} {\bfseries
  04} (2003) 021}, \href{http://arxiv.org/abs/hep-th/0106112}{{\ttfamily
  arXiv:hep-th/0106112}}.

\bibitem{Cotler:2016fpe}
J.~S. Cotler, G.~Gur-Ari, M.~Hanada, J.~Polchinski, P.~Saad, S.~H. Shenker,
  D.~Stanford, A.~Streicher, and M.~Tezuka, ``{Black Holes and Random
  Matrices},'' \href{http://dx.doi.org/10.1007/JHEP05(2017)118}{{\em JHEP}
  {\bfseries 05} (2017) 118}, \href{http://arxiv.org/abs/1611.04650}{{\ttfamily
  arXiv:1611.04650 [hep-th]}}.

\bibitem{Saad:2018bqo}
P.~Saad, S.~H. Shenker, and D.~Stanford, ``{A semiclassical ramp in SYK and in
  gravity},'' \href{http://arxiv.org/abs/1806.06840}{{\ttfamily
  arXiv:1806.06840 [hep-th]}}.

\bibitem{Saad:2019pqd}
P.~Saad, ``{Late Time Correlation Functions, Baby Universes, and ETH in JT
  Gravity},'' \href{http://arxiv.org/abs/1910.10311}{{\ttfamily
  arXiv:1910.10311 [hep-th]}}.

\bibitem{Blommaert:2019hjr}
A.~Blommaert, T.~G. Mertens, and H.~Verschelde, ``{Clocks and Rods in
  Jackiw-Teitelboim Quantum Gravity},''
  \href{http://dx.doi.org/10.1007/JHEP09(2019)060}{{\em JHEP} {\bfseries 09}
  (2019) 060}, \href{http://arxiv.org/abs/1902.11194}{{\ttfamily
  arXiv:1902.11194 [hep-th]}}.

\bibitem{Blommaert:2020seb}
A.~Blommaert, ``{Dissecting the ensemble in JT gravity},''
  \href{http://arxiv.org/abs/2006.13971}{{\ttfamily arXiv:2006.13971
  [hep-th]}}.

\bibitem{Saad:2019lba}
P.~Saad, S.~H. Shenker, and D.~Stanford, ``{JT gravity as a matrix integral},''
  \href{http://arxiv.org/abs/1903.11115}{{\ttfamily arXiv:1903.11115
  [hep-th]}}.

\bibitem{Marolf:2020xie}
D.~Marolf and H.~Maxfield, ``{Transcending the ensemble: baby universes,
  spacetime wormholes, and the order and disorder of black hole information},''
  \href{http://dx.doi.org/10.1007/JHEP08(2020)044}{{\em JHEP} {\bfseries 08}
  (2020) 044}, \href{http://arxiv.org/abs/2002.08950}{{\ttfamily
  arXiv:2002.08950 [hep-th]}}.

\bibitem{Stanford:2020wkf}
D.~Stanford, ``{More quantum noise from wormholes},''
  \href{http://arxiv.org/abs/2008.08570}{{\ttfamily arXiv:2008.08570
  [hep-th]}}.

\bibitem{Blommaert:2019wfy}
A.~Blommaert, T.~G. Mertens, and H.~Verschelde, ``{Eigenbranes in
  Jackiw-Teitelboim gravity},''
  \href{http://arxiv.org/abs/1911.11603}{{\ttfamily arXiv:1911.11603
  [hep-th]}}.

\bibitem{Pollack:2020gfa}
J.~Pollack, M.~Rozali, J.~Sully, and D.~Wakeham, ``{Eigenstate Thermalization
  and Disorder Averaging in Gravity},''
  \href{http://dx.doi.org/10.1103/PhysRevLett.125.021601}{{\em Phys. Rev.
  Lett.} {\bfseries 125} no.~2, (2020) 021601},
  \href{http://arxiv.org/abs/2002.02971}{{\ttfamily arXiv:2002.02971
  [hep-th]}}.

\bibitem{Afkhami-Jeddi:2020ezh}
N.~Afkhami-Jeddi, H.~Cohn, T.~Hartman, and A.~Tajdini, ``{Free partition
  functions and an averaged holographic duality},''
  \href{http://dx.doi.org/10.1007/JHEP01(2021)130}{{\em JHEP} {\bfseries 01}
  (2021) 130}, \href{http://arxiv.org/abs/2006.04839}{{\ttfamily
  arXiv:2006.04839 [hep-th]}}.

\bibitem{Maloney:2020nni}
A.~Maloney and E.~Witten, ``{Averaging over Narain moduli space},''
  \href{http://dx.doi.org/10.1007/JHEP10(2020)187}{{\em JHEP} {\bfseries 10}
  (2020) 187}, \href{http://arxiv.org/abs/2006.04855}{{\ttfamily
  arXiv:2006.04855 [hep-th]}}.

\bibitem{Belin:2020hea}
A.~Belin and J.~de~Boer, ``{Random Statistics of OPE Coefficients and Euclidean
  Wormholes},'' \href{http://arxiv.org/abs/2006.05499}{{\ttfamily
  arXiv:2006.05499 [hep-th]}}.

\bibitem{Anous:2020lka}
T.~Anous, J.~Kruthoff, and R.~Mahajan, ``{Density matrices in quantum
  gravity},'' \href{http://dx.doi.org/10.21468/SciPostPhys.9.4.045}{{\em
  SciPost Phys.} {\bfseries 9} no.~4, (2020) 045},
  \href{http://arxiv.org/abs/2006.17000}{{\ttfamily arXiv:2006.17000
  [hep-th]}}.

\bibitem{Chen:2020tes}
Y.~Chen, V.~Gorbenko, and J.~Maldacena, ``{Bra-ket wormholes in gravitationally
  prepared states},'' \href{http://dx.doi.org/10.1007/JHEP02(2021)009}{{\em
  JHEP} {\bfseries 02} (2021) 009},
  \href{http://arxiv.org/abs/2007.16091}{{\ttfamily arXiv:2007.16091
  [hep-th]}}.

\bibitem{Liu:2020jsv}
H.~Liu and S.~Vardhan, ``{Entanglement entropies of equilibrated pure states in
  quantum many-body systems and gravity},''
  \href{http://dx.doi.org/10.1103/PRXQuantum.2.010344}{{\em P. R. X. Quantum.}
  {\bfseries 2} (2021) 010344},
  \href{http://arxiv.org/abs/2008.01089}{{\ttfamily arXiv:2008.01089
  [hep-th]}}.

\bibitem{Marolf:2021kjc}
D.~Marolf and J.~E. Santos, ``{AdS Euclidean wormholes},''
  \href{http://arxiv.org/abs/2101.08875}{{\ttfamily arXiv:2101.08875
  [hep-th]}}.

\bibitem{Meruliya:2021utr}
V.~Meruliya, S.~Mukhi, and P.~Singh, ``{Poincar\'e Series, 3d Gravity and
  Averages of Rational CFT},''
  \href{http://dx.doi.org/10.1007/JHEP04(2021)267}{{\em JHEP} {\bfseries 04}
  (2021) 267}, \href{http://arxiv.org/abs/2102.03136}{{\ttfamily
  arXiv:2102.03136 [hep-th]}}.

\bibitem{Giddings:2020yes}
S.~B. Giddings and G.~J. Turiaci, ``{Wormhole calculus, replicas, and
  entropies},'' \href{http://dx.doi.org/10.1007/JHEP09(2020)194}{{\em JHEP}
  {\bfseries 09} (2020) 194}, \href{http://arxiv.org/abs/2004.02900}{{\ttfamily
  arXiv:2004.02900 [hep-th]}}.

\bibitem{Stanford:2019vob}
D.~Stanford and E.~Witten, ``{JT Gravity and the Ensembles of Random Matrix
  Theory},'' \href{http://arxiv.org/abs/1907.03363}{{\ttfamily arXiv:1907.03363
  [hep-th]}}.

\bibitem{Okuyama:2019xbv}
K.~Okuyama and K.~Sakai, ``{JT gravity, KdV equations and macroscopic loop
  operators},'' \href{http://dx.doi.org/10.1007/JHEP01(2020)156}{{\em JHEP}
  {\bfseries 01} (2020) 156}, \href{http://arxiv.org/abs/1911.01659}{{\ttfamily
  arXiv:1911.01659 [hep-th]}}.

\bibitem{Belin:2020jxr}
A.~Belin, J.~De~Boer, P.~Nayak, and J.~Sonner, ``{Charged Eigenstate
  Thermalization, Euclidean Wormholes and Global Symmetries in Quantum
  Gravity},'' \href{http://arxiv.org/abs/2012.07875}{{\ttfamily
  arXiv:2012.07875 [hep-th]}}.

\bibitem{Verlinde:2021jwu}
H.~Verlinde, ``{Deconstructing the Wormhole: Factorization, Entanglement and
  Decoherence},'' \href{http://arxiv.org/abs/2105.02142}{{\ttfamily
  arXiv:2105.02142 [hep-th]}}.

\bibitem{Collier:2021rsn}
S.~Collier and A.~Maloney, ``{Wormholes and Spectral Statistics in the Narain
  Ensemble},'' \href{http://arxiv.org/abs/2106.12760}{{\ttfamily
  arXiv:2106.12760 [hep-th]}}.

\bibitem{Betzios:2021fnm}
P.~Betzios, E.~Kiritsis, and O.~Papadoulaki, ``{Interacting systems and
  wormholes},'' \href{http://arxiv.org/abs/2110.14655}{{\ttfamily
  arXiv:2110.14655 [hep-th]}}.

\bibitem{Belin:2021ryy}
A.~Belin, J.~de~Boer, and D.~Liska, ``{Non-Gaussianities in the Statistical
  Distribution of Heavy OPE Coefficients and Wormholes},''
  \href{http://arxiv.org/abs/2110.14649}{{\ttfamily arXiv:2110.14649
  [hep-th]}}.

\bibitem{Saad:2021uzi}
P.~Saad, S.~Shenker, and S.~Yao, ``{Comments on wormholes and factorization},''
  \href{http://arxiv.org/abs/2107.13130}{{\ttfamily arXiv:2107.13130
  [hep-th]}}.

\bibitem{Strominger:1996sh}
A.~Strominger and C.~Vafa, ``{Microscopic origin of the Bekenstein-Hawking
  entropy},'' \href{http://dx.doi.org/10.1016/0370-2693(96)00345-0}{{\em Phys.
  Lett. B} {\bfseries 379} (1996) 99--104},
  \href{http://arxiv.org/abs/hep-th/9601029}{{\ttfamily arXiv:hep-th/9601029}}.

\bibitem{Eberhardt:2018ouy}
L.~Eberhardt, M.~R. Gaberdiel, and R.~Gopakumar, ``{The Worldsheet Dual of the
  Symmetric Product CFT},''
  \href{http://dx.doi.org/10.1007/JHEP04(2019)103}{{\em JHEP} {\bfseries 04}
  (2019) 103}, \href{http://arxiv.org/abs/1812.01007}{{\ttfamily
  arXiv:1812.01007 [hep-th]}}.

\bibitem{Eberhardt:2020bgq}
L.~Eberhardt, ``{Partition functions of the tensionless string},''
  \href{http://dx.doi.org/10.1007/JHEP03(2021)176}{{\em JHEP} {\bfseries 03}
  (2021) 176}, \href{http://arxiv.org/abs/2008.07533}{{\ttfamily
  arXiv:2008.07533 [hep-th]}}.

\bibitem{Eberhardt:2021jvj}
L.~Eberhardt, ``{Summing over Geometries in String Theory},''
  \href{http://dx.doi.org/10.1007/JHEP05(2021)233}{{\em JHEP} {\bfseries 05}
  (2021) 233}, \href{http://arxiv.org/abs/2102.12355}{{\ttfamily
  arXiv:2102.12355 [hep-th]}}.

\bibitem{Saad:2021rcu}
P.~Saad, S.~H. Shenker, D.~Stanford, and S.~Yao, ``{Wormholes without
  averaging},'' \href{http://arxiv.org/abs/2103.16754}{{\ttfamily
  arXiv:2103.16754 [hep-th]}}.

\bibitem{Maldacena:2016hyu}
J.~Maldacena and D.~Stanford, ``{Remarks on the Sachdev-Ye-Kitaev model},''
  \href{http://dx.doi.org/10.1103/PhysRevD.94.106002}{{\em Phys. Rev. D}
  {\bfseries 94} no.~10, (2016) 106002},
  \href{http://arxiv.org/abs/1604.07818}{{\ttfamily arXiv:1604.07818
  [hep-th]}}.

\bibitem{Almheiri:2016fws}
A.~Almheiri and B.~Kang, ``{Conformal Symmetry Breaking and Thermodynamics of
  Near-Extremal Black Holes},''
  \href{http://dx.doi.org/10.1007/JHEP10(2016)052}{{\em JHEP} {\bfseries 10}
  (2016) 052}, \href{http://arxiv.org/abs/1606.04108}{{\ttfamily
  arXiv:1606.04108 [hep-th]}}.

\bibitem{Nayak:2018qej}
P.~Nayak, A.~Shukla, R.~M. Soni, S.~P. Trivedi, and V.~Vishal, ``{On the
  Dynamics of Near-Extremal Black Holes},''
  \href{http://dx.doi.org/10.1007/JHEP09(2018)048}{{\em JHEP} {\bfseries 09}
  (2018) 048}, \href{http://arxiv.org/abs/1802.09547}{{\ttfamily
  arXiv:1802.09547 [hep-th]}}.

\bibitem{Castro:2018ffi}
A.~Castro, F.~Larsen, and I.~Papadimitriou, ``{5D rotating black holes and the
  nAdS$_{2}$/nCFT$_{1}$ correspondence},''
  \href{http://dx.doi.org/10.1007/JHEP10(2018)042}{{\em JHEP} {\bfseries 10}
  (2018) 042}, \href{http://arxiv.org/abs/1807.06988}{{\ttfamily
  arXiv:1807.06988 [hep-th]}}.

\bibitem{Moitra:2019bub}
U.~Moitra, S.~K. Sake, S.~P. Trivedi, and V.~Vishal, ``{Jackiw-Teitelboim
  Gravity and Rotating Black Holes},''
  \href{http://dx.doi.org/10.1007/JHEP11(2019)047}{{\em JHEP} {\bfseries 11}
  (2019) 047}, \href{http://arxiv.org/abs/1905.10378}{{\ttfamily
  arXiv:1905.10378 [hep-th]}}.

\bibitem{Sachdev:2019bjn}
S.~Sachdev, ``{Universal low temperature theory of charged black holes with
  AdS$_2$ horizons},'' \href{http://dx.doi.org/10.1063/1.5092726}{{\em J. Math.
  Phys.} {\bfseries 60} no.~5, (2019) 052303},
  \href{http://arxiv.org/abs/1902.04078}{{\ttfamily arXiv:1902.04078
  [hep-th]}}.

\bibitem{Yang:2018gdb}
Z.~Yang, ``{The Quantum Gravity Dynamics of Near Extremal Black Holes},''
  \href{http://dx.doi.org/10.1007/JHEP05(2019)205}{{\em JHEP} {\bfseries 05}
  (2019) 205}, \href{http://arxiv.org/abs/1809.08647}{{\ttfamily
  arXiv:1809.08647 [hep-th]}}.

\bibitem{Iliesiu:2020qvm}
L.~V. Iliesiu and G.~J. Turiaci, ``{The statistical mechanics of near-extremal
  black holes},'' \href{http://dx.doi.org/10.1007/JHEP05(2021)145}{{\em JHEP}
  {\bfseries 05} (2021) 145}, \href{http://arxiv.org/abs/2003.02860}{{\ttfamily
  arXiv:2003.02860 [hep-th]}}.

\bibitem{Heydeman:2020hhw}
M.~Heydeman, L.~V. Iliesiu, G.~J. Turiaci, and W.~Zhao, ``{The statistical
  mechanics of near-BPS black holes},''
  \href{http://arxiv.org/abs/2011.01953}{{\ttfamily arXiv:2011.01953
  [hep-th]}}.

\bibitem{Mertens:2019tcm}
T.~G. Mertens and G.~J. Turiaci, ``{Defects in Jackiw-Teitelboim Quantum
  Gravity},'' \href{http://dx.doi.org/10.1007/JHEP08(2019)127}{{\em JHEP}
  {\bfseries 08} (2019) 127}, \href{http://arxiv.org/abs/1904.05228}{{\ttfamily
  arXiv:1904.05228 [hep-th]}}.

\bibitem{Witten:2020wvy}
E.~Witten, ``{Matrix Models and Deformations of JT Gravity},''
  \href{http://dx.doi.org/10.1098/rspa.2020.0582}{{\em Proc. Roy. Soc. Lond. A}
  {\bfseries 476} no.~2244, (2020) 20200582},
  \href{http://arxiv.org/abs/2006.13414}{{\ttfamily arXiv:2006.13414
  [hep-th]}}.

\bibitem{Blommaert:2021gha}
A.~Blommaert and J.~Kruthoff, ``{Gravity without averaging},''
  \href{http://arxiv.org/abs/2107.02178}{{\ttfamily arXiv:2107.02178
  [hep-th]}}.

\bibitem{JafferisEOW}
P.~Gao, D.~L. Jafferis, and D.~K. Kolchmeyer, ``{An effective matrix model for
  dynamical end of the world branes in Jackiw-Teitelboim gravity},''
  \href{http://arxiv.org/abs/2104.01184}{{\ttfamily arXiv:2104.01184
  [hep-th]}}.

\bibitem{Mukhametzhanov:2021hdi}
B.~Mukhametzhanov, ``{Factorization and complex couplings in SYK and in Matrix
  Models},'' \href{http://arxiv.org/abs/2110.06221}{{\ttfamily arXiv:2110.06221
  [hep-th]}}.

\bibitem{jackiw1985lower}
R.~Jackiw, ``Lower dimensional gravity,'' {\em Nuclear Physics B} {\bfseries
  252} (1985) 343--356.

\bibitem{teitelboim1983gravitation}
C.~Teitelboim, ``Gravitation and hamiltonian structure in two spacetime
  dimensions,'' {\em Physics Letters B} {\bfseries 126} no.~1-2, (1983) 41--45.

\bibitem{Jensen:2016pah}
K.~Jensen, ``{Chaos in AdS$_2$ Holography},''
  \href{http://dx.doi.org/10.1103/PhysRevLett.117.111601}{{\em Phys. Rev.
  Lett.} {\bfseries 117} no.~11, (2016) 111601},
  \href{http://arxiv.org/abs/1605.06098}{{\ttfamily arXiv:1605.06098
  [hep-th]}}.

\bibitem{Maldacena:2016upp}
J.~Maldacena, D.~Stanford, and Z.~Yang, ``{Conformal symmetry and its breaking
  in two dimensional Nearly Anti-de-Sitter space},''
  \href{http://dx.doi.org/10.1093/ptep/ptw124}{{\em PTEP} {\bfseries 2016}
  no.~12, (2016) 12C104}, \href{http://arxiv.org/abs/1606.01857}{{\ttfamily
  arXiv:1606.01857 [hep-th]}}.

\bibitem{Engelsoy:2016xyb}
J.~Engels\"oy, T.~G. Mertens, and H.~Verlinde, ``{An investigation of AdS$_{2}$
  backreaction and holography},''
  \href{http://dx.doi.org/10.1007/JHEP07(2016)139}{{\em JHEP} {\bfseries 07}
  (2016) 139}, \href{http://arxiv.org/abs/1606.03438}{{\ttfamily
  arXiv:1606.03438 [hep-th]}}.

\bibitem{Dijkgraaf:2018vnm}
R.~Dijkgraaf and E.~Witten, ``{Developments in Topological Gravity},''
  \href{http://dx.doi.org/10.1142/S0217751X18300296}{{\em Int. J. Mod. Phys. A}
  {\bfseries 33} no.~30, (2018) 1830029},
  \href{http://arxiv.org/abs/1804.03275}{{\ttfamily arXiv:1804.03275
  [hep-th]}}.

\bibitem{mirzakhani2007simple}
M.~Mirzakhani, ``Simple geodesics and weil-petersson volumes of moduli spaces
  of bordered riemann surfaces,'' {\em Inventiones mathematicae} {\bfseries
  167} no.~1, (2007) 179--222.

\bibitem{Stanford:2017thb}
D.~Stanford and E.~Witten, ``{Fermionic Localization of the Schwarzian
  Theory},'' \href{http://dx.doi.org/10.1007/JHEP10(2017)008}{{\em JHEP}
  {\bfseries 10} (2017) 008}, \href{http://arxiv.org/abs/1703.04612}{{\ttfamily
  arXiv:1703.04612 [hep-th]}}.

\bibitem{Eynard:2007fi}
B.~Eynard and N.~Orantin, ``{Weil-Petersson volume of moduli spaces,
  Mirzakhani's recursion and matrix models},''
  \href{http://arxiv.org/abs/0705.3600}{{\ttfamily arXiv:0705.3600 [math-ph]}}.

\bibitem{Goel:2020yxl}
A.~Goel, L.~V. Iliesiu, J.~Kruthoff, and Z.~Yang, ``{Classifying boundary
  conditions in JT gravity: from energy-branes to $\alpha$-branes},''
  \href{http://dx.doi.org/10.1007/JHEP04(2021)069}{{\em JHEP} {\bfseries 04}
  (2021) 069}, \href{http://arxiv.org/abs/2010.12592}{{\ttfamily
  arXiv:2010.12592 [hep-th]}}.

\bibitem{Maldacena:2004sn}
J.~M. Maldacena, G.~W. Moore, N.~Seiberg, and D.~Shih, ``{Exact vs.
  semiclassical target space of the minimal string},''
  \href{http://dx.doi.org/10.1088/1126-6708/2004/10/020}{{\em JHEP} {\bfseries
  10} (2004) 020}, \href{http://arxiv.org/abs/hep-th/0408039}{{\ttfamily
  arXiv:hep-th/0408039}}.

\bibitem{Fateev:2000ik}
V.~Fateev, A.~B. Zamolodchikov, and A.~B. Zamolodchikov, ``{Boundary Liouville
  field theory. 1. Boundary state and boundary two point function},''
  \href{http://arxiv.org/abs/hep-th/0001012}{{\ttfamily arXiv:hep-th/0001012}}.

\bibitem{Ponsot:2001ng}
B.~Ponsot and J.~Teschner, ``{Boundary Liouville field theory: Boundary three
  point function},''
  \href{http://dx.doi.org/10.1016/S0550-3213(01)00596-X}{{\em Nucl. Phys. B}
  {\bfseries 622} (2002) 309--327},
  \href{http://arxiv.org/abs/hep-th/0110244}{{\ttfamily arXiv:hep-th/0110244}}.

\bibitem{Mertens:2020hbs}
T.~G. Mertens and G.~J. Turiaci, ``{Liouville quantum gravity -- holography, JT
  and matrices},'' \href{http://dx.doi.org/10.1007/JHEP01(2021)073}{{\em JHEP}
  {\bfseries 01} (2021) 073}, \href{http://arxiv.org/abs/2006.07072}{{\ttfamily
  arXiv:2006.07072 [hep-th]}}.

\bibitem{Mertens:2020pfe}
T.~G. Mertens, ``{Degenerate operators in JT and Liouville (super)gravity},''
  \href{http://dx.doi.org/10.1007/JHEP04(2021)245}{{\em JHEP} {\bfseries 04}
  (2021) 245}, \href{http://arxiv.org/abs/2007.00998}{{\ttfamily
  arXiv:2007.00998 [hep-th]}}.

\bibitem{Hosomichi:2008th}
K.~Hosomichi, ``{Minimal Open Strings},''
  \href{http://dx.doi.org/10.1088/1126-6708/2008/06/029}{{\em JHEP} {\bfseries
  06} (2008) 029}, \href{http://arxiv.org/abs/0804.4721}{{\ttfamily
  arXiv:0804.4721 [hep-th]}}.

\bibitem{Kostov:2002uq}
I.~K. Kostov, ``{Boundary correlators in 2-D quantum gravity: Liouville versus
  discrete approach},''
  \href{http://dx.doi.org/10.1016/S0550-3213(03)00147-0}{{\em Nucl. Phys. B}
  {\bfseries 658} (2003) 397--416},
  \href{http://arxiv.org/abs/hep-th/0212194}{{\ttfamily arXiv:hep-th/0212194}}.

\bibitem{Okuyama:2021eju}
K.~Okuyama and K.~Sakai, ``{FZZT branes in JT gravity and topological
  gravity},'' \href{http://arxiv.org/abs/2108.03876}{{\ttfamily
  arXiv:2108.03876 [hep-th]}}.

\bibitem{Teschner:2000md}
J.~Teschner, ``{Remarks on Liouville theory with boundary},''
  \href{http://dx.doi.org/10.22323/1.006.0041}{{\em PoS} {\bfseries tmr2000}
  (2000) 041}, \href{http://arxiv.org/abs/hep-th/0009138}{{\ttfamily
  arXiv:hep-th/0009138}}.

\bibitem{Witten:2020ert}
E.~Witten, ``{Deformations of JT Gravity and Phase Transitions},''
  \href{http://arxiv.org/abs/2006.03494}{{\ttfamily arXiv:2006.03494
  [hep-th]}}.

\bibitem{Blommaert:2021etf}
A.~Blommaert and M.~Usatyuk, ``{Microstructure in matrix elements},''
  \href{http://arxiv.org/abs/2108.02210}{{\ttfamily arXiv:2108.02210
  [hep-th]}}.

\bibitem{mehta2004random}
M.~L. Mehta, {\em Random matrices}.
\newblock Elsevier, 2004.

\bibitem{brezin1993planar}
E.~Br{\'e}zin, C.~Itzykson, G.~Parisi, and J.-B. Zuber, ``Planar diagrams,'' in
  {\em The Large N Expansion In Quantum Field Theory And Statistical Physics:
  From Spin Systems to 2-Dimensional Gravity}, pp.~567--583.
\newblock World Scientific, 1993.

\bibitem{migdal1983loop}
A.~A. Migdal, ``Loop equations and 1n expansion,'' {\em Physics Reports}
  {\bfseries 102} no.~4, (1983) 199--290.

\bibitem{Dijkgraaf:2009pc}
R.~Dijkgraaf and C.~Vafa, ``{Toda Theories, Matrix Models, Topological Strings,
  and N=2 Gauge Systems},'' \href{http://arxiv.org/abs/0909.2453}{{\ttfamily
  arXiv:0909.2453 [hep-th]}}.

\bibitem{dumitriu2005eigenvalues}
I.~Dumitriu and A.~Edelman, ``Eigenvalues of hermite and laguerre ensembles:
  large beta asymptotics,'' in {\em Annales de l'IHP Probabilit{\'e}s et
  statistiques}, vol.~41, pp.~1083--1099.
\newblock 2005.

\bibitem{Mertens:2017mtv}
T.~G. Mertens, G.~J. Turiaci, and H.~L. Verlinde, ``{Solving the Schwarzian via
  the Conformal Bootstrap},''
  \href{http://dx.doi.org/10.1007/JHEP08(2017)136}{{\em JHEP} {\bfseries 08}
  (2017) 136}, \href{http://arxiv.org/abs/1705.08408}{{\ttfamily
  arXiv:1705.08408 [hep-th]}}.

\bibitem{Mertens:2018fds}
T.~G. Mertens, ``{The Schwarzian theory \textemdash{} origins},''
  \href{http://dx.doi.org/10.1007/JHEP05(2018)036}{{\em JHEP} {\bfseries 05}
  (2018) 036}, \href{http://arxiv.org/abs/1801.09605}{{\ttfamily
  arXiv:1801.09605 [hep-th]}}.

\bibitem{Kitaev:2018wpr}
A.~Kitaev and S.~J. Suh, ``{Statistical mechanics of a two-dimensional black
  hole},'' \href{http://dx.doi.org/10.1007/JHEP05(2019)198}{{\em JHEP}
  {\bfseries 05} (2019) 198}, \href{http://arxiv.org/abs/1808.07032}{{\ttfamily
  arXiv:1808.07032 [hep-th]}}.

\bibitem{Blommaert:2018iqz}
A.~Blommaert, T.~G. Mertens, and H.~Verschelde, ``{Fine Structure of
  Jackiw-Teitelboim Quantum Gravity},''
  \href{http://dx.doi.org/10.1007/JHEP09(2019)066}{{\em JHEP} {\bfseries 09}
  (2019) 066}, \href{http://arxiv.org/abs/1812.00918}{{\ttfamily
  arXiv:1812.00918 [hep-th]}}.

\bibitem{Iliesiu:2019xuh}
L.~V. Iliesiu, S.~S. Pufu, H.~Verlinde, and Y.~Wang, ``{An exact quantization
  of Jackiw-Teitelboim gravity},''
  \href{http://dx.doi.org/10.1007/JHEP11(2019)091}{{\em JHEP} {\bfseries 11}
  (2019) 091}, \href{http://arxiv.org/abs/1905.02726}{{\ttfamily
  arXiv:1905.02726 [hep-th]}}.

\bibitem{Blommaert:2018oro}
A.~Blommaert, T.~G. Mertens, and H.~Verschelde, ``{The Schwarzian Theory - A
  Wilson Line Perspective},''
  \href{http://dx.doi.org/10.1007/JHEP12(2018)022}{{\em JHEP} {\bfseries 12}
  (2018) 022}, \href{http://arxiv.org/abs/1806.07765}{{\ttfamily
  arXiv:1806.07765 [hep-th]}}.

\bibitem{Iliesiu:2021ari}
L.~V. Iliesiu, M.~Mezei, and G.~S\'arosi, ``{The volume of the black hole
  interior at late times},'' \href{http://arxiv.org/abs/2107.06286}{{\ttfamily
  arXiv:2107.06286 [hep-th]}}.

\bibitem{Susskind:2018pmk}
L.~Susskind, \href{http://dx.doi.org/10.1007/978-3-030-45109-7}{``{Three
  Lectures on Complexity and Black Holes},''} SpringerBriefs in Physics.
\newblock Springer, 10, 2018.
\newblock \href{http://arxiv.org/abs/1810.11563}{{\ttfamily arXiv:1810.11563
  [hep-th]}}.

\bibitem{Hsin:2020mfa}
P.-S. Hsin, L.~V. Iliesiu, and Z.~Yang, ``{A violation of global symmetries
  from replica wormholes and the fate of black hole remnants},''
  \href{http://dx.doi.org/10.1088/1361-6382/ac2134}{{\em Class. Quant. Grav.}
  {\bfseries 38} no.~19, (2021) 194004},
  \href{http://arxiv.org/abs/2011.09444}{{\ttfamily arXiv:2011.09444
  [hep-th]}}.

\bibitem{giddings1988loss}
S.~B. Giddings and A.~Strominger, ``Loss of incoherence and determination of
  coupling constants in quantum gravity,'' {\em Nuclear Physics B} {\bfseries
  307} no.~4, (1988) 854--866.

\bibitem{COLEMAN1988867}
S.~Coleman, ``Black holes as red herrings: Topological fluctuations and the
  loss of quantum coherence,''
  \href{https://www.sciencedirect.com/science/article/pii/0550321388901101}{{\em
  Nuclear Physics B} {\bfseries 307} no.~4, (1988) 867--882}.

\bibitem{Das:1989fq}
S.~R. Das, A.~Dhar, A.~M. Sengupta, and S.~R. Wadia, ``{New Critical Behavior
  in $d=0$ Large $N$ Matrix Models},''
  \href{http://dx.doi.org/10.1142/S0217732390001165}{{\em Mod. Phys. Lett. A}
  {\bfseries 5} (1990) 1041--1056}.

\bibitem{Horowitz:2006mr}
G.~T. Horowitz and E.~Silverstein, ``{The Inside story: Quasilocal tachyons and
  black holes},'' \href{http://dx.doi.org/10.1103/PhysRevD.73.064016}{{\em
  Phys. Rev. D} {\bfseries 73} (2006) 064016},
  \href{http://arxiv.org/abs/hep-th/0601032}{{\ttfamily arXiv:hep-th/0601032}}.

\bibitem{Adams:2005rb}
A.~Adams, X.~Liu, J.~McGreevy, A.~Saltman, and E.~Silverstein, ``{Things fall
  apart: Topology change from winding tachyons},''
  \href{http://dx.doi.org/10.1088/1126-6708/2005/10/033}{{\em JHEP} {\bfseries
  10} (2005) 033}, \href{http://arxiv.org/abs/hep-th/0502021}{{\ttfamily
  arXiv:hep-th/0502021}}.

\end{thebibliography}\endgroup
\end{document}